\documentclass{aa}
\usepackage{multirow}
\usepackage{amsmath}
\usepackage[colorlinks,linkcolor=blue,anchorcolor=blue,citecolor=blue]{hyperref}
\usepackage{graphicx} 
\usepackage{lscape}
\usepackage{longtable}
\usepackage{txfonts}
\usepackage{natbib}

\begin{document}

\title{The Chemical Evolution of Galaxy Clusters: Dissecting the Iron Mass Budget of the
Intracluster Medium}

\author{
Ang Liu\inst{1,2,3},
Paolo Tozzi\inst{1},
Stefano Ettori\inst{4,5},
Sabrina De Grandi\inst{6},
Fabio Gastaldello\inst{7},
Piero Rosati\inst{8},
Colin Norman\inst{9}
}

\institute{
INAF - Osservatorio Astrofisico di Arcetri, Largo E. Fermi, I-50122 Firenze, Italy \\
\email{liuang@arcetri.astro.it; ptozzi@arcetri.astro.it}
\and
Department of Physics, Sapienza University of Rome, I-00185 Rome, Italy
\and
Department of Physics, University of Rome Tor Vergata, I-00133, Rome, Italy
\and
INAF - Osservatorio di Astrofisica e Scienza dello Spazio, via Pietro Gobetti 93/3, 40129 Bologna, Italy
\and
INFN, Sezione di Bologna, viale Berti Pichat 6/2, I-40127 Bologna, Italy
\and
INAF - Osservatorio Astronomico di Brera, Via E. Bianchi, 46, I-23807 Merate (LC), Italy
\and
INAF - Istituto di Astrofisica Spaziale e Fisica cosmica di Milano, Italy
\and
Dipartimento di Fisica e Scienze della Terra, Universit\'a degli Studi di Ferrara, via Saragat 1, I-44122 Ferrara, Italy
\and
Department of Physics and Astronomy, Johns Hopkins University, 3400 N. Charles Street, Baltimore, MD 21218, USA
}

\titlerunning{Dissecting the Iron Mass Budget of the
ICM}
 \authorrunning{Liu et al.}

\abstract
   {}
{We study the chemical evolution of galaxy clusters by measuring the iron mass in the ICM 
after dissecting the abundance profiles into different components. }
{We use {\sl Chandra} archival observations of 186 morphologically regular clusters in 
the redshift range [0.04, 1.07].  For each cluster we compute the azimuthally-averaged iron abundance 
and gas density profiles. In particular, we aim at identifying a central peak in the iron 
distribution, associated with the central galaxy, and an approximately constant plateau reaching 
the largest observed radii, possibly associated with early enrichment occurred before and/or 
shortly after the virialization of the cluster. We are able to firmly identify the two components in 
the iron distribution in a significant fraction of the sample, simply relying on the fit of 
the iron abundance profile.  From the abundance and ICM density profiles we compute the iron 
mass included in the iron peak and iron plateau, and the gas mass-weighted iron abundance of the ICM, 
out to an extraction radius of $0.4 \, r_{500}$ and, extending the abundance profile as a constant, 
to $r_{500}$.}
{We find that the iron plateau shows no evolution with redshift.
On the other hand, we find marginal ($<2\sigma$ c.l.) decrease with redshift in the 
iron mass included in the iron peak rescaled by the gas mass. We measure that the fraction of 
iron peak mass is typically a few percent ($\sim 1\%$) of the total iron mass within 
$r_{500}$. Therefore, since the total iron mass budget 
is dominated by the plateau, we find consistently that the global gas mass-weighted iron abundance does 
not evolve significantly across our sample. We are also able to reproduce past claims
of evolution in the global iron abundance, which turn out to be due to the use of cluster 
samples with different selection methods combined to the use of 
emission-weighted instead of gas mass-weighted abundance values.
Finally, while the intrinsic scatter in the iron plateau mass is consistent with zero, 
the iron peak mass exhibits a large scatter, in line with the fact that the 
peak is produced after the virialization of the halo and depends on the formation history 
of the hosting cool core and the strength of the associated feedback processes.}
{We conclude that only a spatially-resolved approach can resolve the issue of the 
iron abundance evolution in the ICM, reconciling the contradictory results 
obtained in the last ten years.  Evolutionary effects below $z\sim 1$ are marginally 
measurable with present-day data, while at $z>1$ the constraints are severely limited by 
the poor knowledge of the high-$z$ cluster population. 
The path towards a full and comprehensive chemical history of 
the ICM necessarily requires the use of high-angular resolution X-ray bolometers 
and a dramatic increase in the statistics of faint, extended X-ray sources.}


\maketitle

\section{Introduction}

Massive galaxy clusters ($M_{500}>10^{14} M_\odot$) are considered as closed boxes that 
retain the past history of their cosmic evolution. The majority of their total mass is in the 
form of dark matter, which contributes 80--90\% of the mass budget. While the stellar mass in 
member galaxies or in a diffuse component only constitutes a minor faction \citep[about 
1--2\% of the total and 6--12\% of the baryonic mass, see][]{2012Lin},  
the baryonic mass is dominated by the intracluster medium (ICM), which is a hot, 
optically-thin diffuse plasma at low densities, in a local collisional equilibrium with 
temperatures of the order of 10$^7$ to 10$^8$ K.  The thermodynamical and dynamical status 
of the ICM is non-trivially linked to the mass accretion history of the dark matter halo, 
the nuclear feedback from the central galaxy, and the star formation processes in the member 
galaxies.  The latter, in particular, leaves its imprint in the ICM as a widespread chemical 
enrichment by heavy elements, mostly produced by supernovae explosions 
in the member galaxies, that can be efficiently measured with X-ray spectroscopy
\citep{2010_Bohringer,mernier2018}, as also supported by simulations 
\citep[see][and references therein]{biffi2018}. Tracing the evolution of metal abundance 
in the ICM can therefore provide useful information to reveal the star formation history 
in cluster galaxies across cosmic time and the process of mixing of the intergalactic medium (IGM) 
with the ICM \citep{2004Boringer,2013dePlaa}.

The abundance of heavy elements (also generically referred as ``metals'') in the ICM can be 
measured through the equivalent width of their emission lines in the X-ray spectrum. In particular, 
iron is the element with the most prominent emission features, and it is therefore the only 
heavy element that has been detected in galaxy clusters up to $z\sim 1.6$ and possibly up to 
$z\sim 2$ \citep{rosati2009,tozzi2013,2015Tozzi,mantz2018} 
thanks to the $K_{\alpha}$  emission line complex 
at 6.7--6.9 keV. The detection of other metals, instead, typically requires high S/N spectra and, 
therefore, is basically limited to lower temperatures ($kT<3$ keV), low redshifts, and central regions 
\citep{degrandi2009,tamura2009,mernier2017}. In this framework, iron is the only element that can 
be robustly used to investigate the spatial distribution in the ICM and the cosmic evolution 
of metals on a timescale of $\sim 10$ Gyr.

Several attempts have been made in the past decades to derive an average cosmic evolution of iron 
abundance in the ICM. After the first attempts \citep[e.g.,][]{mushotzky1997,tozzi2003}, about
ten years ago the first reliable assessment of the cosmic evolution of iron abundance in the 
ICM has been obtained thanks to the exploitation of {\sl Chandra} and XMM-{\sl Newton} archives.  
These works suggested a statistically significant evolution of a factor of 2 in the redshift 
range $0 < z < 1.3$ \citep{2007Balestra,maughan2008,anderson2009}. 
The picture became less clear in recent years, when new analysis showed little or no 
evolution \citep{2015Ettori,mcdonald2016}. In addition, spatially resolved analysis adds further
complications: the results are not only influenced by the radial range used to measure the 
abundance \citep{2012Baldi,2017Mantz}, but also change significantly when using SZ-selected
samples of clusters, instead of the former X-ray selected clusters 
\citep[see][for example]{mcdonald2016}. Moreover, several works have shown that the spatial 
distribution of iron in the central regions evolves significantly with time 
\citep{degrandi2014}, despite this does not necessarily imply a change in the amount of metals in the ICM, but rather a simple redistribution  
\citep{liu2018}. As a consequence, the measurement of iron abundance without 
resolving its spatial distribution can potentially introduce systematic uncertainties as high as 
$\sim$25\% \citep{liu2018}. A further critical aspect is that very little is known on the
distribution of metals at large radii, so that statistical studies are meaningful only 
for radii below $r_{500}$ \citep{2016Molendi}.

We argue that, in order to reach a more clear picture of the evolution of iron in the ICM on 
the basis of current X-ray data archives, we should most efficiently exploit what we know about the 
iron distribution. Both simulations and observations have indicated that the spatial distribution 
of iron in the ICM often appears to be well described as a combination of two main components: a 
peak in the inner regions which is usually centered on the brightest cluster galaxy (BCG), and a 
large-scale component with approximately uniform
\citep{degrandi2001,baldi2007,leccardi2008,sun2009,simionescu2009,
2013Werner,2016tholken,urban2017,Simionescu2017,lovisari2019} 
or slightly decreasing \citep[e.g.,][]{mernier2017,2018biffi} abundance across the cluster.
The BCG is thought to be largely responsible of the iron peak \citep[associated with either 
Type Ia supernovae newly formed in the BCG or/and stellar mass loss in the BCG, 
see][]{degrandi2004,2004Boringer}, while multiple processes, including 
AGN outflow, gas turbulence, galactic winds, ram pressure stripping etc., extract the metal rich
IGM from the member galaxies across the entire lifetime of the cluster, and 
leave their imprints on the distribution of iron in ICM particularly in the densest, 
central regions \citep{kirkpatrick2009,simionescu2009,liu2018}. 
Another minor, but interesting 
component, is a characteristic drop of the iron abundance in the very center, which is associated
both to the mechanical feedback from the AGN and to the iron depletion associated to recent 
star formation events occurring in the BCG \citep[see][]{panagoulia2015,lakhchaura2019,2019Liu}.  
The almost uniform 
large-scale iron plateau, with a typical abundance of $\sim 1/3\, Z_\odot$, is expected to come 
from early star formation in the member galaxies around cosmic noon ($z>2$), therefore, 
before the virialization of the cluster itself \citep[see][]{2017Mantz}. 

In this work, we reconsider the cosmic evolution of iron in the ICM, by performing 
spatially-resolved spectroscopic analysis on a large sample of high-quality {\sl Chandra} 
data to fit the 
iron profile with a double-component model (an iron peak and a plateau), and investigate 
the evolution of the gas mass-weighted iron abundance separately in each component. The paper is 
organized as follows. In Section 2, we describe the selection of cluster sample, and the reduction 
of {\sl Chandra} data.  In Section 3, we investigate the global 
properties of the clusters and the azimuthally-averaged profiles of density, iron abundance, and, 
therefore, iron mass. In Section 4, we discuss the results of our analysis. Our conclusions 
are summarized in Section 5.  Throughout this paper, we adopt the seven-year WMAP cosmology 
with $\Omega_{\Lambda} =0.73 $, $\Omega_m =0.27$, and $H_0 = 70.4$ km s$^{-1}$ Mpc$^{-1}$ 
\citep{2011Komatsu}. Quoted error bars correspond to a 1 $\sigma$ confidence level,
unless noted otherwise.

\section{Sample selection and data reduction}

\begin{figure*}
\begin{center}
\includegraphics[width=0.49\textwidth, trim=30 120 30 130, clip]{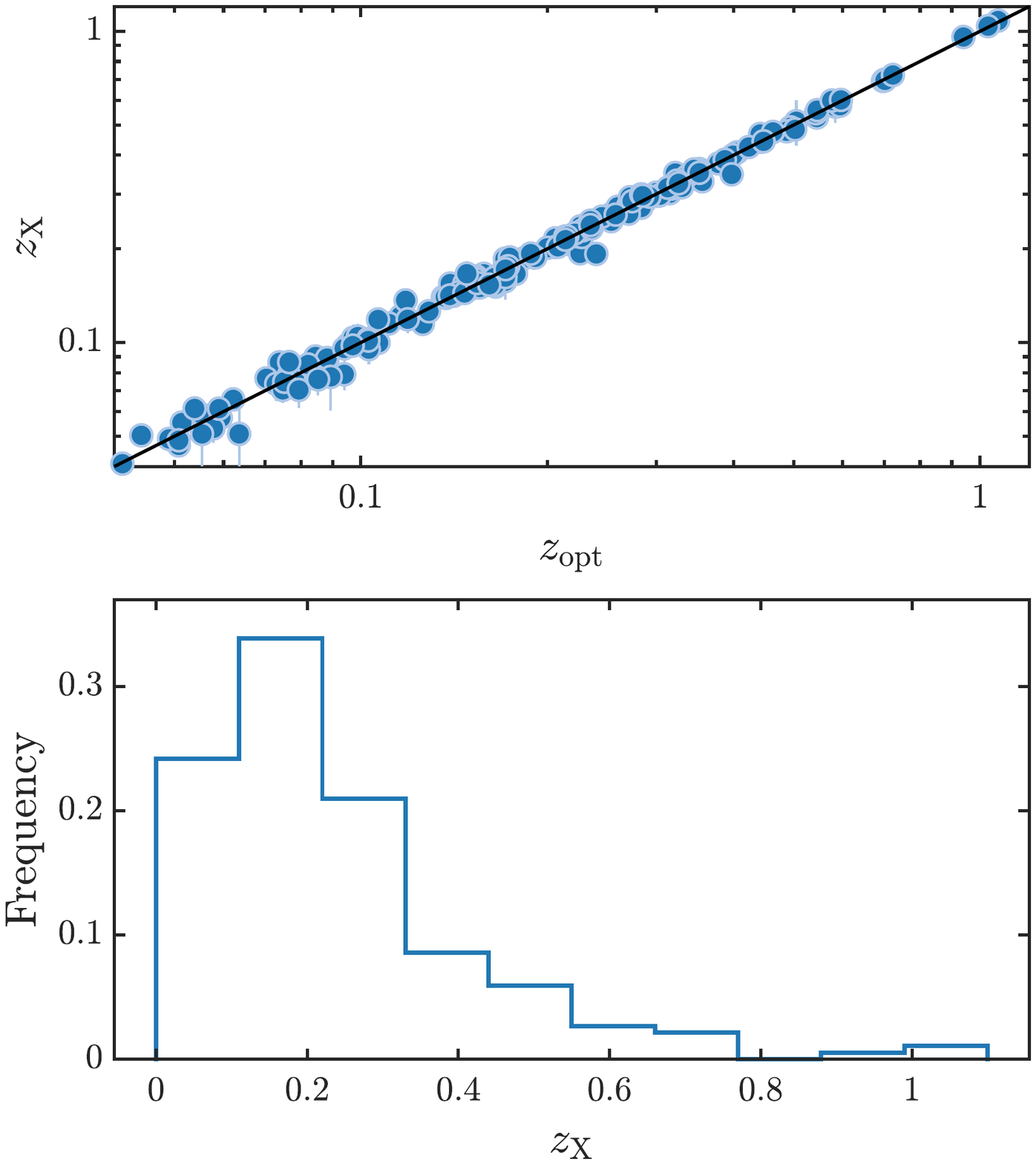}
\includegraphics[width=0.49\textwidth, trim=30 120 30 130, clip]{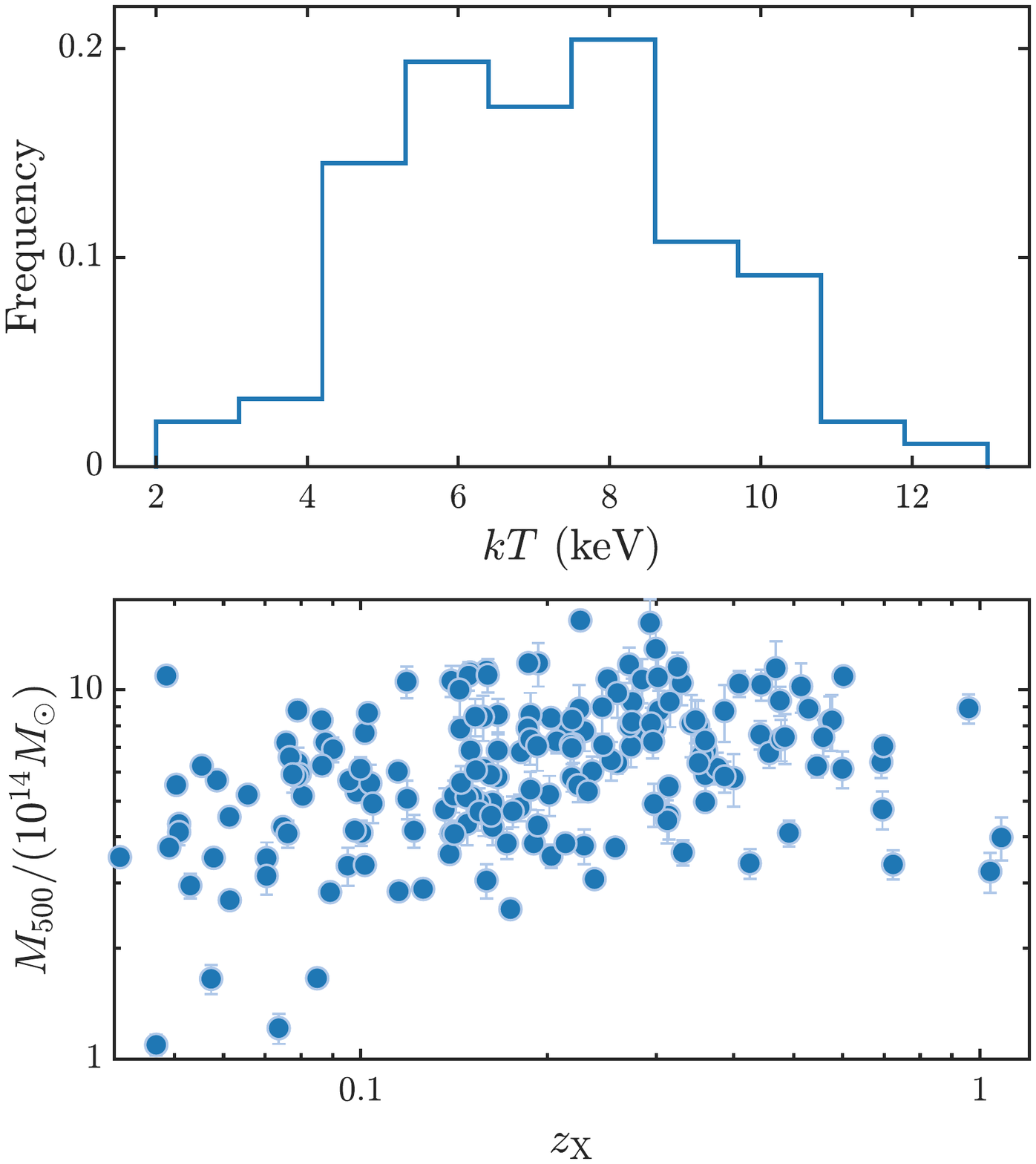}
\caption{General properties of the cluster sample used in this work (186 clusters).  
{\sl Upper left}: X-ray redshift measured in this work compared to the optical redshift 
from the literature. {\sl Lower left}: Distribution of the redshift of the clusters in our sample. 
{\sl Upper right}: Distribution of the emission-weighted (spectroscopic-like) temperature in 
the radial range [0.1, 0.4]$r_{500}$ across the sample. {\sl Lower right}: Distribution 
of $M_{500}$, estimated according to equation 3, plotted versus redshift.}
\label{sample}
\end{center}
\end{figure*}

\subsection{Sample selection}
We start from a complete list of galaxy clusters with public {\sl Chandra} archival observations 
as of February 2019.  Our aim is to resolve the abundance profile and disentangle its spatial 
components under the assumption 
of spherical symmetry, within the largest radius that still allows a robust spectral analysis.
Clearly, the requirement on spherical symmetry puts a strong constraint on the morphology of 
clusters suitable for our analysis.  We select our final sample of clusters on the basis 
of the following criteria. 

First, we require our extraction radius $R_{\rm ext}$ of the iron abundance profiles to be 
entirely covered by the field of view 
of the {\sl Chandra} data. The adopted minimum value for $R_{\rm ext}$ is needed to sample ICM 
regions far enough from the central peak, in order to measure independently the  
large scale plateau. Several studies have shown that $\sim 0.4\, r_{500}$, or $\sim 0.25 r_{200}$, 
is typically well beyond the extension of the iron peak, and reaches the iron plateau 
\citep{urban2017,lovisari2019}. We also find that setting $R_{\rm ext} = 0.4\, r_{500}$ allows a robust
spectral analysis for the large majority of the clusters in our sample.  While for some of them
it would be possible to extend the measurement of the abundance profile out to [0.5--0.6] $r_{500}$, 
this would have a minor impact on the final profile given the large error in the outermost bin. 

On the other hand, we remark that the electron density can be measured out to $r_{500}$ for the large majority of the clusters, 
allowing a proper constraint on the gas mass within $r_{500}$.  Since we are ultimately interested
in the average gas mass-weighted abundance obtained as the ratio of iron mass and total gas mass
within a given radius, and considering that we assume a constant plateau for the abundance at
large radii, we can express our results in terms of gas mass-weighted quantities within $r_{500}$. 
The values of $r_{500}$ used for sample selection are obtained from literature 
\citep{bohringer2007,piffaretti2011}, or estimated from scaling relations 
\citep[e.g.,][]{vikhlinin2006a}. Most of the nearby clusters at
$z<0.05$ are excluded when we apply the criterion on extraction radius.

Second, to produce iron abundance profiles with acceptable quality, we require a number of net 
counts $\ge 5000$ in the 0.5--7 keV energy band and within the extraction radius. This requirement
is needed in order to have at least six independent annuli with more than $\sim 800$ net counts each. 

Third, since we necessarily assume spherical symmetry when deprojecting the azimuthally 
averaged profiles, clusters with clear signatures of non-equilibrium, such as an irregular 
morphology and obvious substructures or mergers (some well known cases are 1E0657-56, 
Abell520, Abell3667), should not be included in the sample. 
Major mergers are observed to affect mostly the inner regions, while at large radii 
often shows a rather flat abundance distribution, similar to relaxed clusters \citep{2019Urdampilleta}.  There are many morphological parameters that can be used
to determine whether a cluster is regular or not, such as the X-ray surface brightness 
concentration \citep{santos2008,cassano2010}, the power ratio \citep{buote1995,buote1996}, 
and the centroid shift \citep{ohara2006,cassano2010,lovisari2017}. 
In this work we adopt the centroid shift parameter, which measures the variance of the 
separations between the X-ray peak and the centroids of emission obtained within a number of 
apertures of different radii:
\begin{equation}
    w = \frac{1}{R_{\rm max}} \times \sqrt{\frac{\sum (\Delta_i - \bar{\Delta})^2}{N-1}},
    \label{centroid}
\end{equation}
where $R_{\rm max}$ is set as $[0.3-1]\,r_{500}$, $N$ is the total number of apertures 
within $R_{\rm max}$, $\Delta_i$ is the separation of the X-ray peak and the centroid computed within 
the $i_{\rm th}$ aperture. The definition of the parameter varies slightly across the literature. 
In this work, we set $R_{\rm max}$ to 0.4$r_{500}$, and the number of apertures to 10. 
The boundary between regular and disturbed clusters adopted in the literature ranges from 0.01 to 0.02 
\citep[see][for example]{ohara2006,cassano2010}. Here we use a relatively loose criterion: 
$w < 0.025$, so that only the most disturbed targets are excluded at this step. Then, we check 
visually the X-ray image of all the clusters that satisfy the centroid-shift criterion
to further identify clusters with a clearly disturbed morphology. 

We note that in this way we are not able to identify major mergers along the line of sight. 
This aspect may be investigated through the redshift distribution of member galaxies, however, 
this goes beyond the goal of this paper.  In addition, unnoticed major merger are mostly 
caught before the first collisions, since they are expected to leave visible feature also 
in the plane of the sky \citep[as expected in the case of a bullet-like cluster seen along 
the line of sight, see][]{2015Liu}.  Therefore, we conclude that the presence of major mergers in 
our final sample is not significant.

Starting from a total of $\sim 500$ targets in the {\sl Chandra} data archive, the sample reduces 
by $\sim50\%$ with the first and second criteria. After the morphology criterion and final check, 
we obtain a final sample consisting of 186 clusters, spreading over a redshift range $0.04<z<1.07$, 
with the bulk of the clusters in the range $0.04<z<0.6$.
We remind that, since the sample is selected from the {\sl Chandra} archive, rather than any 
existing flux-limited or volume-limited catalogs, it has no completeness in mass, or luminosity, 
etc. This aspect may constitute a limitations
for the investigation of the cosmic evolution in the enrichment of the ICM.  
In particular, the requirement on the morphology, with the resulting exclusion of clusters
which experienced recent mergers, would unavoidable alter any selection based on mass or
luminosity. However, the large sample analyzed with a uniform approach is optimal for our main 
scientific goal of identifying potential differences in the evolution of the two components
in the iron distribution, namely the iron peak and plateau.  Possible strategies to 
improve on the sample size and selection will be discussed in Section 4. 

\subsection{Data reduction}
\label{reduction}

Data reduction is performed with {\tt CIAO 4.10}, with the latest release of the {\sl Chandra} 
Calibration Database at the time of writing {\tt (CALDB 4.7.8)}. Unresolved sources within the 
ICM are identified with {\tt wavdetect}, checked visually, and eventually removed. Time
intervals with high background are filtered by performing a 3$\sigma$ clipping of the 
background level. The light curves are extracted in the 2.3--7.3 keV band, and binned with a time interval of 200 s. For clusters with 
multiple observations, we extract the spectrum and 
compute the ancillary response file (ARF) and redistribution matrix file (RMF) for each observation 
separately with the command {\tt mkarf} and {\tt mkacisrmf} (for several observations with the 
temperature of the focal plane equal to -110 K we use {\tt mkrmf} instead). Due to the large extent of the sources and our goal of measuring the low-surface brightness of the 
ICM out to $\sim r_{500}$, the background 
spectrum is extracted from the `blank sky' files, and processed using the 
{\tt blanksky} script (default options have been used with {\tt weight\_method}
``particle'' and {\tt bkgparams=[energy=9000:12000]}). Whenever possible, we also repeat our analysis 
using the 
local background, generated by directly extracting the data from a source-free region on the 
same CCD chip. We confirm that the fitting results using this two backgrounds are in  
good agreement. 

The spectral fits in this work are performed with {\tt Xspec 12.10.1} \citep{1996Arnaud} 
using C-statistics \citep{cash1979}. The AtomDB version is 3.0.9. All the abundance values 
in this paper are relative to the solar values of \citet{asplund2009}. To measure the iron abundance in a projected annulus, the emission of the 
ICM within this annulus is fitted with a double-{\tt vapec} thermal plasma emission 
model \citep{smith2001} for a better fit to the multiple-temperature structure \citep[see][]{kaastra2004}. It has also been shown that the use of two temperatures is 
sufficient to remove the systematics associated to the thermal structure of the ICM, while 
the inclusion of more thermal components do not provide significant improvements
\citep{2016Molendi}.  The metal 
abundances of the two {\tt vapec} components are linked. The abundances of O, Ne, Mg, and Al, which are mostly ejected by core-collapse supernovae, are independent from 
the Fe abundance and linked together. Other prominent metals are linked to Fe, while the abundance 
of He is always fixed to solar value. Due to the high temperature of the clusters in our 
sample and the relatively low S/N of the data, in most of the cases we are not able to obtain 
constraints on the abundance of the elements produced by core-collapse supernovae, at least
not at a confidence level comparable to that of the iron abundance.  For this reason, we do not 
discuss metals other than iron in this paper. Galactic hydrogen absorption is described by the
model {\tt phabs} \citep{phabs1992}, where the Galactic column density $n_{\rm H}$ at the cluster 
position is initially set as $n_{\rm H,tot}$ from \citet{2013Willingale}, which takes into 
account not only the neutral hydrogen, but also the molecular and ionized hydrogen that may 
bias the spectral fitting if not considered properly \citep{lovisari2019}. When fitting the 
global emission, we set the $n_{\rm H}$ free to vary below a very loose upper limit 
at 10$\times n_{\rm H,tot}$, and measure the best-fit 
$n_{\rm H,free}$. This value is then adopted as the input $n_{\rm H}$ in further 
spatially-resolved analysis, but it is allowed to fluctuate within its 1$\sigma$ 
statistical confidence interval, or $\pm50\%$ if its uncertainty is lower than 50\%. 
We will discuss the impact of the $n_{\rm H}$ value on our results in Section 4.

To determine the X-ray center of each cluster, we smooth the 0.5--7~keV image of the 
extended emission (after removing point sources) with a Gaussian kernel with FWHM = 3$\arcsec$, 
and find the position of the brightest pixel. This is a very quick and efficient method to identify 
the X-ray centroid for relaxed, cool-core clusters.  In the case of a very low surface 
brightness also in the central regions, a more robust method is to perform a 0.5--7 keV 
band photometry within a circle with a fixed radius (typically $\sim 40 $ kpc), and choose 
the position that maximize the net counts. Clearly, having removed the clusters with 
irregular morphology, any change in the X-ray centroid within the uncertainties has a 
negligible impact on the final results.

\section{Imaging and Spectral Analysis}

\subsection{Global properties: redshift, temperature, $r_{500}$, and concentration}

\begin{figure}
\begin{center}
\includegraphics[width=0.49\textwidth, trim=13 50 10 70, clip]{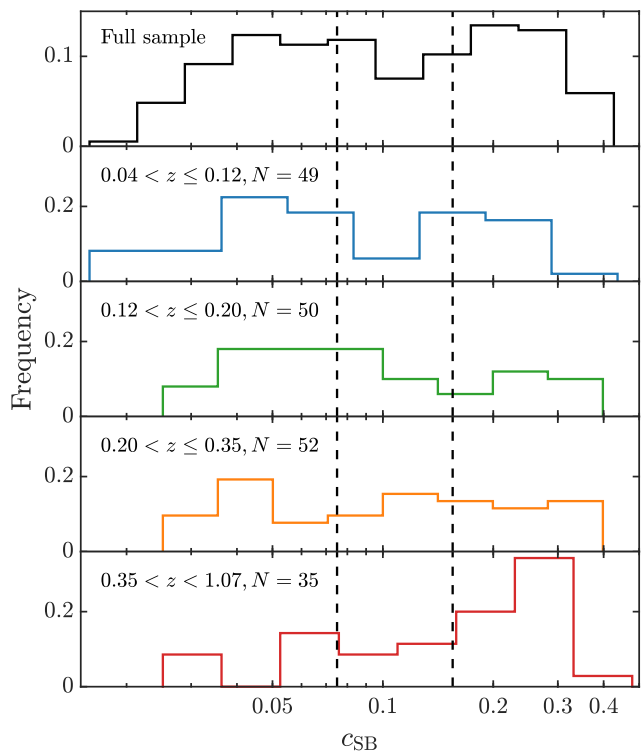}
\caption{Distribution of the surface brightness concentration $c_{\rm SB}$ of the clusters in the full sample and in 4 independent redshift bins with roughly the same number of clusters ($N$).
The vertical dashed lines indicate the threshold for non-cool-core and weak cool-core clusters: 
$c_{\rm SB} = 0.075$ and $c_{\rm SB} = 0.155$. }
\label{csb}
\end{center}
\end{figure}

We first derive the global properties of the clusters, including the X-ray redshift, the global 
temperature, the value of $r_{500}$ and $M_{500}$. The X-ray redshift is measured by fitting the 
spectrum of the global emission within the radius maximizing the signal to noise ratio in 
the 0.5--7 keV band image. Among the 186 clusters in our sample, 184 have optical spectroscopic 
redshifts published in the literature. In Figure \ref{sample} (top-left panel) 
we compare the X-ray and optical redshifts, and find that the {\sl rms} of 
$(z_{\rm X}-z_{\rm opt})$ is slightly lower than the average statistical uncertainty 
in the redshift measurements, implying a good agreement  between $z_{\rm X}$ and $z_{\rm opt}$.
Therefore, we fix the redshift at the 
best-fit X-ray value in the following analysis. In some clusters, the difference in X-ray and 
optical redshifts may slightly influence the measurement of abundance, but this influence is 
rather small and negligible for our purpose \citep[e.g.,][]{liu2018}. In general, our sample
contains a large fraction of low redshift clusters, with $\sim$70\% clusters at $z<0.3$, 
and only less than ten clusters at $z>0.6$ (see  bottom-left panel of Figure \ref{sample}). 
This is mostly due to the requirement on the minimum number of net counts.

The global temperature $\langle kT\rangle$ \citep[or ``spectroscopic-like'' temperature, 
see][]{2004Mazzotta} is measured by fitting the spectrum extracted in the region
$0.1 r_{500} < r < 0.4r_{500}$, a choice that is often adopted to obtain temperature values that 
more closely trace an ideal virial value, avoiding the effect of the cool
core, when present. We use a single-temperature {\tt apec} model, therefore $\langle kT\rangle$ is an
emission-weighted value resulting from the range of temperatures present in the explored radial range.
We find that the values of  $\langle kT\rangle$ range from 4 to 12 keV, with a minority of clusters
with $\langle kT\rangle< 4$ keV (see top-right panel of Figure \ref{sample}). 

To estimate $r_{500}$, we use the average relation described in \citet{vikhlinin2006a}, which has 
been widely adopted in literature \citep{2012Baldi,liu2018,2019Mernier}:
\begin{equation}
 r_{500} = \frac{0.792}{hE(z)} \left( \frac{\langle kT \rangle}{5~{\rm KeV}} \right)^{0.527} {\rm Mpc},
\label{r500}
\end{equation}
where $E(z) = (\Omega_{\rm m}(1+z)^{3}+\Omega_{\rm \Lambda})^{0.5} $. The global temperature
$\langle kT \rangle$ and $r_{500}$ are evaluated iteratively until converged. The total mass 
within $r_{500}$ is also estimated from the scaling relation in \citet{vikhlinin2006a}, or, 
equivalently, can be written as:
\begin{equation}
 M_{500} = \frac{4\pi}{3}r_{500}^3\cdot 500\rho_{\rm c}(z),
\label{m500}
\end{equation}
where $\rho_{\rm c}(z)=3H^2(z)/8\pi G$ is the critical density at cluster's redshift. 
Our sample spans a mass range of [1, 16] $\times 10^{14}M_{\odot}$, with
only four clusters with  $M_{500} < 2 \times 10^{14}M_{\odot}$ (see bottom-right panel of
Figure \ref{sample}).  We also note {\sl a posteriori} that $r_{500}$ is within the 
ACIS-I or ACIS-S\footnote{All the clusters observed with ACIS-S only have $z>0.08$.} 
field of view except for ten clusters, where the field of view covers 
only a radius of $\sim 0.6 \, r_{500}$. In this cases the ICM density profile up to $r_{500}$ is 
obtained by extrapolating the profile beyond $0.6\, r_{500}$, an approximation that may not be 
extremely accurate but it may introduce only a few percent uncertainties in less than 5\% of our sample, 
well below the statistical errors.

Since cool-core and non-cool-core clusters are significantly different in both the 
abundance and spatial distribution of iron in the ICM, we estimate the fraction of 
cool-core clusters in our sample with the surface brightness concentration 
$c_{\rm SB}$ \citep{santos2008,santos2010}, defined as the ratio of the fluxes observed
within 40 kpc and 400 kpc:
\begin{equation}
    \centering
    c_{\rm SB} \equiv \frac{{\rm S}(40 {\rm kpc})}{{\rm S}(400 {\rm kpc})}.
    \label{eq_csb}
\end{equation}
The fluxes in equation \ref{eq_csb} are computed in the 0.5--2 keV band, 
and are estimated directly from the net count rate after considering the 
``beheading effect'', due to the $K$-correction that depends on redshift and the minimum
temperature observed in the core \citep[see][for more details]{santos2010}.
We remark that the concentration parameter is a simple and reliable parameter to classify the
cool-core strength, which is, in reality, a definition that involve complex physics 
\citep{hudson2010}. A bimodal distribution can be seen in the top panel of Figure \ref{csb}, 
which reflects the bimodality of cool-core and non-cool-core clusters as already
investigated in other properties like pseudo-entropy
\citep[see][for example]{Sanderson2009,hudson2010}. Clearly, a more robust 
classification of cool-core and non-cool-core clusters should rely on more 
diagnostics, e.g., central cooling time, temperature drop, etc. However,
since this is not the main focus of this paper, we will not make further analysis on 
the cool-core properties of the clusters, but merely investigate the global fraction of 
cool cores in our sample. Using $c_{\rm SB} < 0.075$ and $c_{\rm SB} > 0.155$ as the 
thresholds between non-cool-core/weak cool-core, and weak/strong cool-core clusters, 
respectively \citep[see][]{santos2008}, we find that 72 clusters in
our sample are non-cool-core clusters, while 46 and 68 are weak- and  
strong- cool-core clusters. These numbers correspond to a percentage of 
38.7\%, 24.7\% and 36.6\% of non-cool-core, weak-cool-core and strong-cool-core clusters, 
respectively.

Interestingly, the balance between cool-core and non-cool-core clusters in our sample is 
redshift dependent.  In the lower panels of Figure \ref{csb} we show how the bimodality disappears
at $z>0.2$, while the cool-core clusters become dominant in the range $z>0.35$.  Given 
the coarse redshift binning, this is not in contradiction with previous claims on the 
dearth of cool-core clusters at $z>0.7$ \citep[see][]{santos2008}, considering that we
have only 7 clusters at $z>0.7$.  In addition, we note that the requirement on the S/N 
slightly favors CC clusters as the redshift increases.  Therefore, no claim can be made on the
evolution of cool cores with cosmic time with the current sample.

Overall, the fraction of $\sim 61$\% of clusters hosting a cool core in our sample, is in line with 
what is usually found in X-ray selected samples, 
such as MACS, where \citet{rossetti2017} found a cool-core fraction of $(59\pm 5)$\%, but is 
significantly higher than the fraction found in SZ selected samples 
\citep[$\sim$30\% for $\sl Planck$ clusters as found in ][]{rossetti2017}. This discrepancy, 
which is robust against differences in the detailed definition of cool-core, is the 
well known ``cool core bias'' \citep[e.g.,][]{eckert2011,2017Andrade}, and may affect
the overall thermal and chemical properties of a sample. 
In general, our sample shares the same core properties as other X-ray selected samples, despite 
it includes a sizeable fraction of SZ-selected clusters. 

\subsection{Azimuthally-averaged profiles of electron density, iron abundance, and iron mass}

We now measure the azimuthally-averaged profiles of gas density and iron abundance, and 
consequently the iron mass cumulative profile. While accurate deprojection is always mandatory 
for density profiles, we choose to use only the projected profiles for iron abundance. The 
reason for this choice is twofold. First, since the typical metallicity variation across a 
cluster is usually smaller than a factor of $\sim 3$ ($\sim Z_{\odot}$ at the iron peak to 
$\sim Z_{\odot}/3$ in the outskirts), projection effect has an actually mild impact on the 
measured abundance in most of the cases. Second, deprojection on metallicity usually requires much 
more photons but results in a much larger error in single measurement. If the cluster deviates 
from perfect spherical symmetry, which is in fact very common, deprojection induces extra 
uncertainty, which can not be properly assessed. For these reasons we adopt deprojected profiles 
of density, and projected profiles of iron abundance, a procedure that is commonly adopted
in recent papers dealing with ICM abundance \citep[see, e.g.,][]{mernier2017,lovisari2019}. 
The potential impact of this assumption is discussed in Section 4.

Each iron abundance profile contains 6--13 radial bins out to the extraction radius 
$R_{\rm ext}\sim 0.4 \, r_{500}$, 
roughly corresponding to $\sim 0.25\, r_{200}$. As previously discussed, this extraction radius 
is chosen on the basis of the expected iron plateau in most of the clusters, which is typically
reached at these radii \citep[see][]{urban2017}. The inner and outer radii are adjusted to ensure 
that each bin encloses similar number of net photons. The minimum net photons in 0.5--7 keV 
energy band within each bin is 800, and can reach $>$ 20000 in some bright clusters with 
very deep observations. The spectrum of each bin is fitted with a double-{\tt vapec} model, with independent temperatures and linked abundance, as described in Section 2. 

With this modelization, we can efficiently remove the bias on the best-fit abundance value when the temperature gradient 
is significant within the spatial bin \citep[see][]{2016Molendi}, especially in the center of cool-core 
clusters. In fact, in most cases we find no significant difference between the iron abundance 
obtained by fitting with a double {\tt vapec} model and a single {\tt vapec} or {\tt apec} model. 
The use of double-temperature has little or no impact on metallicity outside the cool core. 
Despite this, for simplicity we do not change the spectral-fitting strategy with radius, 
and we use a metallicity-linked double-{\tt vapec} model to fit the spectral both within and 
outside the cool core. On the other hand, no attempt is made of considering different abundance 
values associated with different gas phase within a projected bin, 
since it is not possible to investigate such an effect with present-day data. 
In fact, a relevant effect would be given by correlated fluctuations in the ICM density and 
abundance on small ($\sim $ kpc) scale, 
an occurrence which has been never observed and is not expected. The only exception is given by the
galactic coronae around BCGs \citep[see][]{2001Vikhlinin} and 
presence of low-surface brightness infalling clumps at large radii, which has been
treated in dedicated works and is not expected to affect radii smaller than 
$r_{500}$ \citep[see][]{2015Eckert}. Therefore, we conclude that the
assumption of a constant abundance in each projected bin is accurate for our science
goals, and it provides a robust description of the actual azymuthally-averaged abundance profile.

\begin{figure}
\begin{center}
\includegraphics[width=0.49\textwidth, trim=15 125 20 135, clip]{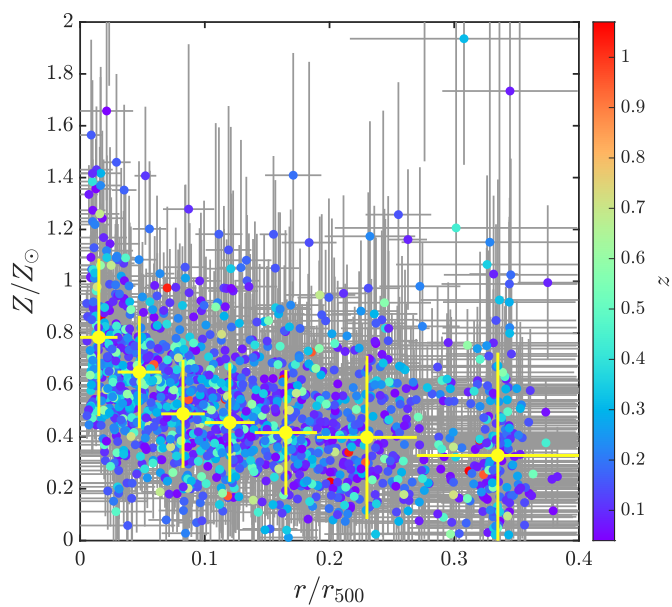}
\caption{Projected iron abundance profiles for all the clusters in our sample. Each point, 
color coded by redshift, is the best-fit value in the corresponding radial bin.  For 
clarity error bars are shown in light grey.  The yellow points show the sample-average 
abundance and {\sl rms} within seven radial bins. }
\label{abun}
\end{center}
\end{figure}

The projected iron abundance profiles of all the clusters in our 
sample are plotted in Figure \ref{abun}, where the yellow points show the sample-average in 
seven radial bins. We confirm that, on average, the iron peak appears at radii $<0.1 r_{500}$, 
while a plateau, or a very weakly decreasing profile, is evident at radii $>0.2 r_{500}$.

We fit the measured iron abundance profiles with a double-component model. The first component
is a $\beta$ model to fit the iron peak, while the second component is a constant representing
the iron plateau:
\begin{equation}
    \centering
    Z = Z_{\rm peak}\cdot\left[1+\left(\frac{r}{r_0}\right)^2\right]^{-\alpha}+Z_{\rm plateau}.
    \label{abun_model}
\end{equation}

\noindent
The central drop component is not considered if not in the few cases where the innermost 2--3 
bins are significantly lower than the outer bins. In this way we do not force this component 
to be used when the statistical significance is low. In fact, a systematic study of the
iron drop is feasible only for nearby clusters \citep{2019Liu}, while a search 
throughout our sample would be dominated by noise. Despite this, the few cases
where a central drop improves significantly the fit are discussed in Section 3.4.

\begin{figure*}
\begin{center}
\includegraphics[width=0.49\textwidth, trim=5 135 30 145, clip]{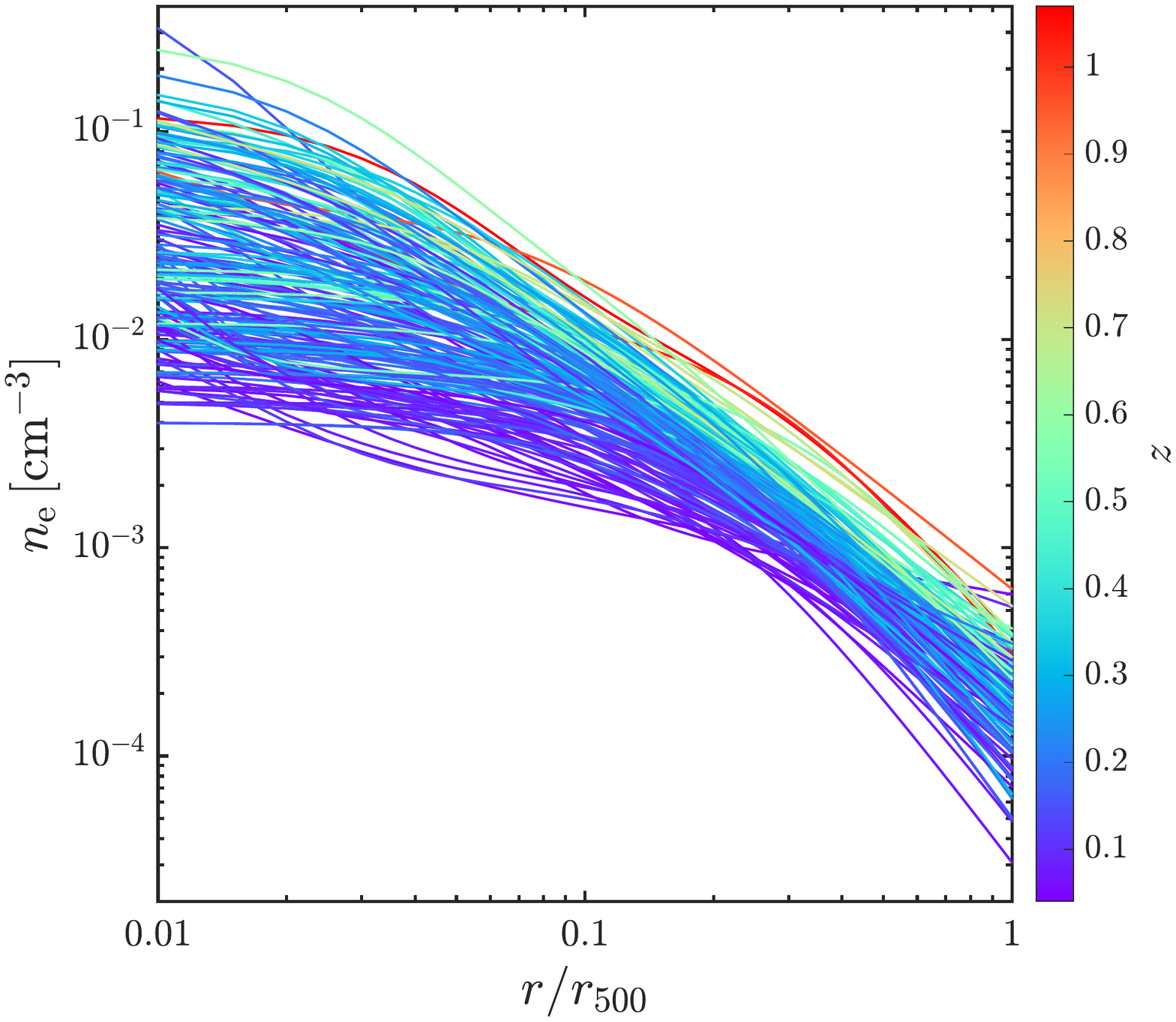}
\includegraphics[width=0.49\textwidth, trim=5 135 30 145, clip]{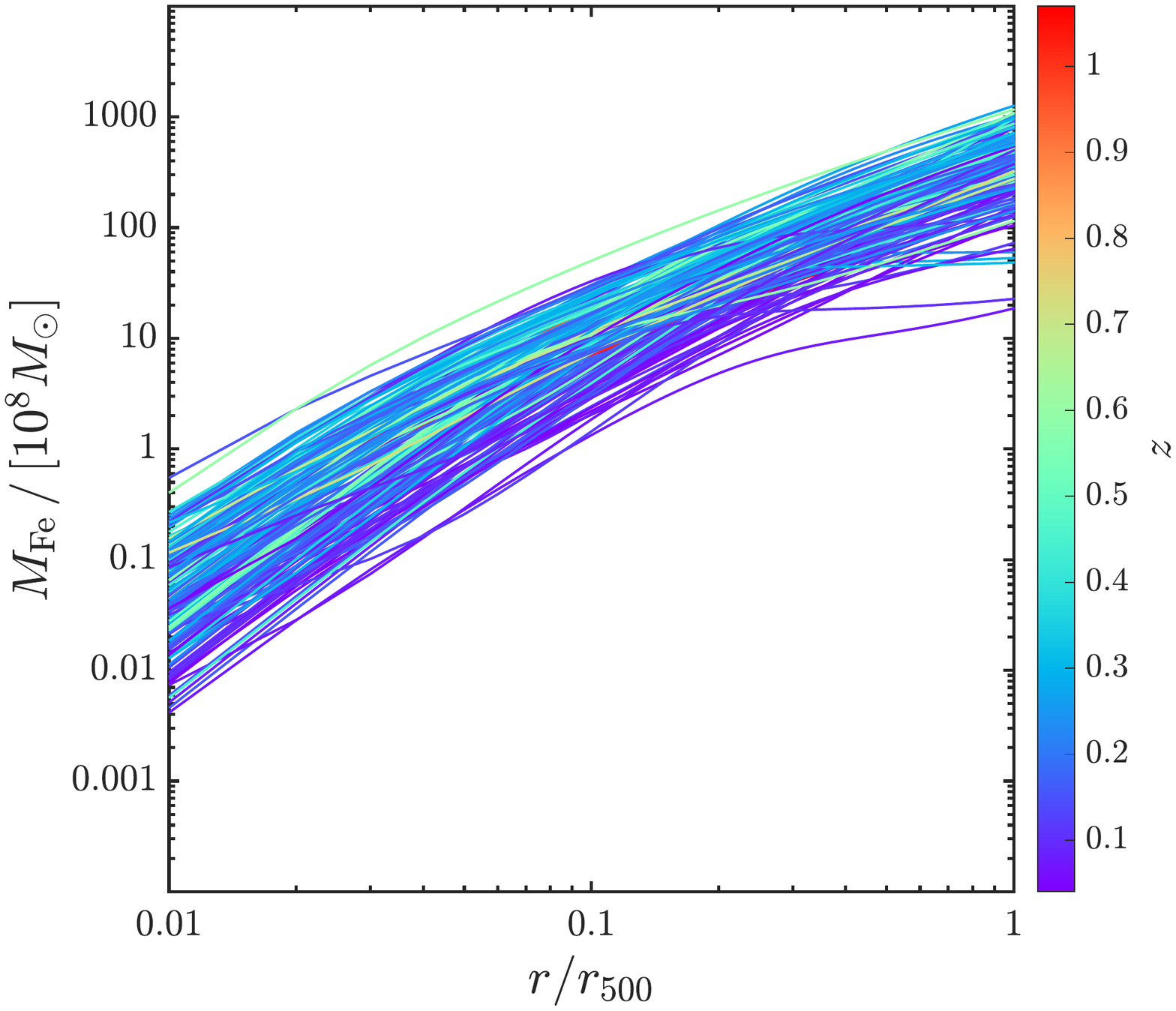}
\caption{{\sl Left panel:} best-fit double-$\beta$ model of the deprojected density profiles, 
color-coded by redshift, for all the clusters in our sample. {\sl Right panel:} total iron 
mass profiles obtained by convolving the gas density profile with the best fit abundance profile
for each cluster in our sample. At $r>0.4\, r_{500}$ the iron mass is computed
extrapolating the constant plateau up to $r_{500}$.
 }
\label{ne_femass}
\end{center}
\end{figure*}

For the electron density profiles, we extend the maximum extraction 
radius to $\sim r_{500}$, and adopt a lower criterion of net photons in each bin in order to 
increase the spatial resolution. The spectrum in each bin is deprojected using the 
{\tt dsdeproj}\footnote{http://www-xray.ast.cam.ac.uk/papers/dsdeproj/} routine 
\citep{sanders2007,2008Russell}, which deprojects a spectrum directly by subtracting the 
geometrically rescaled count rates of the foreground and background emission. The deprojected spectrum is then fitted with a single {\tt apec} model. Electron density 
is derived directly from the geometrically scaled normalization parameter of the best-fit model:
\begin{equation}
    \centering
    {\tt norm} = \frac{10^{-14}}{4\pi [D_{\rm A}(1+z)]^2} \int
    n_{\rm e}n_{\rm p}{\rm d}V,
\end{equation}
where $z$ is the redshift of the cluster, $D_{\rm A}$ is the corresponding angular diameter 
distance, $V$ is the volume of the emission region. $n_{\rm e}$ and $n_{\rm p}$ ($n_{\rm H}$) 
are the number densities of electron and proton. The ICM gas density is then computed by 
$\rho_{\rm gas} = n_{\rm e}m_{\rm p}A/Z$, where $m_{\rm p}$ is proton mass, $A$ and $Z$ are 
the average nuclear charge and mass of the ICM. For ICM with $\sim 1/3$ solar abundance, 
$A \approx 1.4$ and $Z \approx 1.2$, and therefore $n_{\rm e} \approx 1.2n_{\rm p}$.

The deprojected electron density profiles are fitted with a double-$\beta$ model, which can 
produce reasonable fit to the central density peak when a cool core is present. In the 
literature, the usual 
form of a double-$\beta$ model used  to describe density profiles, where 
density is computed from the surface brightness, is the square root of the quadratic sum of 
two $\beta$ model components \citep[e.g.,][]{ettori2000,hudson2010,Ettori2013}.
Instead, the density in this work is measured directly from the deprojected spectrum, thus we 
simply adopt a double-$\beta$ model as a linear summation of two $\beta$ model components, that reads:
\begin{equation}
    \centering
    n_{\rm e}(r) = n_{01}\cdot\left[1+\left(\frac{r}{r_{01}}\right)^2\right]^{-3\beta_1 /2} + n_{02}\cdot\left[1+\left(\frac{r}{r_{02}}\right)^2\right]^{-3\beta_2 /2}.
    \label{ne_model}
\end{equation}

\noindent 
For completeness, we also repeat the fit of the density profiles using the more conventional 
form of the quadratic sum of two $\beta$ models, and find that the results are in very good
agreement with those obtained using Equation \ref{ne_model}. 

The best fits for the electron density profile $n_{\rm e}(r)$ obtained with the double-$\beta$ 
model are shown in the left panel of Figure \ref{ne_femass}. 
From the iron abundance $Z \equiv [n_{\rm Fe}/n_{\rm H}]/[n_{\rm Fe}^{\odot}/n_{\rm H}^{\odot}]$, 
assuming the same solar abundance used in the {\tt Xspec} spectral fits 
$[n_{\rm Fe}^{\odot}/n_{\rm H}^{\odot}] = 3.16\times 10^{-5}$ \citep[from][]{asplund2009}, 
we then derive the cumulative profile of the Fe mass $M_{\rm Fe}(<r)$, as shown in the 
right panel of Figure \ref{ne_femass}. We remind that, since our iron abundance profiles extend 
only up to $0.4\, r_{500}$, at larger radii the iron mass is computed extrapolating the constant 
plateau up to $r_{500}$, differently from the gas mass that is obtained from the data extending 
up to $\sim r_{500}$. Therefore, the cumulative mass value above $0.4\, r_{500}$ depends on the 
assumption of a constant plateau at any radius. We note there that if the large scale iron distribution
is, instead, a shallow power law, we may overestimate the total iron mass.  Unfortunately, the
measurement of the shape of the large scale iron profile (i.e., adopting a power-law instead of a 
constant plateau) across our sample is not within our reach.  The assumption of a constant plateau
is a clear limitation of our approach.  A possible way out, but only for a minority of our sample,
is to combine {\sl Chandra} and XMM-{\sl Newton} data, a strategy that will be briefly mentioned 
in the Discussion Section.

\subsection{The identification of two components in $Z_{\rm Fe}$ profiles}

\begin{figure}
\begin{center}
\includegraphics[width=0.49\textwidth, trim=10 120 30 140, clip]{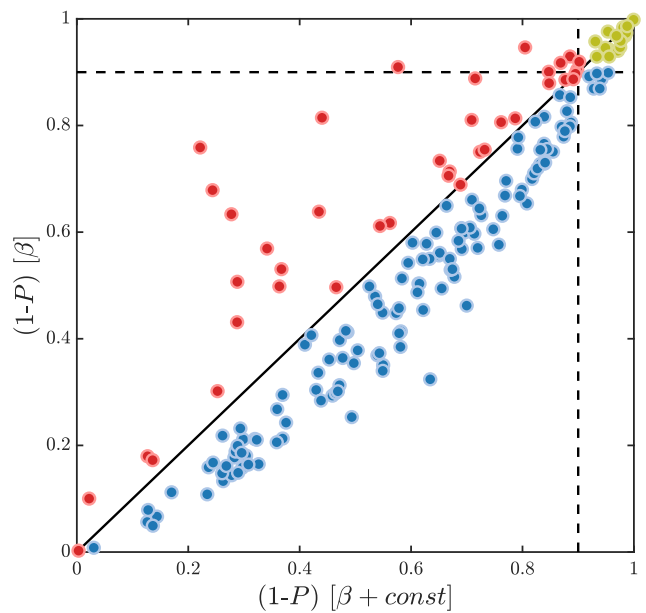}
\caption{Probability of rejection of the abundance profiles with and without the iron plateau. The 
dashed lines mark $(1-P)=0.90$, hence the yellow points shows clusters for which both the single- and 
double-component models are rejected at $>90\%$ c.l.  
Clusters colored in red favor the double-component model, 
while both models provide similar quality fits to the clusters colored in blue. In these cases, the
double-component model returns a slightly lower goodness, because of the inclusion of 
an additional parameter in the fit. }
\label{compare}
\end{center}
\end{figure}

\begin{figure*}
\begin{center}
\includegraphics[width=0.32\textwidth, trim=10 120 30 145, clip]{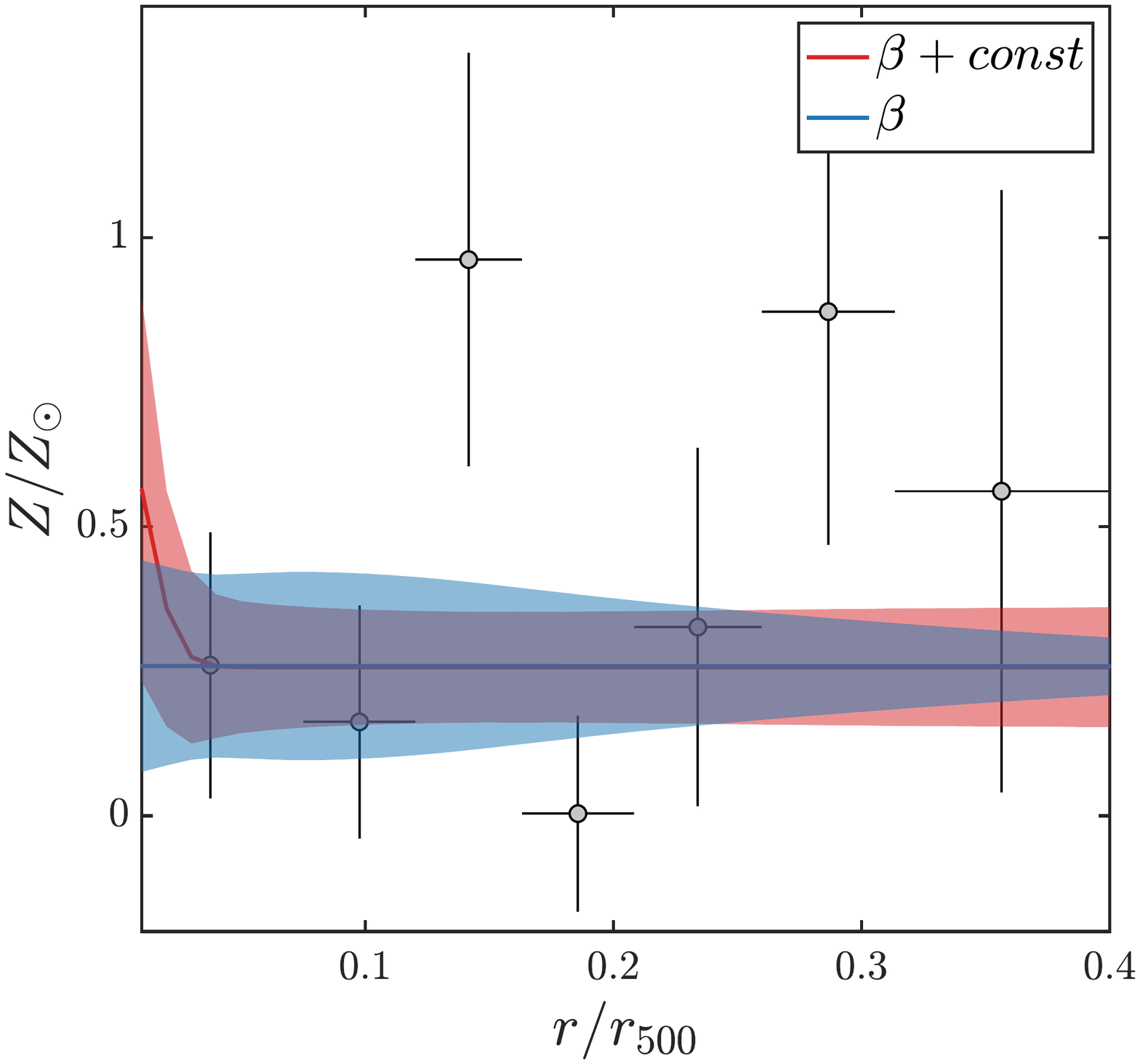}
\includegraphics[width=0.32\textwidth, trim=10 120 30 145, clip]{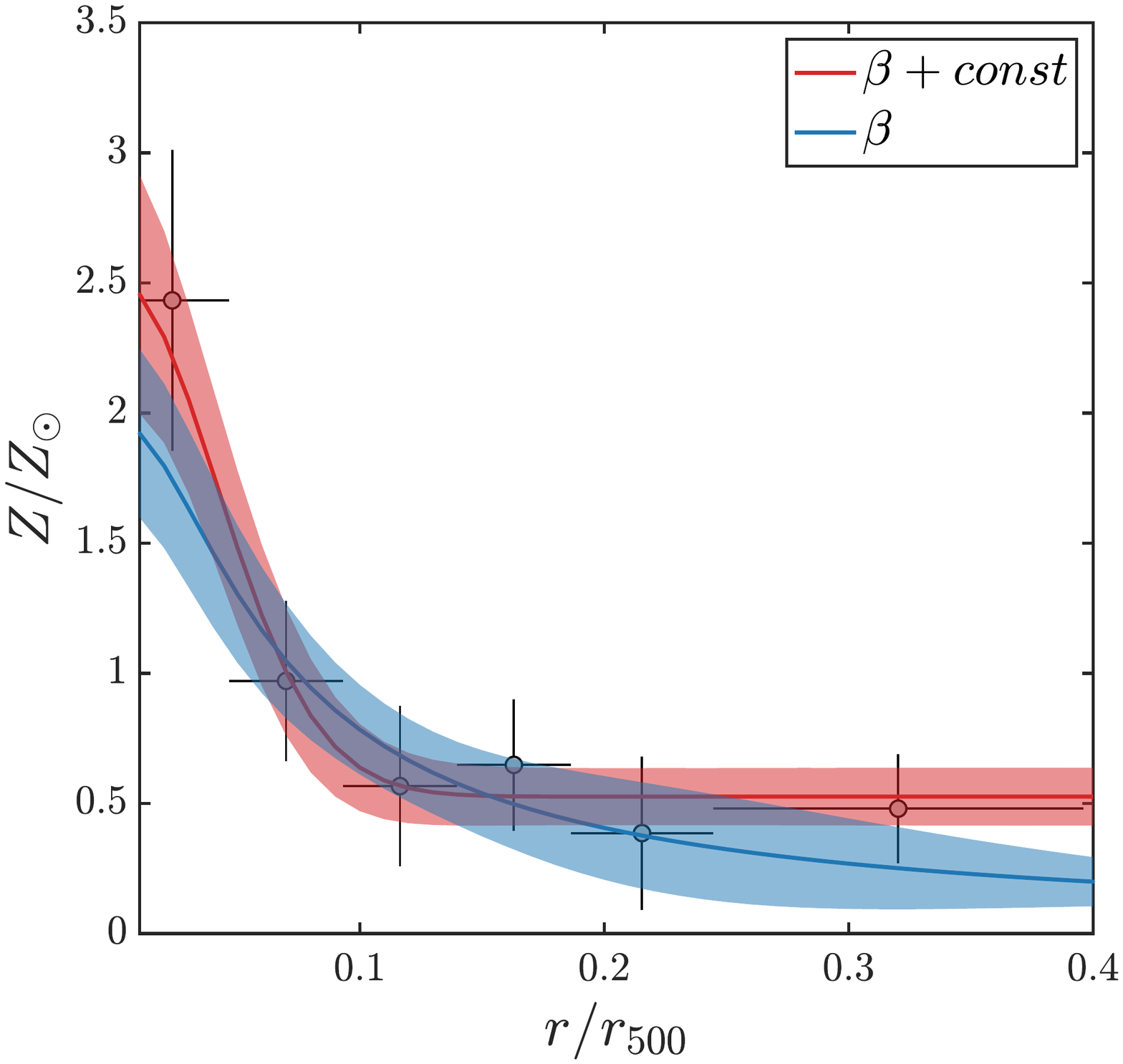}
\includegraphics[width=0.32\textwidth, trim=10 120 30 145, clip]{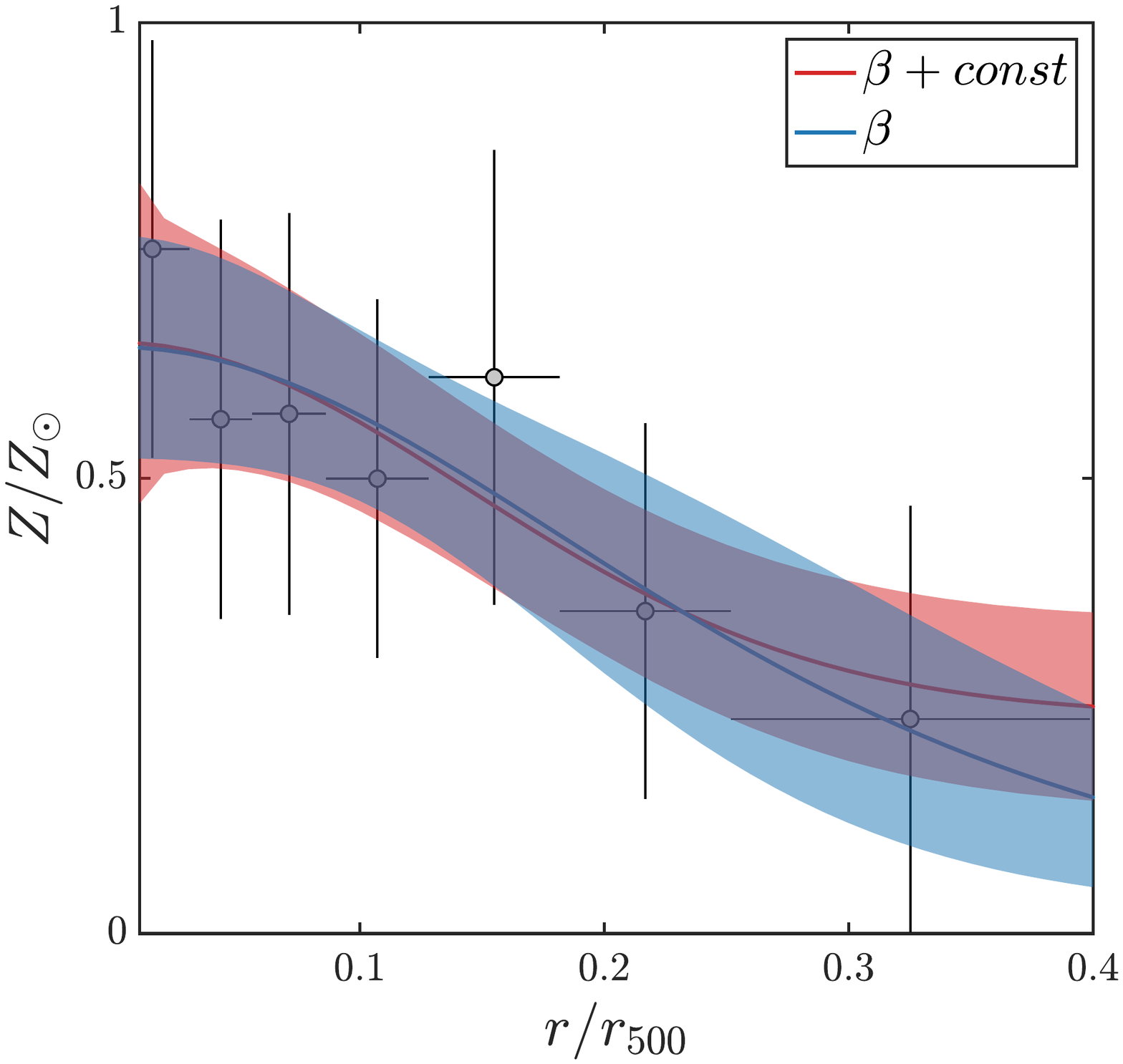}
\caption{Examples of abundance profiles with the corresponding double-component (red) 
and single-component (blue) best-fit models. From left to right: Abell 2050, CLJ1415+3612, 
and PSZ2 G241.77-24.00. The three examples are extracted from the yellow, red, and blue dots 
in Figure \ref{compare}, which are clusters that: cannot be fitted with either model (yellow);  
favor the double-component model (red); and can be fitted with both models 
in the observed radial range (blue). The best-fit values and uncertainties (of all the curve fittings in this paper, unless noted otherwise) are obtained using the MCMC tool of \citet{Foreman2013}.}
\label{compare_example}
\end{center}
\end{figure*}

A necessary step before proceeding in our analysis is to check whether the use of 
a double-component model is statistically preferred to a simpler model. In other words, 
we want to assess the relevance of the two components not only on the basis of theoretical 
premises, but also from a blind fit of the measured abundance profiles.  The relevance 
of this check is twofold. First, the distribution of iron in the ICM is sensitive to many dynamical 
processes, such as the outflow of central AGN, and large-scale sloshing. Some of these 
processes have relatively weak impact on the global morphology, but may significantly 
affect the distribution of iron. In these cases, the iron profile may not follow our idealized
pattern of a central peak plus a constant plateau, even in the case of a rather 
regular morphology. Second, due to the relatively small extraction radii of the profiles 
(0.4$r_{500}$), the iron plateau may not be well identified in cases where the iron 
peak has a large extension. Therefore, we repeat the fit of the iron abundance profiles, 
using a single-$\beta$ model, without the iron plateau, and compare the goodness of the fit, 
estimated by the $P$-value of the two models. We note that a $\beta$ model can 
provide an accurate description also in the case of a power-law behavior, which is
typically obtained with small values for the core radius and the radial slope.  
Moreover, we are aware that the fit with a single-$\beta$ model clearly predicts a rapidly declining metallicity value in the regions
at radii $>0.4r_{500}$ not sampled by our data.  This is in contradiction 
with the current and sparse knowledge about the ICM metallicity in the few clusters where outskirts
have been properly studied \citep[e.g.,][]{urban2017,mernier2018}.
As we already stressed, we have
no control on the actual abundance profile at large radii, so a constant plateau is an
assumption of our modelization.  In any case, the goal of this statistical test 
is to evaluate the robustness of our description on the basis of the data without 
using any prior.

We plot in Figure \ref{compare} the (1-$P$) value, which indicates the confidence level at which
the fit is rejected. If we consider the 90\% c.l. as our tolerance threshold, we find 16 clusters, 
shown as yellow circles in Figure \ref{compare}, for which both models are formally rejected.
We check the images of the 16 clusters with these peculiar iron abundance profiles, and find 
no obvious signs of a disturbed morphology. Despite that, the iron profile appears to be 
dominated by significant intrinsic scatter between different annuli, making it impossible to 
fit the profile with a smoothly varying function. As we mentioned, there are various 
processes that can result in these peculiar profiles, e.g., unnoticed mergers, major AGN outflows,
core-sloshing in different scales, and projected gas clumps in cluster outskirts, among others. 
A concrete diagnosis on the physical reasons of the peculiar distribution of iron in 
these cases requires a more in-depth and case-by-case study of the dynamics of each 
cluster, which goes well beyond the goal of this paper.  Therefore, we decide to 
exclude these 16 clusters in our following analysis, and focus on a sample of the 
remaining 170 clusters. 

In 39 clusters (colored in red), the double-component model provides a smaller $(1-P)$ value. 
Despite in several cases the difference is not dramatic, we find that at least 
in 1/5 of the sample the use of a double component in the iron distribution provides a 
significantly better fit, after considering the additional parameter.
In the remaining 131 clusters, the profiles can be well fitted with both models within a 
confidence level of $>90\%$.  In these cases, the double-component model returns a 
larger $(1-P)$ value because it has an additional free parameter and therefore
a larger number of degrees of freedom. 

One example for each of these three classes is shown in 
Figure \ref{compare_example}. The left panel shows a noisy iron abundance profile, that can 
be hardly reconciled with any smoothly-varying azimuthal function of the kind we consider here.  
In the central panel, the data clearly show a flat plateau that cannot be fitted with a single-$\beta$ model.  In some other cases, the central iron profile is better described by a broad 
bump rather than a well-defined peak, so that the plateau does not stand out clearly in the data.  
This situation is shown in the right panel, where it is not possible to differentiate 
statistically between the two models, and the abundance profile itself is indistinguishable 
from a simple power-law, at least in the explored range.  For completeness, we repeat the same test, 
but using a power-law instead of a $\beta$ model with no plateau. The results are very close to 
what we obtained in Figure \ref{compare}. 

In general, we conclude that the $\beta$ model with a constant plateau is statistically preferred 
with respect to the use of a single $\beta$ model or power-law for a significant fraction 
of our sample. Clearly, a more complete modelling of the 
profiles would be a $\beta$ model plus an transitional power-law, constrained to have a mild 
slope, plus a constant plateau in the external regions. A slow decrease is actually expected in 
some modelization of the iron distribution \citep[see][]{mernier2017,2018biffi} 
and can be used to describe an intermediate 
regime where the iron distribution, far from the core, is still slowly decreasing before 
reaching the flat plateau associated to the pristine, uniform enrichment. However, 
the quality of data we use in this work is clearly not sufficient to assess the 
presence of this transitional component between the iron peak and plateau. We will dedicate 
a future work on a more extended modelization of the iron distribution, mostly in the
perspective of the future X-ray missions (Tozzi et al., in preparation).

Finally, we also inspected the distribution of the size of the iron peak. Differently from 
what we have done in \citet{liu2018}, we can now directly compute an effective size of the
iron peak as $r_{\rm Fe} = r_0 \cdot \sqrt{2^{1/\alpha} -1 }$, where $r_0$ and $\alpha$ are the 
two best-fit parameters that fully characterize the shape of the peak.  In principle, 
the global distribution of $r_{\rm Fe}$ reflects different physical phenomenon, including 
effects of past/recent mergers that erased the peak or smoothed it into a broad bump, 
and the broadening effects of the AGN feedback from 
the BCG plus minor mergers. In practice, it is impossible to disentangle the two phenomena. 
However, the broadening of the iron peak due to AGN feedback can be investigated by selecting the
stronger cool cores, which are most likely the oldest one where no major merger has recently occurred. In this case, the typical size can be a way to parameterize the age of the peak through the 
broadening effect of the AGN feedback.  This is what we have done in 
\citet{liu2018} on a sample with bright and strong cool core, reaching the conclusion that
the size of the iron peak of CC clusters is actually increasing 
by a factor of three with cosmic time in the redshift range $0.1<z<1$. If we consider the CC
clusters in our sample (defined as usual as those with $c_{\rm SB}>0.075$) we find the same trend but
weaker, consistent with the fact that, having a sample with less concentrated cores, we 
are including a wider range of ages for the observed iron peak, with younger peaks being narrower. 
The trend we obtain is an average increase of a factor of 2 (from 0.025 to 0.05 r$_{500}$) in the redshift range $0.05<z<0.6$, still consistent with what we have found in \citet{liu2018}. 

To summarize, we find that the choice to fit the abundance distribution with a $\beta$ model 
plus a constant plateau is a good compromise between a comprehensive physical modelization and 
the data quality, and it is adequate to effectively describe a large sample of clusters observed 
with {\sl Chandra} with a wide range in mass, redshift, and exposure time.  While a two-component 
model is physically motivated and favoured by the data, more sophisticated approach 
are not able to extract more information.  Ideally, we should try to have more handle on the abundance
profile at large radii.  However, due to the limited field of view of ACIS, but mostly 
because of the rapidly decreasing signal, this is unfeasible.  In fact, as we already mentioned, 
despite that the surface brightness is detected up to $r_{500}$ in most of our clusters, 
at radii larger than $0.4\, r_{500}$ the spectral analysis would be strongly affected by 
uncertainties in the background subtraction.  There are two ways to tackle this issue. 
The first is to use 
XMM-{\sl Newton} for the clusters that have been observed with both instruments, exploiting the 
$\sim 5 \times$ larger collecting efficiency, and the larger field of view. However, 
the discrepancy in the temperature measurements between {\sl Chandra} and XMM-{\sl Newton}
\citep[e.g.,][]{Schellenberger2015} increases the complexity of such a combined analysis.
The second method is to wait for the X-ray micro-calorimeter {\it Resolve} onboard 
XRISM\footnote{The X-ray Imaging 
and Spectroscopy Mission (XRISM), formerly named the ``X-ray Astronomy Recovery Mission'' (XARM), 
is a JAXA/NASA collaborative mission, with ESA participation 
\citep[see][and references therein]{2018Guainazzi}, expected 
to be launched in 2022. }, able to identify the 
iron line thanks to the $\sim 10 \times$ larger spectral resolution, in external regions 
of nearby clusters where angular resolution is not an issue, despite this will require
a sizeable investment of observing time due to the limited grasp of the bolometer.
The first method goes beyond the goal 
of this paper and it is deferred to a further work on the entire {\sl Chandra} and 
XMM-{\sl Newton} archives.  The second approach is definitely a time-consuming
but promising way to use XRISM to  attack this problem \citep[see][for more XRISM science related to clusters]{2014Kitayama}, as we will mention also in the discussion Section.

\subsection{The effect of the central iron drop}

\begin{figure}
\begin{center}
\includegraphics[width=0.49\textwidth, trim=20 125 30 145, clip]{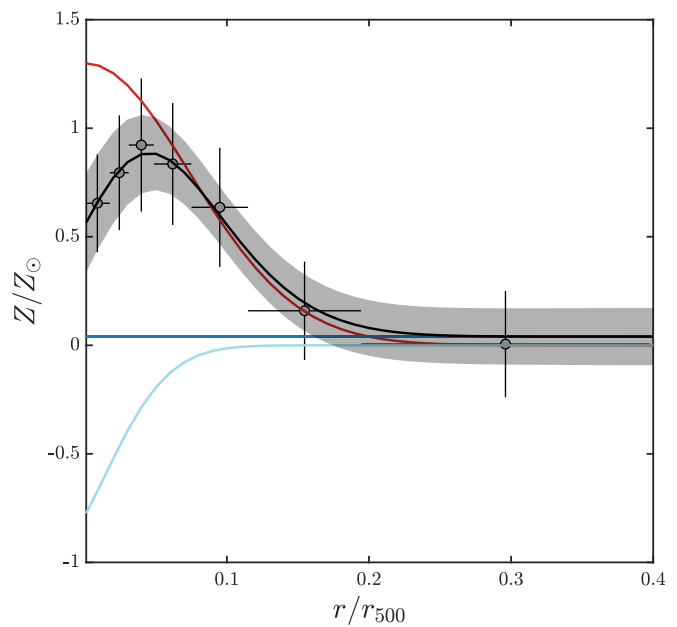}
\caption{The iron abundance profile of MACSJ0242.5-2132 with the best-fit showing a 
pronounced central iron drop. The red, cyan, and blue curves are the iron peak, 
iron drop, and iron plateau, respectively, as described by equation \ref{drop_model}.}
\label{drop_example}
\end{center}
\end{figure}

The central iron drop observed in a few clusters is a significant feature 
with a typical scale of $\sim 10$ kpc \citep{panagoulia2015,2019Liu,lakhchaura2019}. 
It requires a high spatial resolution and a high S/N
to be detected, and has a negligible impact on the iron mass in most cases. 
We do not perform here a systematic investigation of the iron drop in our sample, due to the lack 
of signal. Instead, we proceed first by identifying about 20 profiles that, after a visual inspection,
may show a central drop.  Typically, this occurs when the first and/or second bin shows an abundance
value lower than the value of the second/third bin at $\sim 2 \sigma$.  Then, we repeat the 
fit to the abundance profile including the central drop, allowing for this 
component in our fit in the form of a ``negative Gaussian'' as follows:
 
\begin{equation}
    \centering
    Z = Z_{\rm peak}\cdot \left[1+\left(\frac{r}{r_0}\right)^2\right]^{-\alpha}-a\cdot{\rm exp}\left[\frac{-(r-\mu)^{2}}{2\sigma^2}\right]+Z_{\rm plateau}.
    \label{drop_model}
\end{equation}

\noindent
This modelization of the central drop is simpler than the one used in \citet{2019Liu} due to the lower quality
of the profiles, and it has been used also in \citet{mernier2017}.  Then, we collect all the cases 
where the improvement of the $\chi^2$ formally corresponds to a confidence level of 90\%. 
In the end, we do find a central iron drop in 8 out of 186 clusters. An example is shown in 
Figure \ref{drop_example}. The typical size of the iron drop measured in these 8 clusters is 
$\sim$ [0.05--0.1]$r_{500}$, significantly larger than what has been found in nearby clusters and groups 
\citep{panagoulia2015,2019Liu}. However, this is probably due to the relatively low resolution 
of the profiles we have in this work, which masks the small-scale iron drops, leaving only 
the large-scale ones. For these clusters, the final iron peak component is therefore computed by 
considering the ``hole'' in the iron distribution.  Clearly, the amount of mass removed by the drop 
is limited to less than 10\% of the total iron mass in the peak, and it is often compensated 
by the re-adjustment of the iron peak profile, so that the impact on our final results is 
negligible. Nevertheless, we stress that the presence of a central 
drop in the iron distribution is an important component to be included when more detailed 
profiles will be available, not only for its effect on the total iron budget, but also for its
physical relevance.  The effects of feedback and of dust depletion, which are responsible of 
the iron drop, are indeed expected to be present at least since $z\sim 1$.

\begin{figure*}
\begin{center}
\includegraphics[width=0.49\textwidth, trim=20 125 30 145, clip]{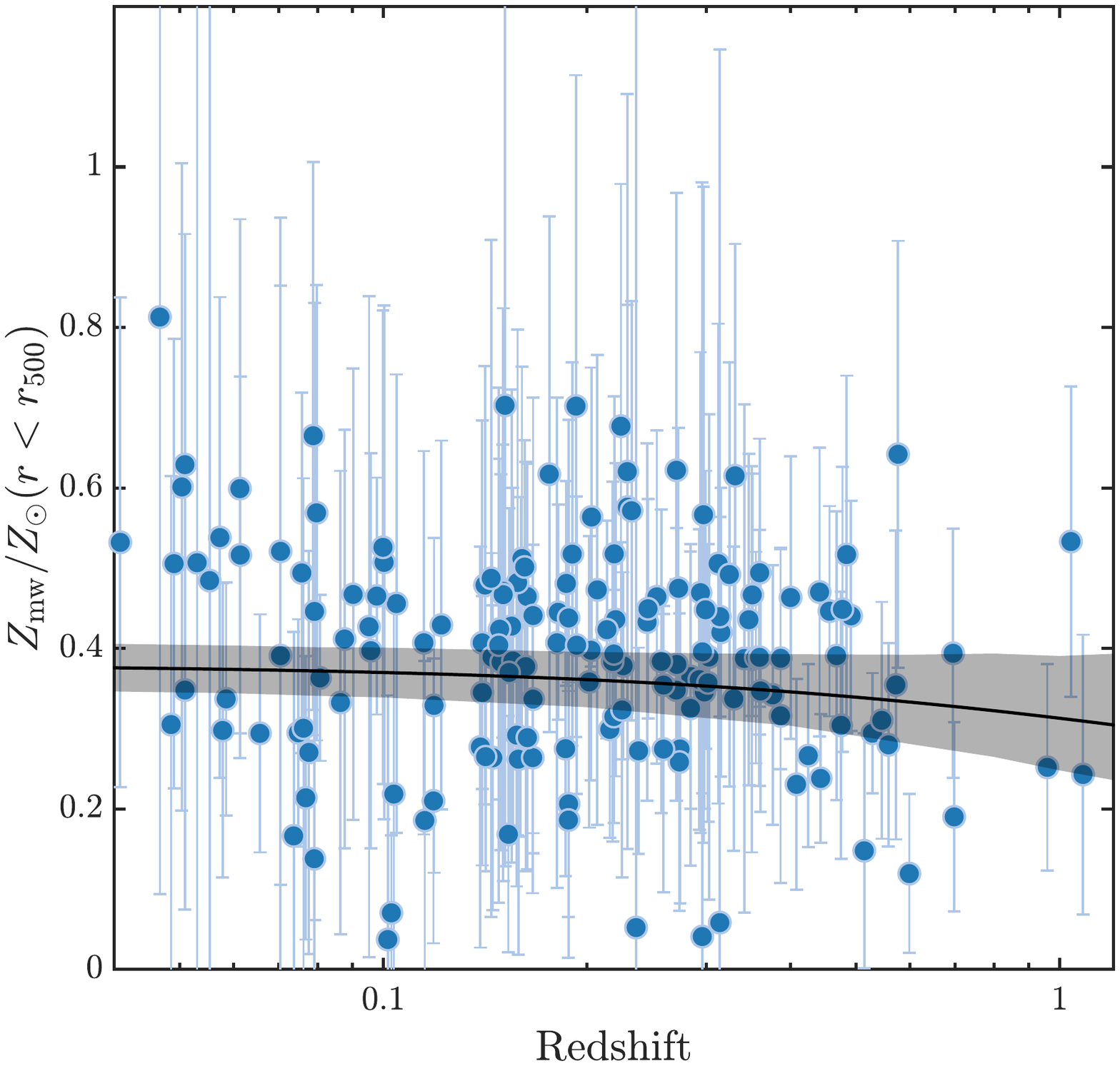}
\includegraphics[width=0.49\textwidth, trim=20 125 30 145, clip]{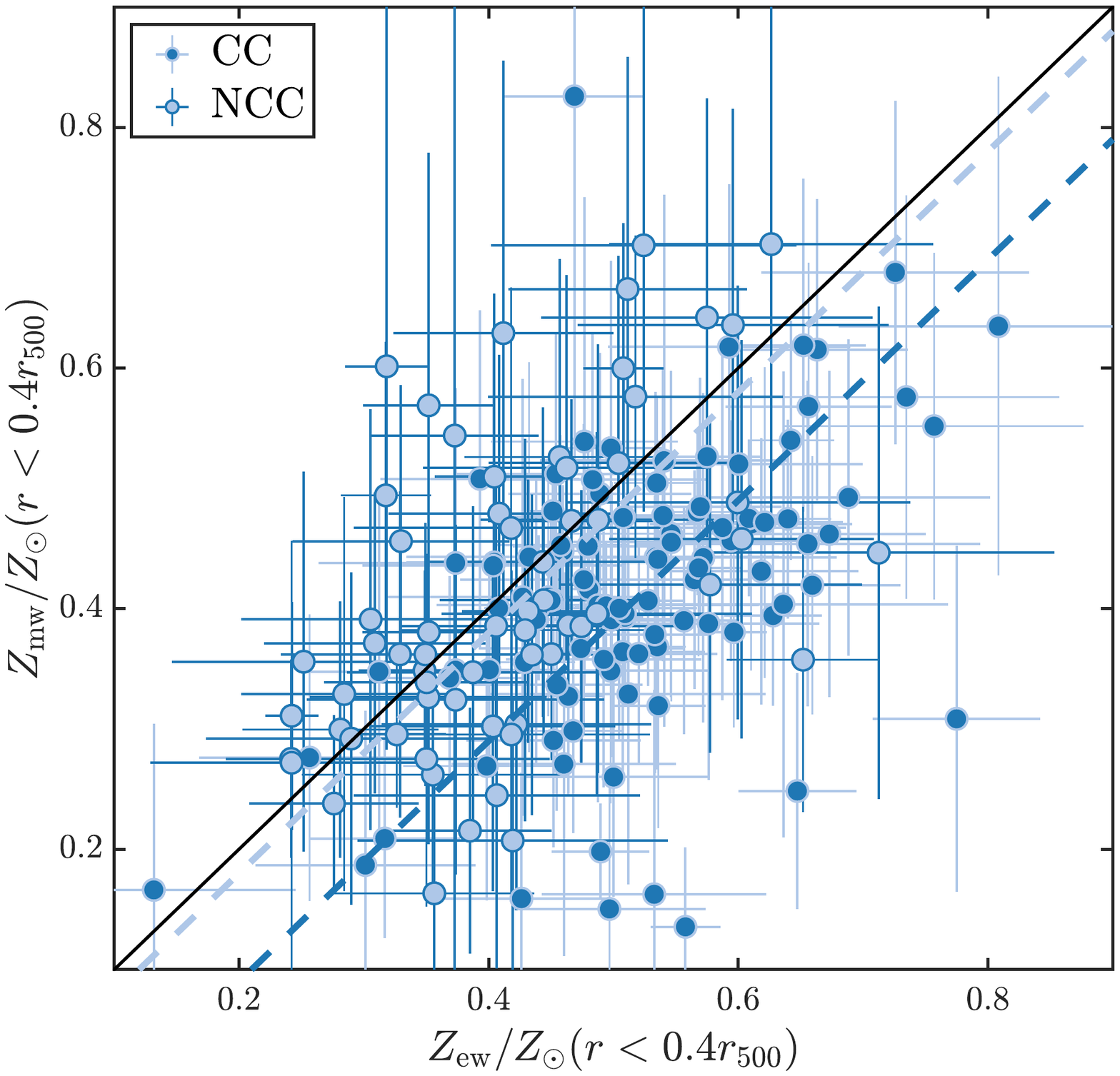}
\caption{{\sl Left panel}: the correlation between the average, gas mass-weighted iron abundance 
(within $r_{500}$) and the redshift of all the clusters. The black curve and shaded area show the 
best-fit function $Z_{\rm mw} = Z_{\rm mw,0}\cdot (1+z)^{-\gamma_{\rm mw}}$ with 
$Z_{\rm mw,0}=(0.38\pm0.03)\, Z_\odot$ and $\gamma_{\rm mw}=0.28\pm0.31$. {\sl Right panel}: 
the gas mass-weighted abundance within $0.4\, r_{500}$ plotted against the emission-weighted value in the 
same radial range. The solid line corresponds to $Z_{\rm mw} = Z_{\rm ew}$.
Dashed lines show the average relation for cool-core and non-cool-core clusters, 
$Z_{\rm mw}^{\rm CC}=Z_{\rm ew}-0.11$ and $Z_{\rm mw}^{\rm NCC}=Z_{\rm ew}-0.02$. }
\label{mw_abun}
\end{center}
\end{figure*}

\subsection{Gas mass-weighted iron abundance}

As already mentioned in Section 3.2, we first compute the average gas mass-weighted iron 
abundance, defined as $Z_{\rm mw}\equiv \sum (Z_{\rm Fe}^{i}\cdot 
M^{i}_{\rm gas}$)/$\sum M_{\rm gas}^{i}$, without making any distinction between the two 
components.  Here the index $i$ runs over the annuli.  In addition, where we are at radii
$r>0.4\, r_{500}$, we simply have $Z_{\rm Fe}^{i}=Z_{\rm plateau}$.
\citet{liu2018} have shown that gas mass-weighted value is more appropriate than the 
emission-weighted value in quantifying the average abundance of iron in the ICM, because the latter, 
despite much easier to measure\footnote{A global value can be obtained for a number of net
counts in the 0.5--7 keV band as low as 1000.}, can be affected by a significant bias 
in cool-core clusters. We note that the gas mass-weighted abundance is by definition different
from a truly mass-weighted abundance as that obtained from numerical simulations, for instance.  
The point is that we assume a smooth ICM (i.e., not clumped) distribution within the angular
scales resolved in the bins of our spectral analysis.  Differently, in the presence of 
significant unresolved clumps, the emission-weighted value over-represents the cooler ICM.
Therefore, excluding the presence of significant clumpiness, the observed gas mass
can be considered an accurate estimate of the true gas mass, and the product of the emission-weighted abundance measured in each radial bin of the spectral analysis by the gas mass 
in that spherical shell can be considered a reliable proxy of the true gas mass-weighted abundance.

We show the gas mass-weighted iron abundance 
within $r_{500}$ as a function of redshift in the left panel of 
Figure \ref{mw_abun}.  From a visual inspection, a good guess is to assume a value constant
with redshift.  If we compute the root mean square value around the mean over the redshift, 
or the {\sl raw} scatter as defined in \citet{2009Pratt}, we find that both quantities are 
comparable to the average statistical error.  This implies that the intrinsic scatter, 
which is beyond any doubt present in a complex quantity such as $Z_{\rm mw}(r<r_{500})$, is
negligible with respect to the measurement uncertainty.  Considering that the uncertainty on the 
redshift is not relevant here, we can safely search for a best-fit function by a simple 
$\chi^2$ minimization.  If we fit the $Z_{\rm mw}-z$ relation with a simple power-law defined as $Z_{\rm mw} = Z_{\rm mw,0}\cdot (1+z)^{-\gamma_{\rm mw}}$, we obtain 
the best-fit parameters $Z_{\rm mw,0}$=(0.38$\pm 0.03) \, Z_\odot$ 
and $\gamma_{\rm mw}$=0.28$\pm 0.31$, consistent with no evolution of $Z_{\rm mw}$ across our sample. 

Limited by the extraction radius of our iron abundance profiles, the comparison between the gas mass- and 
emission-weighted abundances is only possible within $0.4\, r_{500}$. The last quantity is 
simply obtained fitting with a metallicity-linked double-temperature {\tt vapec} model to the 
total emission within the same radius. In the right panel of 
Figure \ref{mw_abun} we compare the two quantities.  We find that, on 
average, the emission-weighted abundance 
within 0.4$r_{500}$ is higher than the gas mass-weighted value by $\sim 18$\%. 
We note that this is slightly lower than that found in \citet{liu2018}, where 
$(Z_{\rm ew}-Z_{\rm mw})/Z_{\rm mw}\approx 0.25$. However, this is expected, 
because most of the clusters in \citet{liu2018} host a strong cool core, and the investigated 
radius is 0.2$r_{500}$, thus more affected by the iron peak. We expect that 
the difference between gas mass- and emission-weighted abundance will further decrease with larger 
extraction radius and higher fraction of non-cool-core clusters in the sample. 
This effect is more evident if we split our samples in two halves, populated
by cool-core and non-cool-core clusters adopting as a threshold $c_{\rm SB}=0.075$. 
We find that this discrepancy becomes 22\% for cool-core clusters, and drops to only 4\% for 
non-cool-core clusters. A simple fit to the distribution with a 
linear function $Z_{\rm mw}=Z_{\rm ew}-\delta Z$ gives $\delta Z=0.11 \, Z_\odot$ for 
cool-core clusters, and $\delta Z=0.02 Z_\odot$ for non-cool-core clusters. The different 
behaviors of cool-core and non-cool-core clusters reflect
the effect of the iron peak on the measurement of emission-weighted abundance. 

The average emission-weighted abundance within 0.4$r_{500}$ of cool-core clusters 
in our sample is (0.51$\pm0.01)\, Z_\odot$, significantly higher than that of 
non-cool-core clusters: 
(0.41$\pm 0.01) \, Z_\odot$. This difference has been already noticed in several other works 
\citep[e.g.,][]{degrandi2001,degrandi2004}. We find that this difference is significantly
reduced albeit still marginally significant when considering the average gas mass-weighted
abundance, which turns out to be ($0.41\pm 0.01)  \, Z_\odot$ and $(0.38\pm 0.02)  \, Z_\odot$ 
for cool-core and non-cool-core clusters, respectively. 
These results confirm that the difference in iron abundance between cool-core and non-cool-core
clusters is largely due to the use of emission-weighted abundance, while it almost
disappears when using gas mass-weighted values, which are representative of the true 
iron mass content.  At the same time, a residual difference in the average, gas mass-weighted 
abundances shows that the effect is not entirely due to the different ICM distribution, but
it may be due to a slightly larger amount of iron in cool-core clusters, further strengthening 
the hypothesis of two different physical origins for the iron peak and the iron plateau.  

\subsection{The properties of the iron plateau and iron peak}

\begin{figure*}
\begin{center}
\includegraphics[width=0.49\textwidth, trim=5 135 30 145, clip]{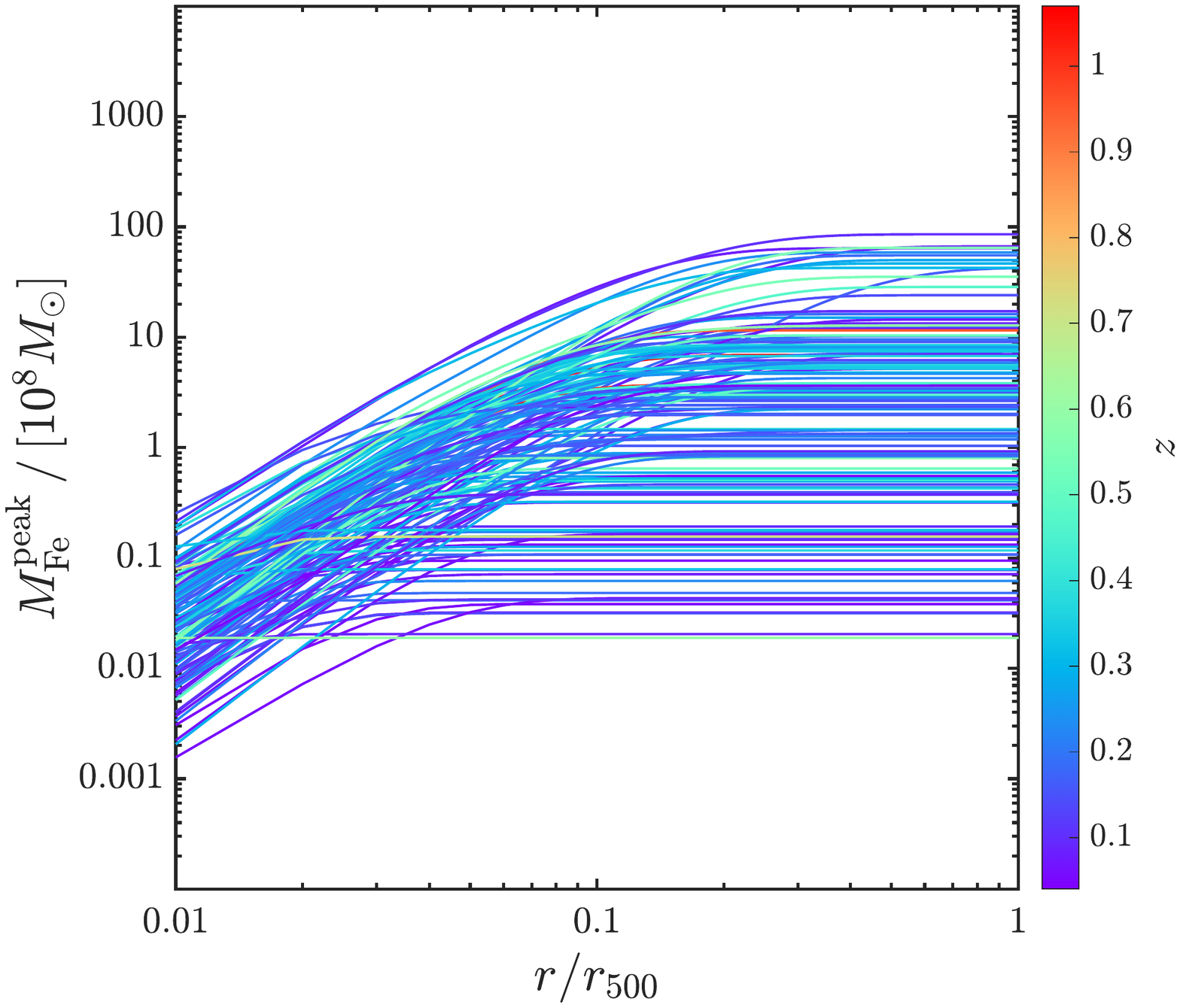}
\includegraphics[width=0.49\textwidth, trim=5 135 30 145, clip]{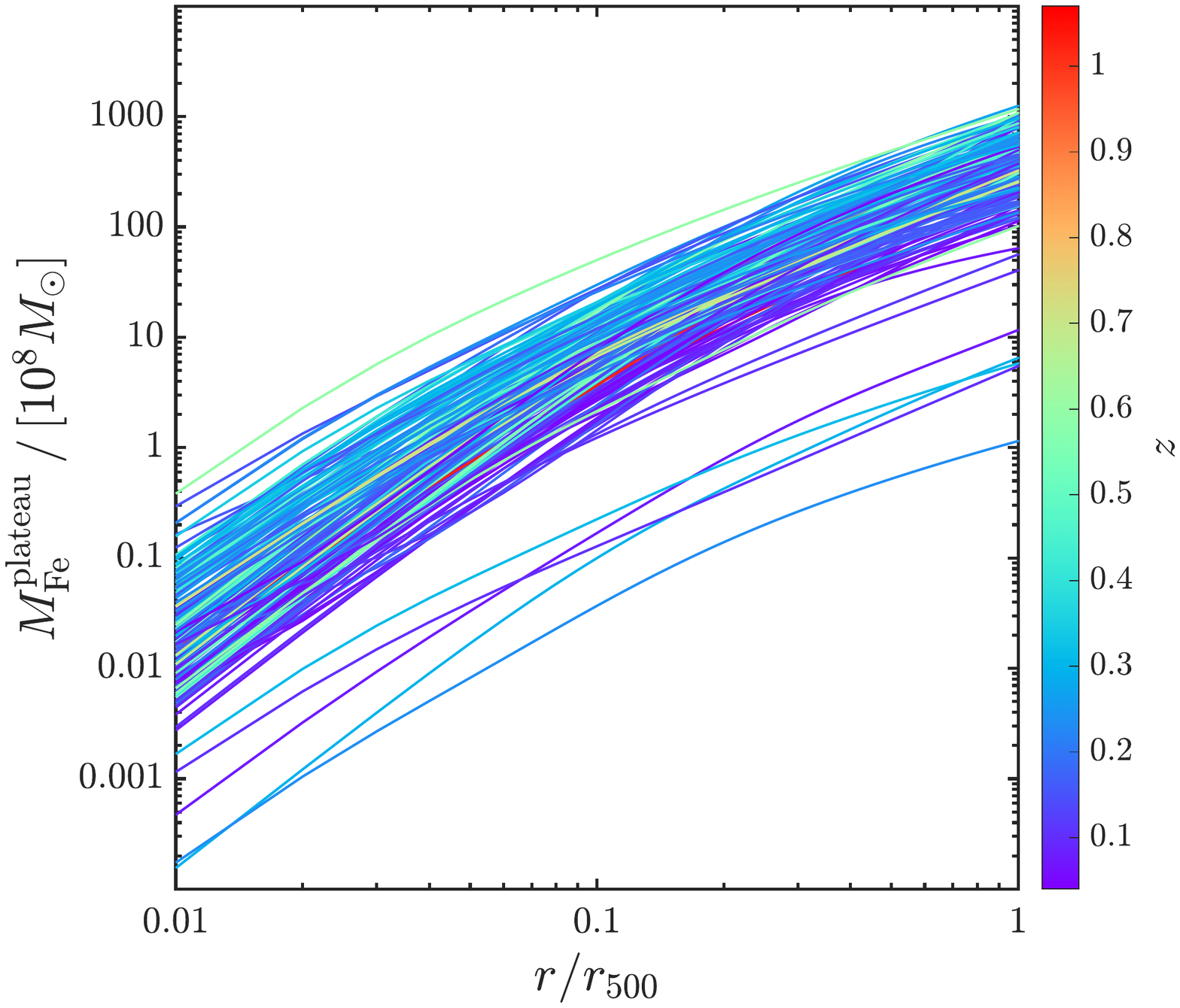}
\caption{{\sl Left panel:} Cumulative iron mass profiles corresponding to the peak 
component, color-coded by redshift for all the 170 clusters with regular abundance profiles
considered in this work.  {\sl Right panel:} as in the left panel, but for the plateau
component. }
\label{mass2}
\end{center}
\end{figure*}

\begin{figure}
\begin{center}
\includegraphics[width=0.49\textwidth, trim=0 110 10 130, clip]{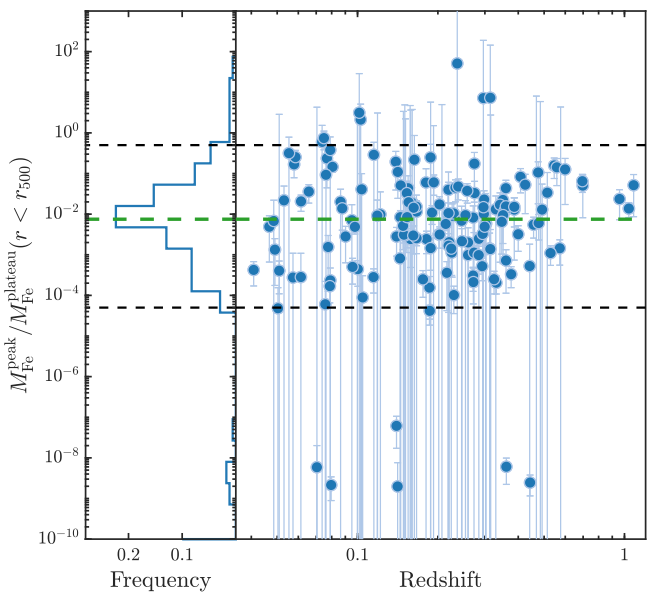}
\caption{Distribution of the ratio of iron peak mass to iron plateau mass within $r_{500}$, and 
the correlation with redshift. The green dashed line indicates the weighted average at $\sim$0.008. 
The black dashed lines mark the $[5\times10^{-5}, 0.5]$ range roughly corresponding to $>90$\% of 
the clusters symmetrically distributed around the central value (as shown in the left-side panel). }
\label{peak_const_hist}
\end{center}
\end{figure}

\begin{figure*}
\begin{center}
\includegraphics[width=0.49\textwidth, trim=20 120 30 145, clip]{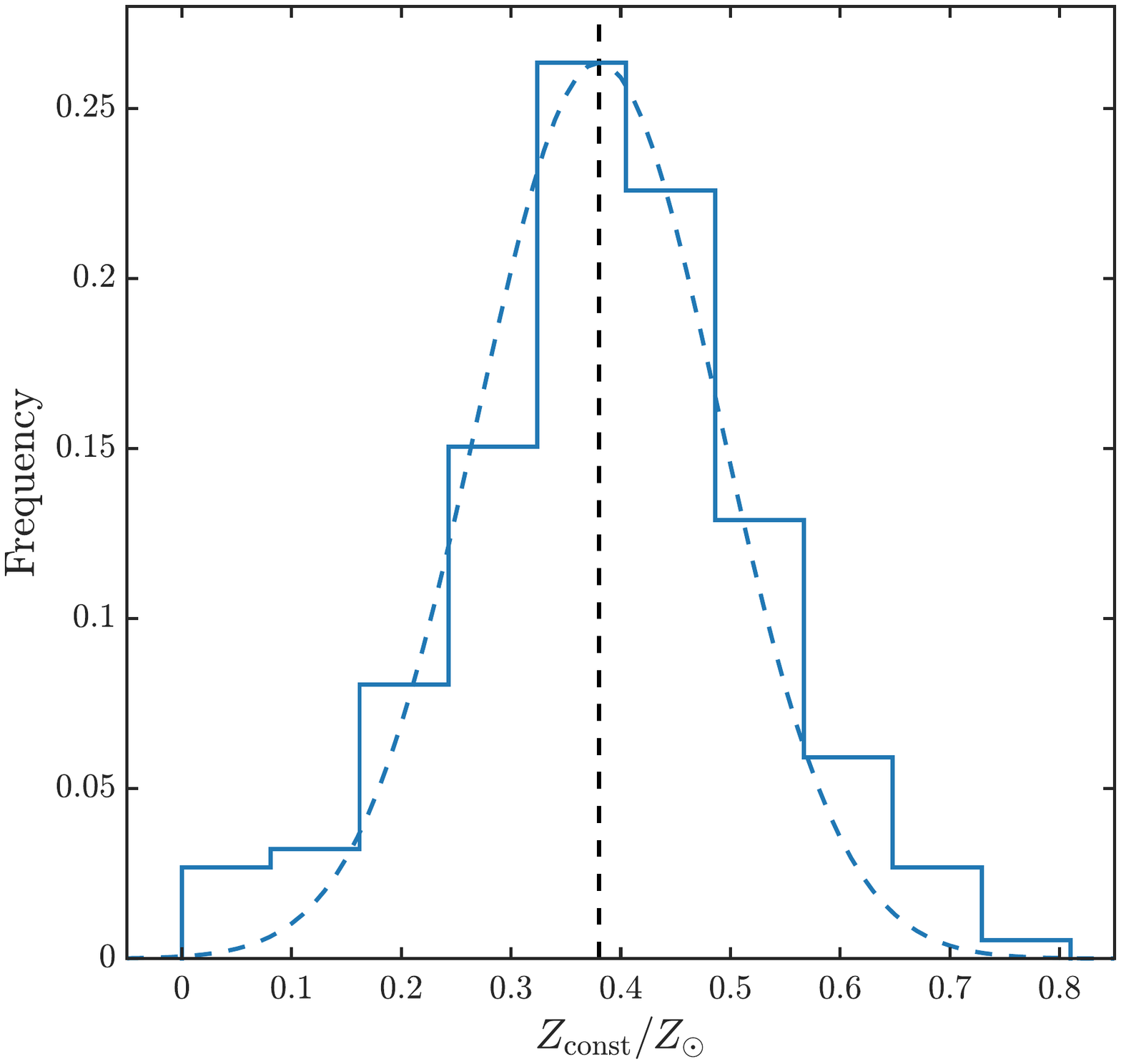}
\includegraphics[width=0.49\textwidth, trim=20 120 30 145, clip]{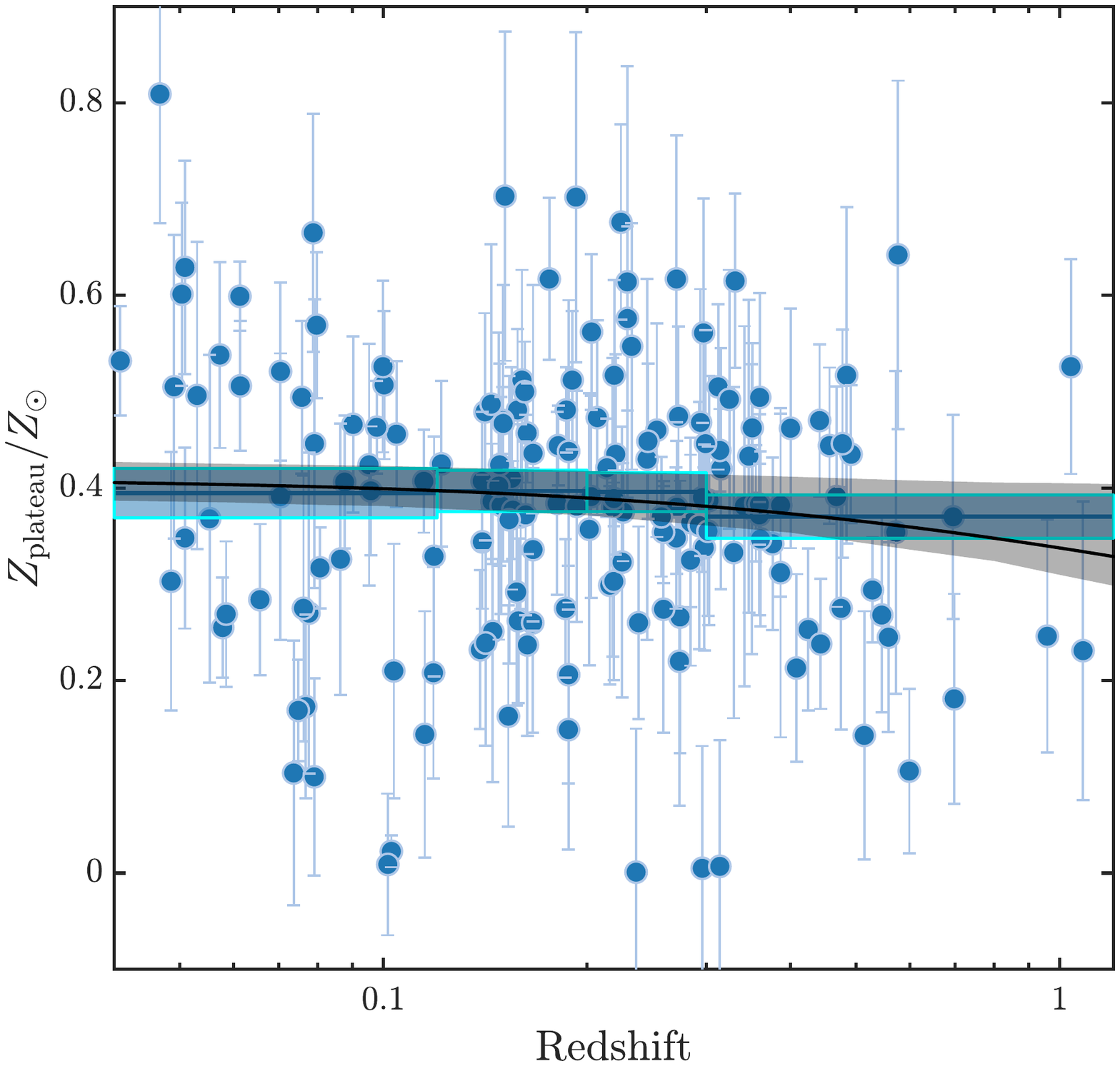}
\includegraphics[width=0.49\textwidth, trim=20 120 30 145, clip]{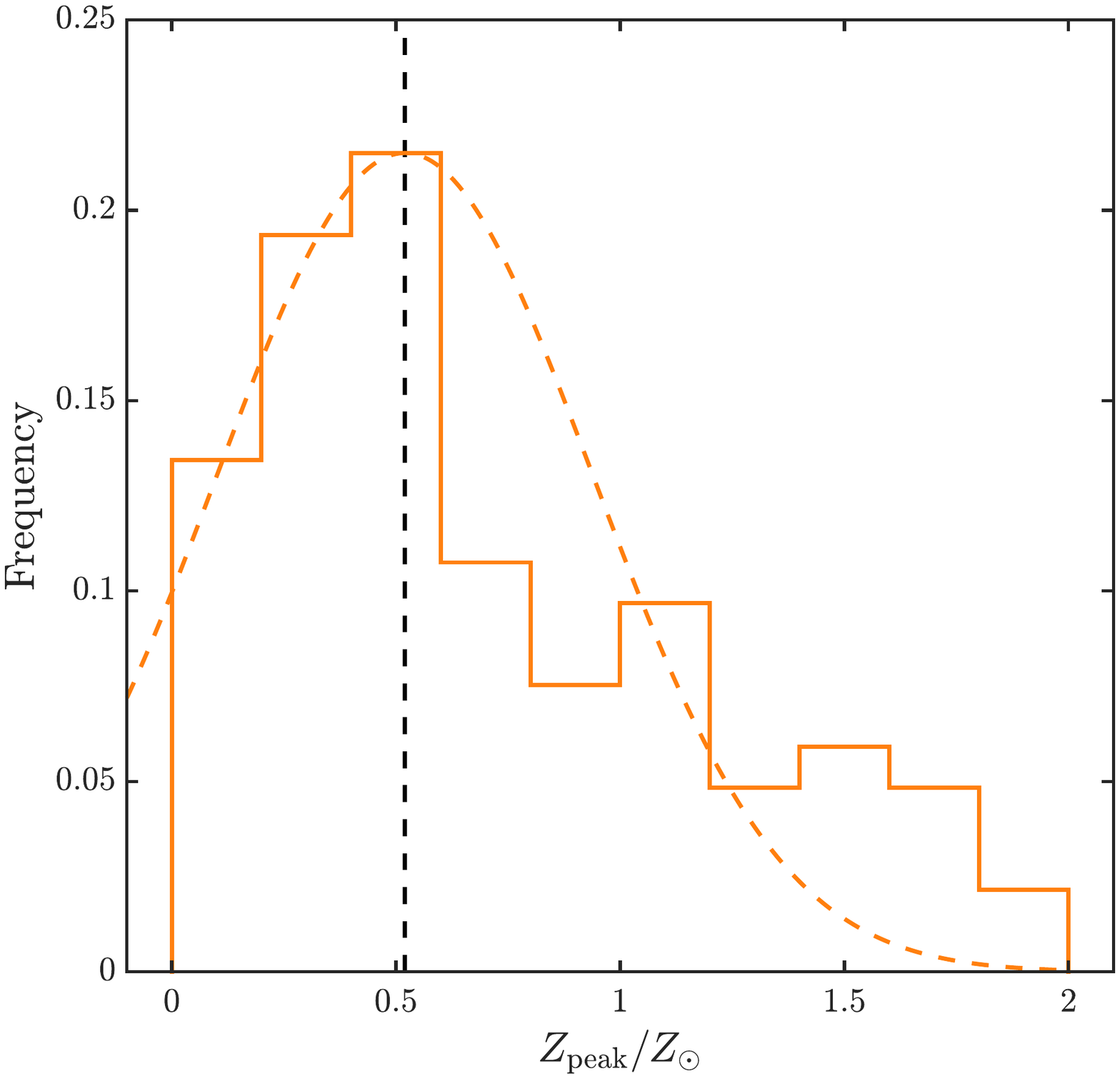}
\includegraphics[width=0.49\textwidth, trim=20 120 30 145, clip]{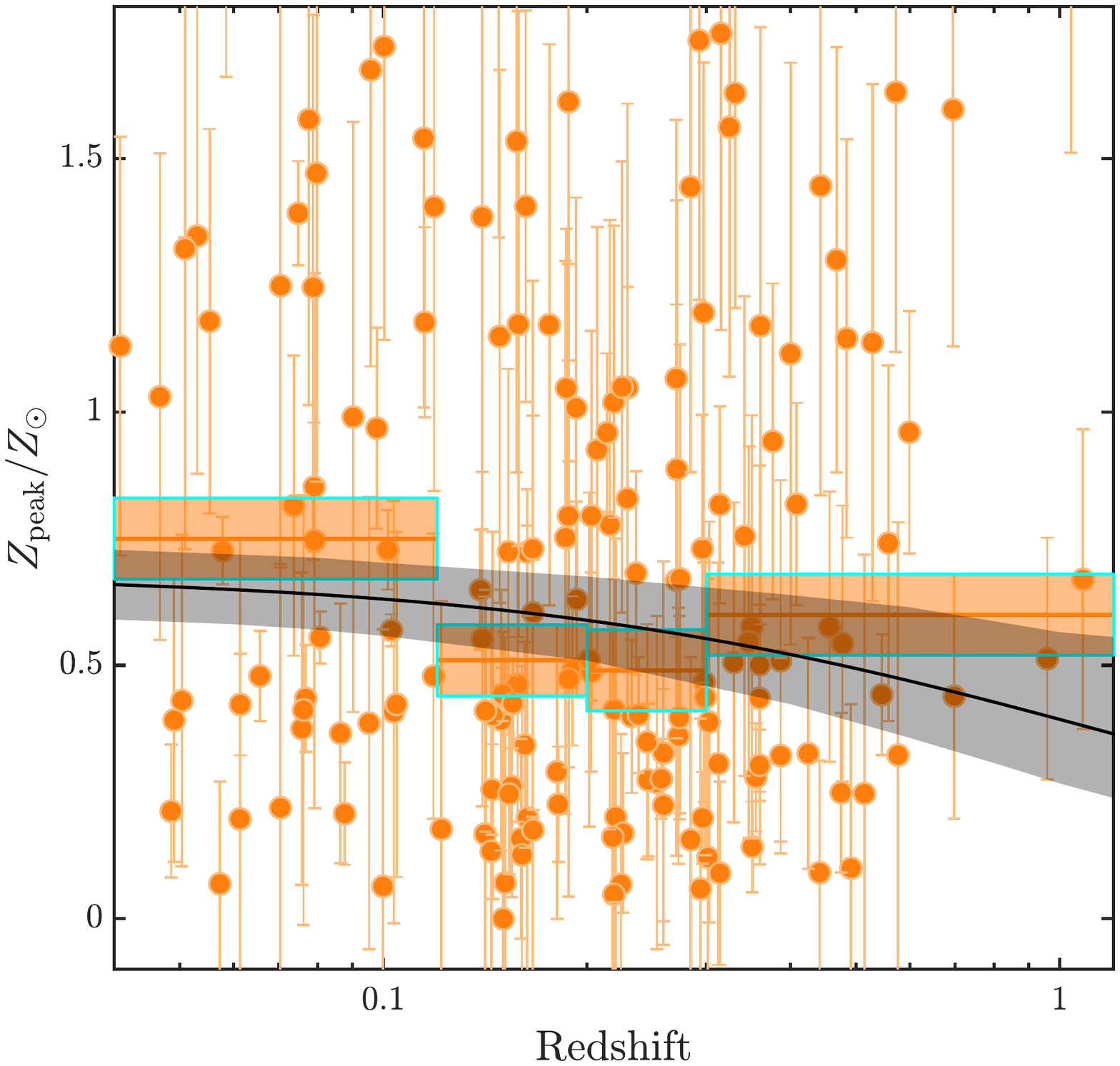}
\caption{{\sl Upper left}: Distribution of the abundance of the iron plateau component $Z_{\rm plateau}$. 
The dashed line indicates the weighted average value  $\langle Z_{\rm plateau} \rangle = 0.38\, Z_\odot$. The dashed curve shows a normalized Gaussian with $\sigma=0.11\, Z_{\odot}$, corresponding to the average statistical error, and $\mu=0.38\, Z_{\odot}$, corresponding to the weighted average value.
{\sl Upper right}: The abundance of the iron plateau plotted against cluster redshift. The black curve 
and shaded area show the best-fit function 
$Z_{\rm plateau} = Z_{\rm plateau,0}\cdot (1+z)^{-\gamma_{\rm plateau}}$ 
with $Z_{\rm plateau,0}=(0.41\pm0.02)\, Z_\odot$ and $\gamma_{\rm plateau}=0.21\pm0.18$, which are obtained by fitting the weighted average values and uncertainties of the four bins shown as blue solid lines and shaded areas. 
{\sl Lower left}: Distribution of the normalization of the iron peak component $Z_{\rm peak}$. 
The dashed line indicates the weighted average value  $\langle Z_{\rm peak} \rangle = 0.52\, Z_\odot$. The dashed curve shows a normalized Gaussian with $\sigma=0.42\, Z_{\odot}$, corresponding to the average statistical error, and $\mu=0.52\, Z_{\odot}$, corresponding to the weighted average value.
{\sl Lower right}: The normalization of the iron peak component $Z_{\rm peak}$ plotted against cluster 
redshift. The black curve and shaded area show the best-fit function 
$Z_{\rm peak} = Z_{\rm peak,0}\cdot (1+z)^{-\gamma_{\rm peak}}$ with 
$Z_{\rm peak,0}=(0.68\pm 0.07)\, Z_\odot$ and $\gamma_{\rm peak}=0.79\pm0.53$.}
\label{Z1}
\end{center}
\end{figure*}

\begin{figure*}
\begin{center}
\includegraphics[width=0.49\textwidth, trim=20 120 30 145, clip]{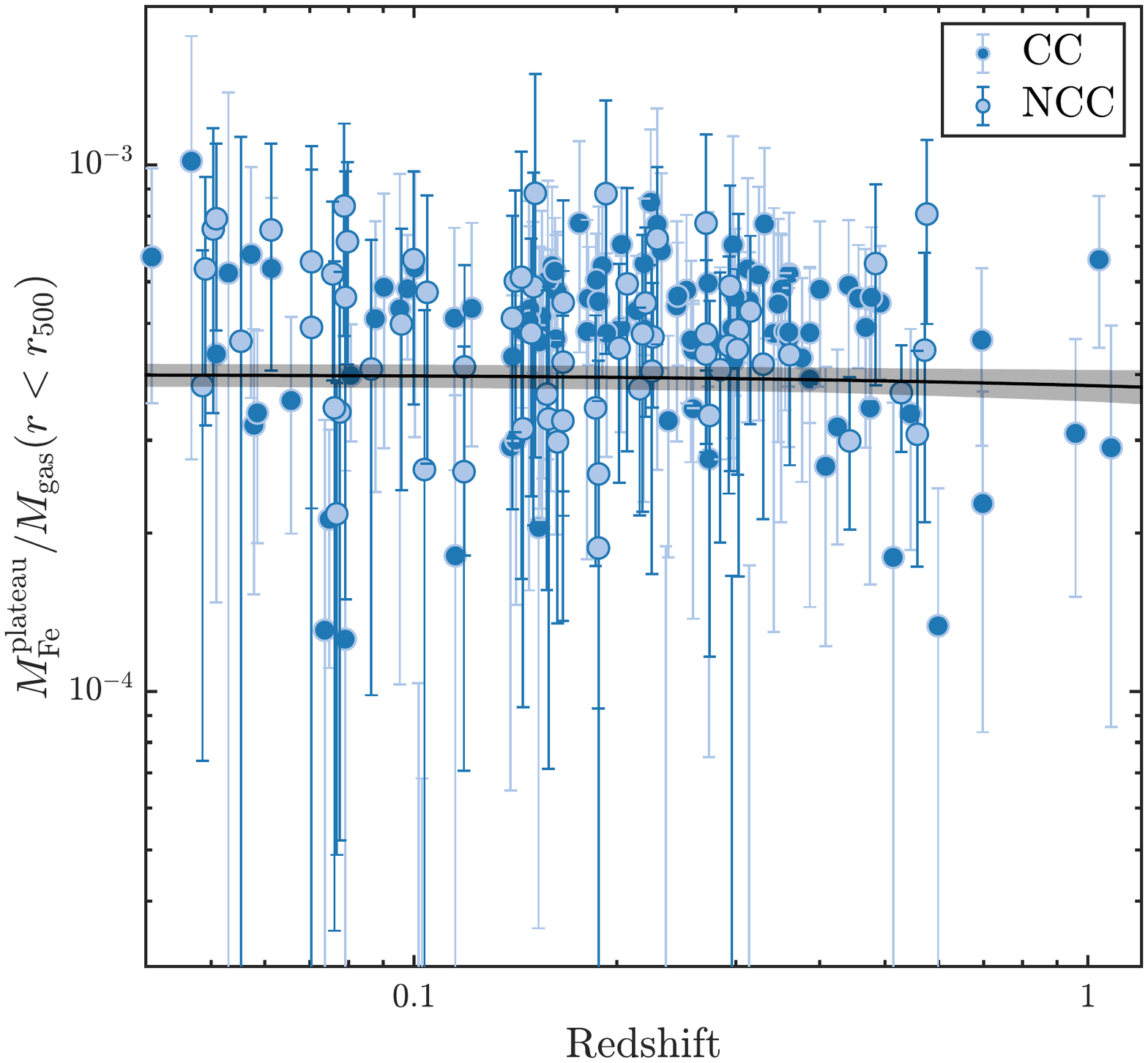}
\includegraphics[width=0.49\textwidth, trim=20 120 30 145, clip]{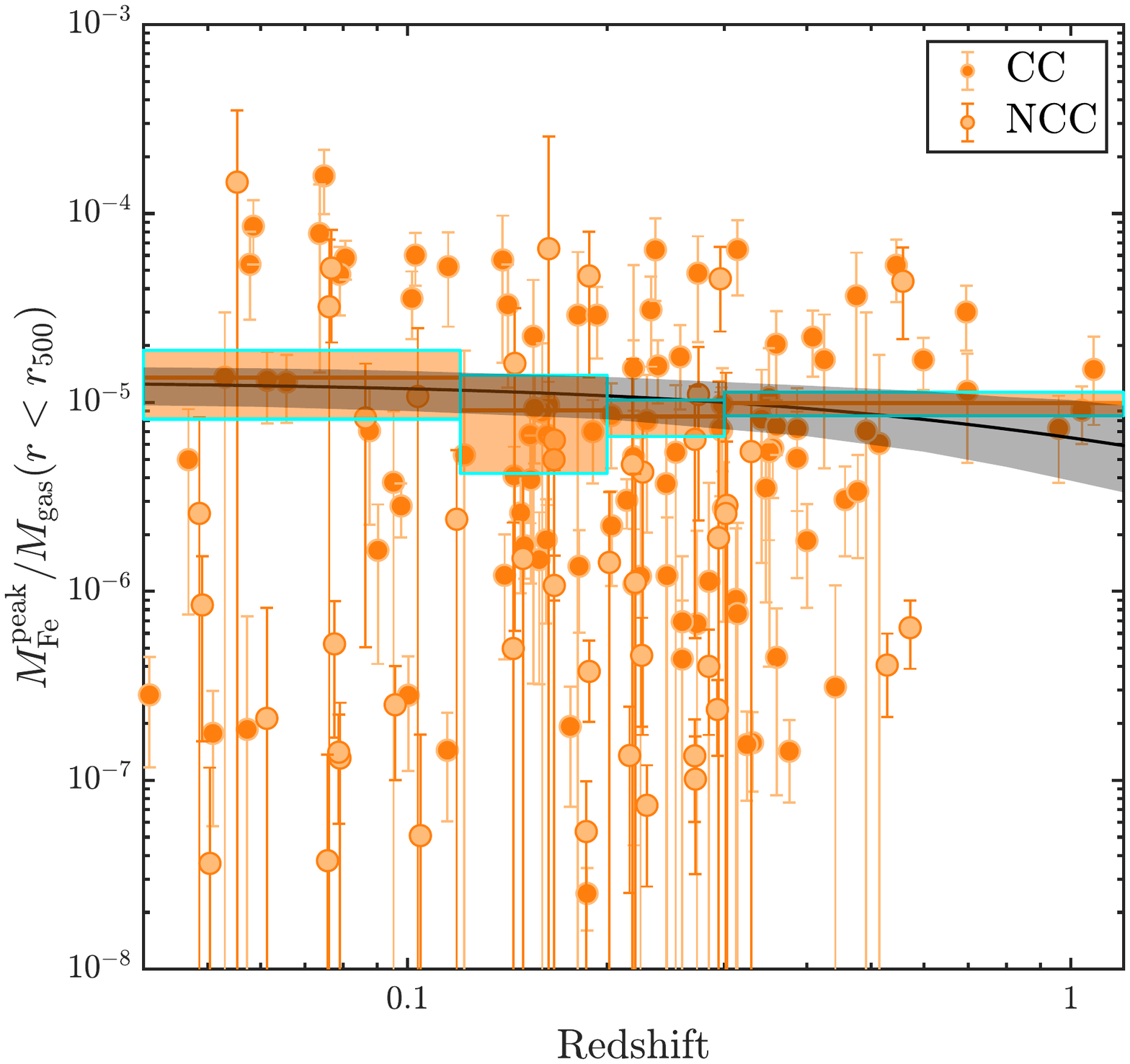}
\includegraphics[width=0.49\textwidth, trim=20 120 30 145, clip]{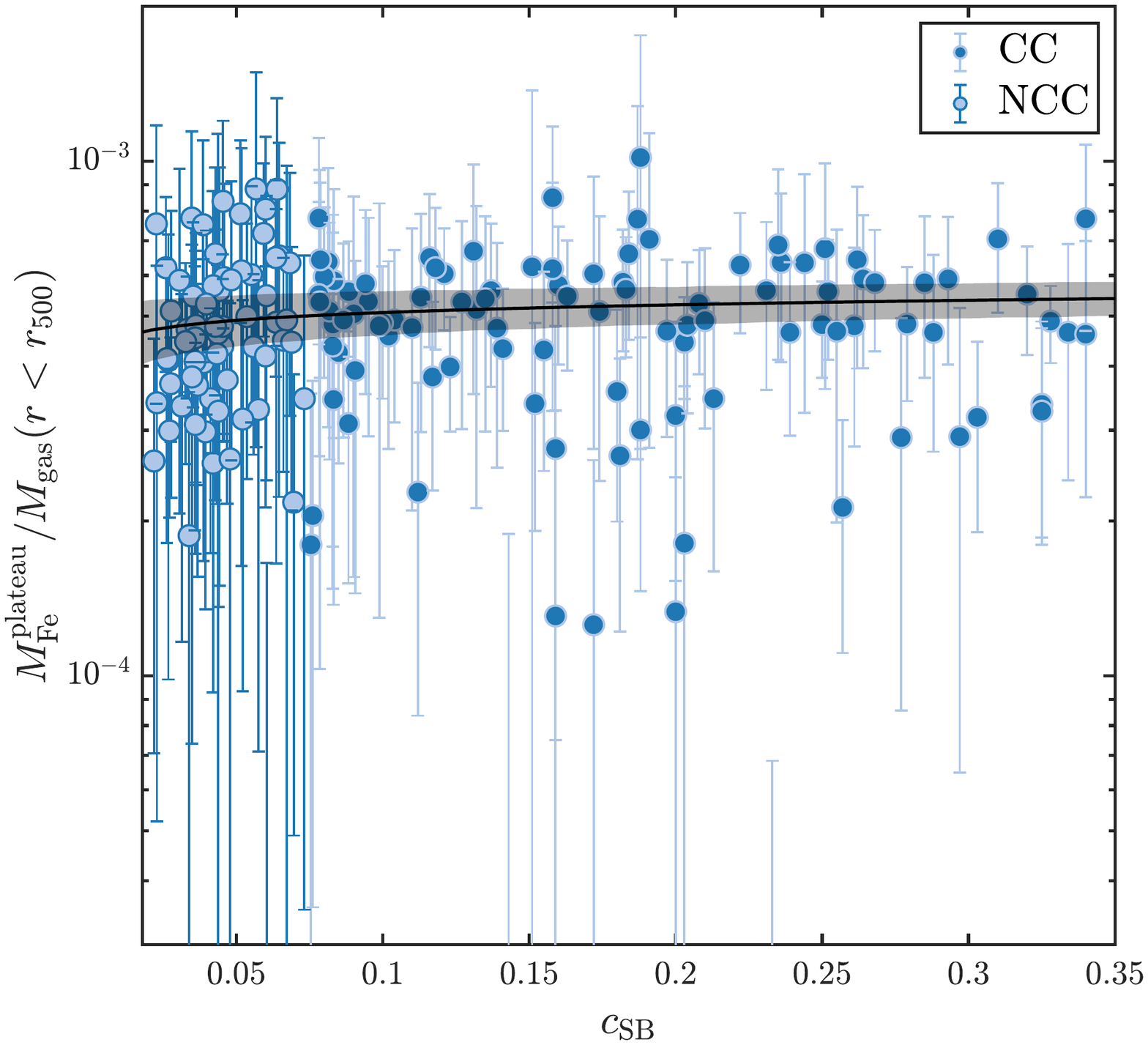}
\includegraphics[width=0.49\textwidth, trim=20 120 30 145, clip]{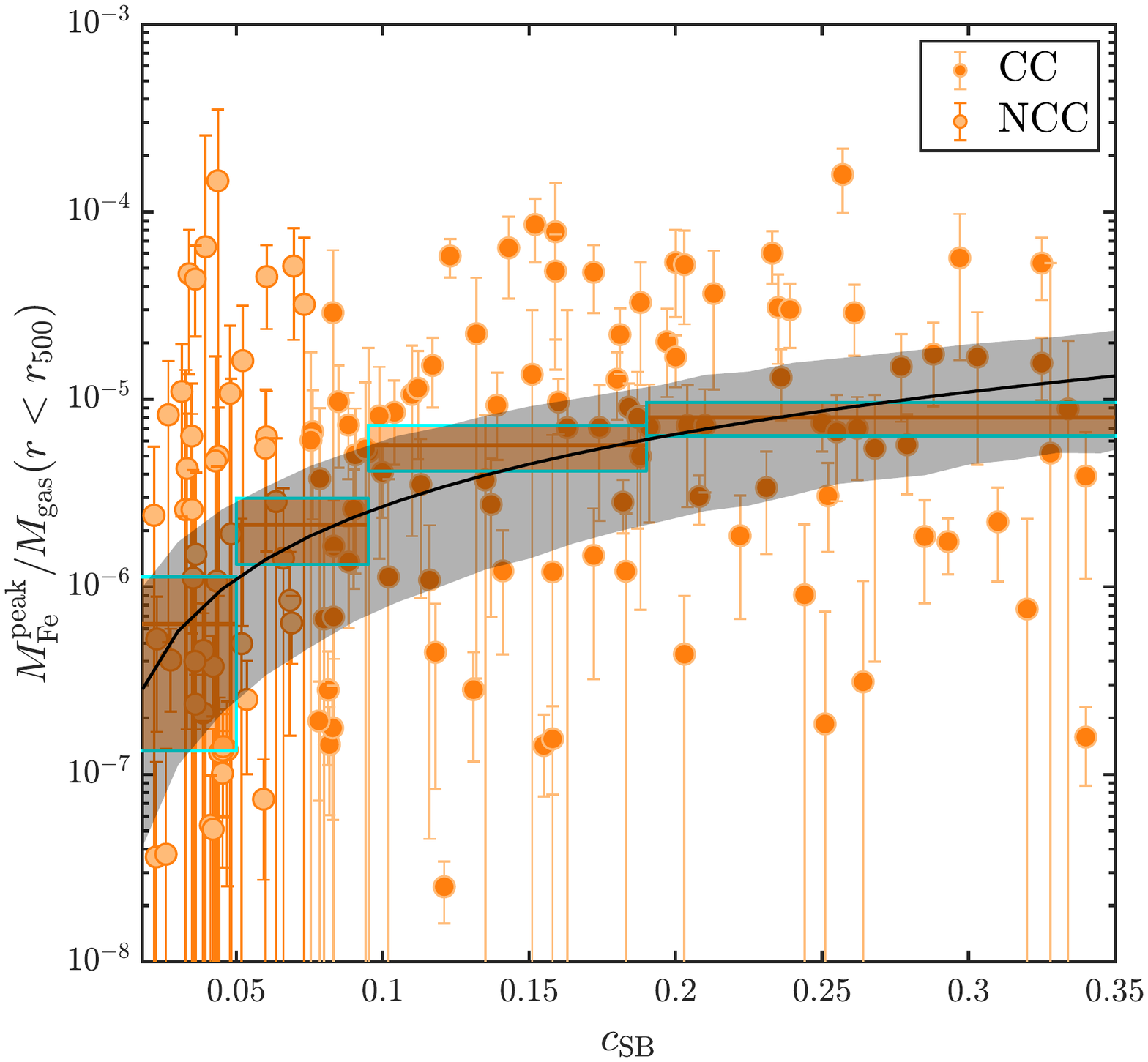}
\caption{{\sl Upper} panels: The ratio of iron mass of the plateau (left) and the peak (right) 
to the gas mass within $r_{500}$ versus cluster redshift. The black curves show the best-fit 
functions $M_{\rm Fe}^{\rm plateau}/M_{\rm gas} = 4.0\times10^{-4}\cdot (1+z)^{-0.07}$ and 
$M_{\rm Fe}^{\rm peak}/M_{\rm gas} = 1.3\times10^{-5}\cdot (1+z)^{-1.00}$.  {\sl Lower} 
panels: The correlation between the iron mass of the two components and the surface 
brightness concentration $c_{\rm SB}$. The black curves show the best-fit functions 
$M_{\rm Fe}^{\rm plateau}/M_{\rm gas} = 5.7\times10^{-4}\cdot (c_{\rm SB})^{0.05}$ 
and $M_{\rm Fe}^{\rm peak}/M_{\rm gas} = 5.4\times10^{-5}\cdot (c_{\rm SB})^{1.32}$. 
Shaded area indicates the 1 $\sigma$ confidence interval of the best-fit model. For the iron peak, the best-fits are  obtained by fitting the weighted average values and uncertainties of the four bins shown as orange solid lines and shaded areas. }
\label{ratio}
\end{center}
\end{figure*}

In this section we analyze the profiles of iron abundance and iron mass by resolving the two 
components, namely the iron plateau and the iron peak. From Figure 
\ref{mass2}, one can immediately assess the contributions of the two components to the 
iron mass budget. In Figure \ref{peak_const_hist} we show the distribution of the ratio of iron 
peak mass to iron plateau mass within $r_{500}$, and also the correlation of the ratio with redshift. No redshift-dependence of the ratio is found from Figure \ref{peak_const_hist}. Despite that the ratio for most clusters are 
distributed within the range [5$\times10^{-5}$, 0.5], and centered at 0.008, 
we find clusters with extremely low iron peak mass. We check the spectral fits and profile fits 
for these cases, and find consistently that the clusters with low iron peak mass are  
non-cool-core clusters which host no or very weak iron peak in the center. A small number of 
clusters show $M^{\rm peak}_{\rm Fe}/M^{\rm plateau}_{\rm Fe}>0.1$; in these cases the central iron distribution 
is broad and slowly declining, so that it is ascribed mostly to the central peak. These cases would 
probably be better described by a third component in the form of a shallow power-law, however, the
quality of the data makes it impossible to identify such additional component.  In these cases the
iron mass in the peak should not be associated to the BCG, but rather to the mix of the two components
that appears as a broad bump.  This is admittedly a limitation of the method, since it is impossible to
spatially separate the two components when the central peak has been smeared out.  

We also notice, from Figure \ref{peak_const_hist} and also the right panel of Figure \ref{mass2}, that a few clusters have a very low
iron plateau. We check the profiles of these clusters, and 
find that this is mostly driven from one or more measurements of very low abundance 
in the outskirts, probably due to the low S/N of the data. A bias toward low abundance values in 
the outer regions has been noticed, and it has been shown that it can be removed by excluding
the 0.9--1.3 keV rest-frame band (corresponding to the iron L band emission complex, S. Molendi 
private communication). However, due to the limited signal of the spectra in the outer bins, 
we are not able to verify this effect nor the robustness of these low measurements. 
If, in these cases, we fix the iron plateau to some value, e.g., $0.2 \, Z_\odot$, we find that the
ratio between the iron peak and iron plateau becomes, by construction, consistent with the 
average value of the sample, while the fit to the profiles are still good due 
to the poor statistical weight of the low-abundance data points in outskirts. 
This, in fact, implies that we have 
a very loose constraint on the iron plateau in several clusters, which is, nevertheless, 
already accounted for in the uncertainty of the fitting result. Given the low number 
of clusters with $Z_{\rm plateau} \sim 0$ (5 out of 170), 
these cases do not require a change of our fitting strategy nor have an impact on our final results.

We then check the normalization of the iron plateau ($Z_{\rm plateau}$) across the sample. In the 
upper panels of Figure \ref{Z1} we present the distribution of $Z_{\rm plateau}$ and its relation 
with redshift. From the upper left panel of Figure \ref{Z1} one can immediately observe that the distribution of $Z_{\rm plateau}$ fits well to a symmetrical Gaussian. This implies that the plateau is made up of many additive processes all acting independently, which is quite consistent with the picture that a wide range of randomly sampled galaxies eject out metals that are all adding up in the plateau ICM. We compute the weighted average of $Z_{\rm plateau}$ (where the weights are defined
$w_i=1/\sigma_i^2$) and find $\langle Z_{\rm plateau} \rangle = 0.38\, Z_\odot$, thus consistent with a 1/3 solar abundance of the ICM in cluster outskirts, a value that has been commonly reported by many works \citep[see][for example]{1977Serlemitsos,1996Mushotzky,2013Simionescu,2016Molendi,urban2017}.  
The arithmetic mean is also $\langle Z_{\rm plateau} \rangle = 0.38\, Z_\odot$, while the total 
scatter with respect to the mean is $0.14\, Z_\odot$.  Since the average statistical error is 
$0.11\, Z_\odot$, we can estimate the intrinsic scatter assuming 
$\sigma^2_{\rm tot}=\sigma^2_{\rm stat}+\sigma^2_{\rm intr}$, obtaining $\sigma_{\rm intr} = 0.09\, Z_\odot$.
The intrinsic scatter of the plateau normalization, therefore, is lower but not 
negligible with respect to the statistical uncertainty. This is shown in the upper-left
panel of Figure \ref{Z1}, where the histogram of the best-fit values of $Z_{\rm plateau}$ is shown
with a Gaussian centered on  $\langle Z_{\rm plateau} \rangle$ and with width equal to 
the average statistical error.  This implies that the intrinsic fluctuations in the plateau, 
which are naturally expected, amount to $\sim 25$\% of the average plateau value. 
This is consistent with a roughly uniform enrichment at high-$z$ at least
in the massive cluster range.

These results hold under the assumption of a constant plateau normalization as a function of 
redshift.  If we now focus on the evolution, from the top-right panel of Figure \ref{Z1} 
we can immediately notice 
the absence of a significant correlation with redshift. 

Given the presence of a significant scatter in $Z_{\rm plateau}$, we should not fit 
the $Z_{\rm plateau}-z$ relation with a simple $\chi^2$ minimization. To describe the properties 
of the iron plateau as a function of redshift, we decide to focus on four bins of redshift 
with a similar number of clusters, namely $z<0.12$, $0.12<z<0.2$, $0.2<z<0.3$, and $z>0.3$, 
with about 42 points each.  We inspect the histogram distribution of the $Z_{\rm plateau}$ values 
in each redshift bin, and verify that the weighted mean $\langle Z_{\rm plateau,z} \rangle$ closely
traces the peak of the distribution.  Then, we are allowed to use a $\chi^2$ minimization on the four bins to fit the behavior of the $Z_{\rm plateau}$ distribution with redshift.
We adopt an empirical function 
$Z_{\rm plateau} = Z_{{\rm plateau},0}\cdot (1+z)^{-\gamma_{\rm plateau}}$, and obtain 
$Z_{{\rm plateau},0} = (0.41\pm 0.02)\, Z_\odot$ and $\gamma_{\rm plateau} = 0.21\pm 0.18$, 
suggesting no evolution with redshift. This corroborates the hypothesis that the
plateau is dominated by the contribution from a pristine and uniform enrichment, possibly 
occurred before the virialization of the main halo. 

Then, we focus on the normalization of the iron peak $Z_{\rm peak}$. In the lower panels of 
Figure \ref{Z1}, we show the statistic of $Z_{\rm peak}$ and the $Z_{\rm peak}-z$ distribution. 
Unlike the $Z_{\rm plateau}$, whose distribution is well approximated by a Gaussian, 
the distribution of $Z_{\rm peak}$ is closer to a power law, suggesting that the underlying process
may be described by random jumps, such as intense star formation events associated to 
intermittent cooling flows and responsible for the creation and ejection of metals. 
The weighted average value of $Z_{\rm peak}$ is $\langle Z_{\rm peak} \rangle = 0.52\, Z_\odot$, 
with a {\sl rms} dispersion of $0.49\, Z_\odot$. Since the average statistical error is 
$0.42 \, Z_\odot$, the estimated intrinsic (and symmetric) scatter
is $\sim 0.26 \, Z_\odot$.  This is clearly seen in the bottom-left panel of Figure \ref{Z1}, 
where the Gaussian centered on the weighted mean and representing the width of the statistical
uncertainties, fails in describing the right side of the distribution.  The reason is that the
high-$Z_{\rm peak}$ values represent a population of clusters which experienced a relatively
low number of minor and major mergers, so that the central regions evolved undisturbed for several 
Gyr with the late, BCG-related iron piling-up in the core.  Ideally, all massive clusters 
should show a high iron peak if the mass growth is smooth, but in reality the stochastic
merger events reset the thermodynamic and chemical properties of the cores, creating the
distribution of properties we actually observe.  

The estimate of the intrinsic scatter is based on a double assumption: a symmetric intrinsic 
scatter, and a constant $\langle Z_{\rm peak}\rangle$ value with redshift. We can immediately see 
from Figure \ref{Z1} that the first assumption is not met. To test the evolution of 
$\langle Z_{\rm peak}\rangle$, we use a $\chi^2$ minimization on the four redshift bins as in the 
previous case.  Using the same empirical function for $Z_{\rm plateau}$, we obtain 
$Z_{\rm peak,0} = (0.68\pm 0.07)\, Z_\odot$ and $\gamma_{\rm peak} = 0.79 \pm 0.53$. This 
result, despite the large scatter of $Z_{\rm peak}$ across the sample, is consistent with an 
increase of $\sim 75\%$ from $z\sim 1$ to low-redshift, but it is also consistent with 
no evolution within less then $2\sigma$. Also, we need to bear in mind that, despite our sample 
spans a redshift range $0.04<z<1.1$, the weight of high-$z$ ($z>0.6$) clusters is limited, and 
the fit shown in the lower-right panel of Figure \ref{Z1} is actually driven by the data points 
at redshifts below 0.6.  In any case, a mild, positive evolution with cosmic time, if confirmed, 
supports a different origin of the iron peak, more recent in time and associated with 
the central BCG and epochs after the cluster virialization ($z<1$). In other words, the 
observed iron peaks are consistent with being formed within the cluster {\sl in situ} 
around the BCG, increasing in strength from redshift $\sim 1$ to local as the feedback 
cycle associated to the BCG creates short but intense period of star formations, with the 
associated creation and diffusion of iron. Considering the redshift distribution of our sample, 
this evolution, if any, is occurring on a time scale of about 5 Gyr, corresponding to 
the interval $0.05<z<0.6$. 

Finally, we explore the evolution of iron in terms of the iron mass of the two components. 
In Figure \ref{ratio}, we plot this two components within $r_{500}$ divided by the gas mass 
within the same radius against cluster redshift. We find that $M_{\rm plateau}/M_{\rm gas}$ 
appears to be distributed around an average value with a scatter entirely consistent with the 
statistical uncertainty, and, therefore, we can investigate its redshift dependence directly 
with a $\chi^2$ minimization.  On the contrary, the quantity $M_{\rm peak}/M_{\rm gas}$ 
shows a significant intrinsic scatter, and we adopt the same 
strategy as before, consisting in fitting the weighted mean $\langle M_{\rm peak}\rangle$
in four redshift bins.  Using the same function $X=n\cdot (1+z)^{-\gamma}$, we obtain 
$M_{\rm Fe}^{\rm plateau}/M_{\rm gas}=(4.0\pm0.2)
\times 10^{-4} \times (1+z)^{-0.07\pm 0.07}$ for the iron plateau, and 
$M_{\rm Fe}^{\rm peak}/M_{\rm gas}=(1.3\pm0.3)\times 10^{-5} \times (1+z)^{-1.00\pm 0.61}$ 
for the iron peak.  Therefore we confirm that the plateau does not seem to evolve significantly 
in this redshift range, well consistent with an early ($z>2$) and uniform enrichment. 
On the other hand, the iron peak mass shows some hint of an increase with cosmic time. 
This growth is not statistically significant, similarly to that observed in the peak 
normalization.  If confirmed, we can interpret this trend, regardless of its large 
uncertainty, large scatter, and the 
incompleteness of our sample particularly at high-$z$, as an average increase of $\sim 100\%$ 
of the amount of iron produced and/or released within the clusters in the central 
region at $z<1$. 

The relatively recent origin of the iron peak and its strong dependence on the intermittent 
star formation history in the BCG, coupled to the stochastic merger events, are corroborated by the large observed scatter, particularly if compared to the
scatter of the plateau.  For the iron plateau mass (divided by gas mass), the total 
scatter turns out to be $\sim 90$\% of the average measurement (1 $\sigma$) uncertainty, 
therefore implying no intrinsic scatter.  Once again, this suggests an early and uniform 
enrichment.  On the other hand, the distribution of the iron peak mass has a scatter 
several times higher than the statistical uncertainty: $\sigma_{\rm tot}\sim 6\times 
\sigma_{\rm stat}$.  This fact is due not only to a large diversity in the history of star 
formation episodes in the BCG responsible for the iron mass excess, abut also to the 
widely distributed dynamical age of the core, which strongly affects the ICM mass associated with 
the peak and, therefore, strongly amplifies the scatter with respect to the $Z_{\rm peak}$ 
distribution. 

We know that the fraction of mass included in the iron peak is only $\sim 1$ 
percent of the total iron mass in the ICM within $r_{500}$. 
Therefore, the evolution of iron in the ICM is dominated by the 
amplitude of the iron plateau. The lack of evolution in the iron plateau, therefore, 
drives the total (peak plus plateau) iron abundance to be almost constant with redshift, 
as we already show in Figure \ref{mw_abun}.  Our approach also demonstrates that 
considering only the global abundance is not adequate to properly constrain the evolution 
and the physical origin of the (at least) two different components of the iron distribution. 

We also investigate the possible difference in the behaviors of cool-core ($c_{\rm SB}\ge 0.075$) 
and non-cool-core ($c_{\rm SB}<0.075$) clusters. As can be seen in the top left panel of 
Figure \ref{ratio}, cool-core and non-cool-core clusters do not show a different behavior
when considering the iron plateau. However, we find from the upper right panel of Figure 
\ref{ratio} the expected result that non-cool-core clusters tend 
to be lower than cool-core clusters in the iron peak mass. To further explore the link 
between the iron peak and the presence of a cool core, we also 
investigate the correlation between the iron to gas mass ratios in the two components 
with the surface brightness concentration $c_{\rm SB}$, as shown in the lower panels 
of Figure \ref{ratio}. We fit the distribution with a simple power-law, 
$M_{\rm Fe}/M_{\rm gas}=A\cdot (c_{\rm SB})^B$. For the iron plateau, we obtain 
$A=(5.7\pm 0.4)\times 10^{-4}$ and $B=0.05\pm 0.03$.  As expected, the strength of the cool core
does not affect much the properties of the iron distribution on large scales. 
For the iron peak, we again divide the sample into four bins as before: 
$c_{\rm SB}<0.05$, $0.05<c_{\rm SB}<0.095$, $0.095<c_{\rm SB}<0.19$, and $c_{\rm SB}>0.19$, 
and fit the weighted average of the four bins. We obtain $A=(5.4\pm 3.0)\times 10^{-5}$ 
and $B=1.32\pm 0.33$, which implies a significant correlation between the amount of 
iron in the peak and the strength of the cool core.  This is largely expected, since 
the presence of a peak in the iron distribution has been always observed in association 
with a strongly peaked surface brightness, as already mentioned \citep{degrandi2004}. 
Another way to express this result is in terms of the
correlation of metallicity and density (or anticorrelation of metallicity and entropy)
observed in the cluster center. This result is in line with the 
general picture that the iron peak is associated with the formation and evolution of 
the cool core, and it is mainly created {\sl in situ} thanks to the periodic starburst 
occurring in the BCG.  Here, a still missing link is the connection of the total amount of 
iron in the peak with the integrated star formation history consistent with the old 
stellar population of BCG.  It is well known that episodic and recurrent starbursts occur in 
BCG, however a quantitative assessment of the amount of freshly ($z<1$) produced iron in the 
BCG has never been accurately measured.  The connection between the BCG star formation
history and the iron peak will be investigated in a forthcoming paper (Liu et al. in preparation).

\begin{figure*}
\begin{center}
\includegraphics[width=0.49\textwidth, trim=20 120 30 145, clip]{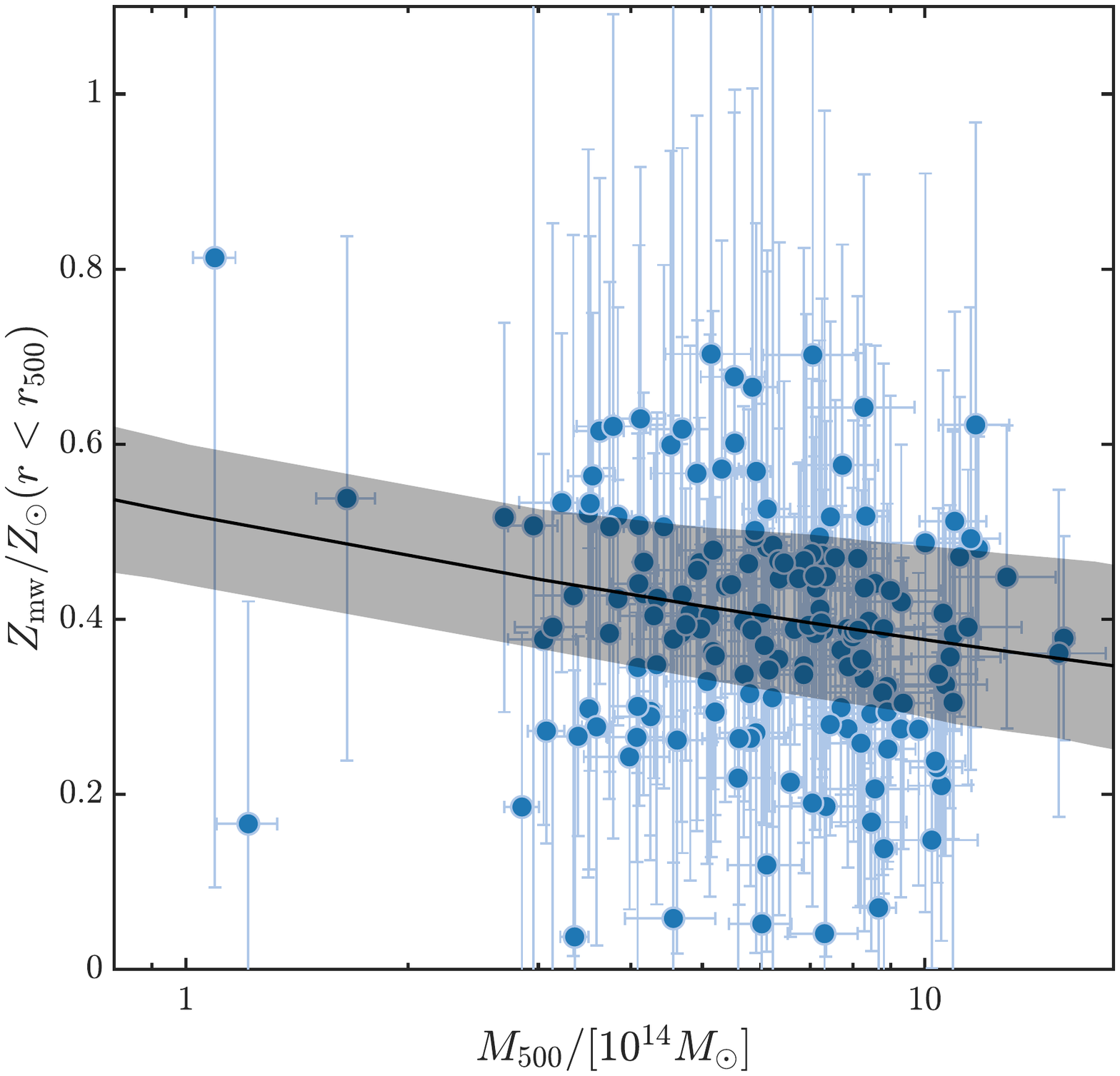}
\includegraphics[width=0.49\textwidth, trim=20 120 30 145, clip]{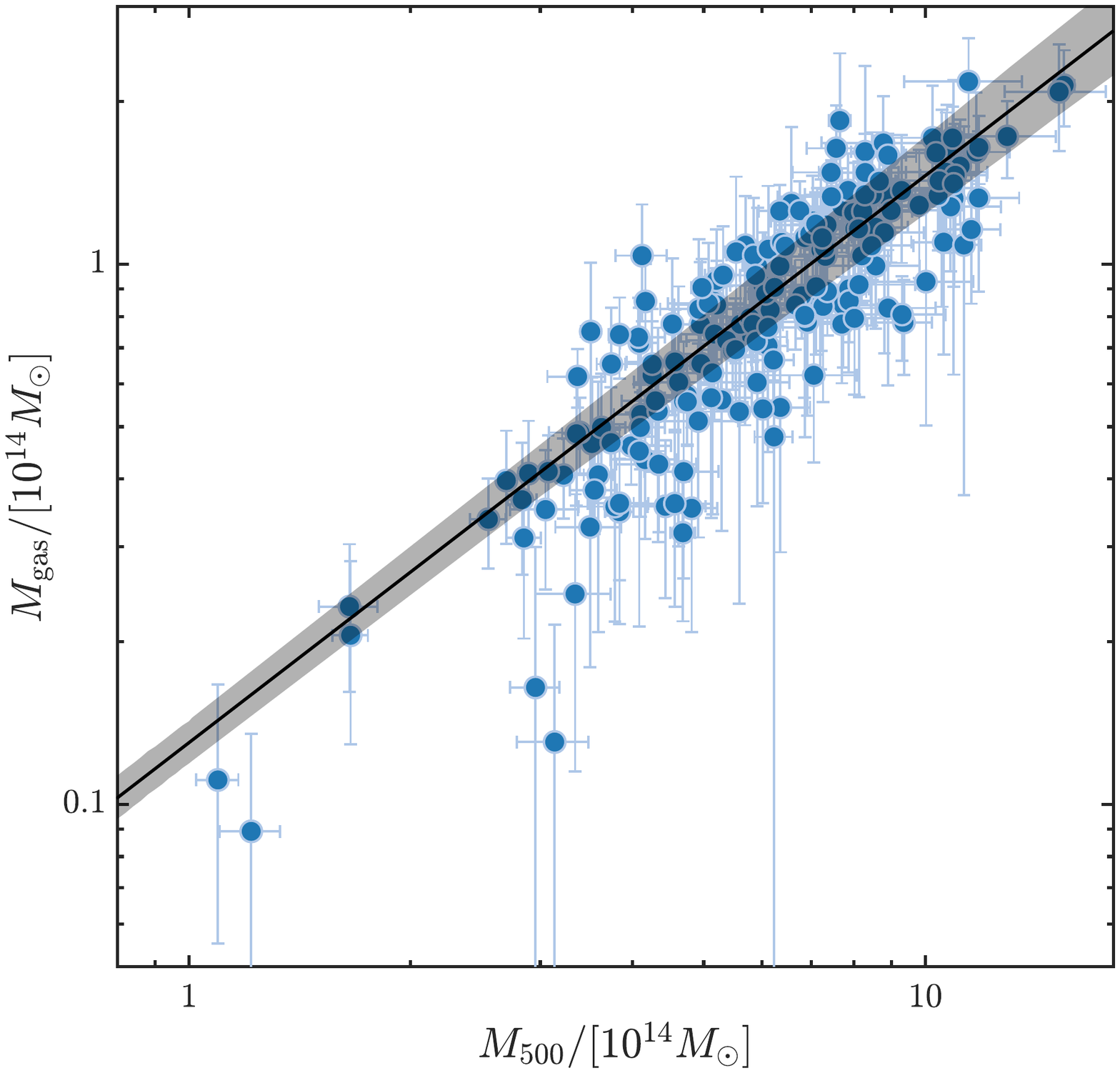}
\caption{{\sl Left}: The correlation between gas mass-weighted abundance within $r_{500}$ and the 
total mass $M_{500}$ for all the clusters in the sample. The black curve and shaded area denote 
the best fit function and the 1 $\sigma$ confidence interval: 
$Z_{\rm mw}=(0.52\pm0.08)\cdot(M_{500}/10^{14}M_{\odot})^{-0.14\pm0.09}$. 
{\sl Right}: The correlation between the gas mass and $M_{500}$. The black curve and shaded 
area denote the best fit function and the 1 $\sigma$ confidence interval: 
$M_{\rm gas}/10^{14}M_\odot=(0.131\pm0.012)\cdot(M_{500}/10^{14}M_{\odot})^{1.05\pm0.06}$.}
\label{mass_abun}
\end{center}
\end{figure*}

\subsection{The correlation between $M_{500}$ and iron abundance}

In Figure \ref{mass_abun} we explore the relation between the gas mass-weighted abundance within 
$r_{500}$ and $M_{500}$, the total mass of the cluster within $r_{500}$. As usual, we use a 
simple power-law 
$Z_{\rm mw}=Z_0\cdot (M_{500}/10^{14}M_{\odot})^{-\alpha}$. We perform a linear regression 
for the ${\rm log}(Z)-{\rm log}(M)$ relation requiring a minimization of the orthogonal distance 
of the points from the best-fit relation, thus considering uncertainties in both
quantities.  The best-fit gives 
$Z_0=(0.52\pm 0.08) Z_\odot$ and $\alpha=0.14\pm 0.09$.  This implies that in the 
mass range [3--10]$\times 10^{14} M_\odot$, the global gas mass-weighted abundance change 
only by less than 20\%, being slightly higher at lower halo masses. Given the small mass 
range probed here, 
this is comparable to the correlation of stellar mass with halo mass found by \citet{2012Lin}.  
Taken at face value, this relation would imply a rapidly increasing average abundance at lower
masses (below $2\times 10^{14} M_\odot$ into the group regime), a range which is not 
explored here, in the assumption that
the entire stellar-mass budget is contributing to the chemical enrichment of the ICM.  
We must bear in mind though, that \citet{2012Lin} do not include the contribution of the 
intracluster light \citep[see][and references therein]{2014Presotto}, that can be larger at 
high masses, flattening the stellar mass - halo mass relation.   

In the right panel of Figure \ref{mass_abun} we also show the correlation between the gas 
mass (within $r_{500}$) and the total mass $M_{500}$. Also in this case the statistical
errors in both quantities are considered. The best-fit function of 
$M_{\rm gas}/10^{14}M_{\odot}=A\cdot (M_{500}/10^{14}M_{\odot})^{B}$ gives 
$A=0.131\pm 0.012$ and $B=1.05 \pm 0.06$.  The slope we find here is still consistent with 
the value of $1.13 \pm 0.03$ found by \citet{2012Lin}, which implied 
that smaller mass halos have slightly less ICM within $r_{500}$ compared to the most massive 
clusters. However, despite the agreement, we do not find statistically significant evidence 
for this trend in our sample.
If we include also a dependence on redshift of the form $(1+z)^{\gamma_3}$ we find 
$A=0.130\pm 0.010$, $B=1.04\pm 0.06$ and $\gamma_3=0.08\pm 0.21$, again in good agreement 
with \citet{2012Lin} and with previous claims by \citet{2009Vikhlinin}.  However, we remark that 
the observed evolution of $f_{\rm ICM}$ depends on cosmology \citep[see][]{allen2011}, and 
therefore we do not discuss possible physical implications for the trend found here.

A comprehensive discussion on the correlations between the integrated quantities
(global metallicity, temperature, mass) is postponed to a forthcoming work.

\section{Discussion}

In this Section we discuss several aspects, ranging from the control of the systematics in 
our spectral analysis\footnote{We do not discuss two potential sources of systematics such 
as the calibration of the X-ray instruments and the plasma code used to fit the data, both of which 
are expected to affect the iron abundance at the level of few percent 
\citep[see][and references therein]{2016Molendi}.}, to the physical interpretation of our results, 
and a direct comparison with previous works.  We start from this last aspect, which is relevant 
here since the original motivation of this work was the contradictory results obtained in the 
last 10 years on the iron evolution in the ICM.

\subsection{Comparison with previous works}

\begin{table}
\caption{\label{gamma} Slope of the relation between average abundance and redshift from the literature, compared
to the value found in this work. $r (r_{500})$ shows the radial range used in each work. }
\begin{center}
\begin{tabular}[width=0.5\textwidth]{lccc}
\hline
Reference & $r (r_{500})$ & Sample & $\gamma$ \\
\hline
(1) & 0.15--1  & 111 & $1.63\pm0.35$ \\
(2) & 0--0.15  & 70 & $1.60\pm 0.22 $ \\
(2) & 0.15--0.4  & 83 & $0.70\pm 0.32 $ \\
(2) & $>0.4$  & 68 & $0.26\pm 0.61 $ \\
(3) & 0--1  & 153 & $0.41\pm0.25 $ \\
(3) & 0.15--1  & 153 & $0.03\pm0.06 $ \\
(4) & 0--0.1  & 186 & $0.14\pm0.17 $ \\
(4) & 0.1--0.5  & 245 & $0.71\pm0.15 $ \\
(4) & 0.5--1  & 86 & $0.30\pm0.91 $ \\
This work & 0--1  & 170 & $0.28\pm0.31 $ \\
\hline
\end{tabular}
\end{center}
\tablebib{
(1) \citet{maughan2008}; (2) \citet{2015Ettori}; (3) \citet{mcdonald2016}; (4) \citet{2017Mantz}.
}
\end{table}

\begin{figure}
\begin{center}
\includegraphics[width=0.49\textwidth, trim=20 120 30 150, clip]{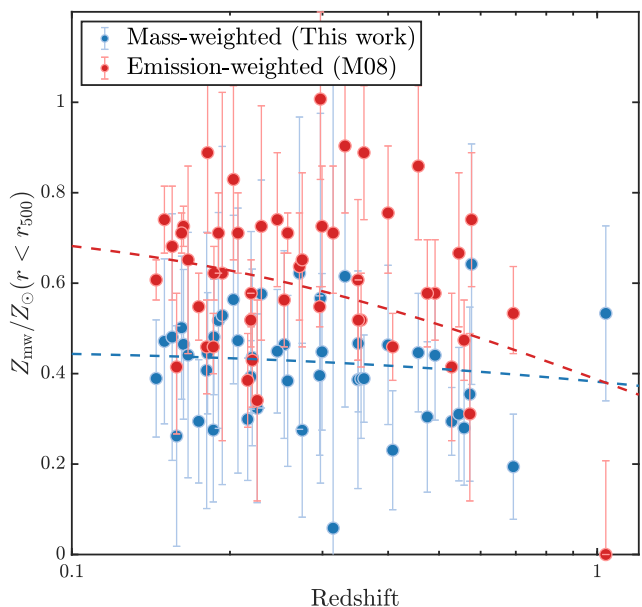}
\caption{Comparison between the gas mass-weighted abundance measured in this work, and the 
emission-weighted abundance provided by \citet{maughan2008}, for the overlapping 48 clusters. 
The reference solar abundance for the data points of \citet{maughan2008} has been adjusted 
from \citet{1989Anders} to \citet{asplund2009}, for a direct comparison with our results. }
\label{m08}
\end{center}
\end{figure}

Since the beginning of {\sl Chandra} and XMM era, a number of attempts have been made to explore 
the cosmic evolution of ICM metal abundance. Although early works such as \citet{tozzi2003} 
and \citet{2007Balestra} are likely to suffer from the limited size of cluster sample and the 
low S/N of the data, more recent works based on large samples and deeper observations 
present contradictory results, and no obvious solution could be found to explain
the observed discrepancies. Both the selection of the sample and the extraction radius 
used to measure the ICM abundance vary in these works, which increases the difficulty in 
comparing the previous results. Another issue is that all these previous results are obtained 
from emission-weighted measurements, thus might be influenced by the fraction of cool-core 
clusters in the sample. This effect, as we know, would be mitigated, if not eliminated, once the gas
mass-weighted abundances are used. Given these caveats, for the sake of comparison with the literature
we consider only the evolution measurements based on the core-excised clusters, namely, using
a radial range of 0.1 (or 0.15)--1 $r_{500}$. The functional form $Z\propto (1+z)^{-\gamma}$ 
is commonly used to quantify the evolution, thus we can directly compare the slope 
$\gamma$. We list the measurements of $\gamma$ in different works in Table \ref{gamma}, 
and compare them to the result we obtained by fitting the distribution of the gas
mass-weighted abundance within $r_{500}$ in this work. 

In general, we find that our results for the evolution of the gas mass-weighted abundance 
($\gamma=0.28\pm 0.31$) are consistent with most of the previous literature, based on 
emission-weighted measurement, with some noticeable exception. We first consider the results
of \citet{2015Ettori}, where significant and strong evolution is found in the innermost regions
$r<0.1\, r_{500}$, weaker and much less significant evolution is found at larger radii. 
While the evolution at small radii may be reconciled with our $\sim2\sigma$ evolution in the peak
component, we reconsider their results and average the [0.15--0.4] $r_{500}$ and $>0.4\, r_{500}$ bins, 
finding $\gamma=0.48\pm 0.34$, which is now consistent with our results.
We also compare with the results of \citet{mcdonald2016}, where $\gamma=0.03\pm0.06$ 
for [0.15--1] $r_{500}$, with a $\sim 2 \sigma$ evolution when considering the full 
[0--1] $r_{500}$ range.  Again, given the expected effects of the emission-weighted analysis, we 
are fairly consistent with their results.

More difficult to reconcile are the results by \citet{2017Mantz}, where 
no evolution is found in the very [0--0.1] and in the [0.5-1] $r_{500}$ radial range, while
a mild but very significant evolution is found for the range [0.1--0.5] $r_{500}$.
If we try to average their [0.1--0.5] $r_{500}$ and [0.5--1] $r_{500}$ bins, however, we find 
$\gamma=0.51\pm 0.46$ which is again consistent with our findings.  
The point here is that a spatially-resolved analysis focusing on fixed radial bins normalized 
to $r_{500}$ may not be adequate to  properly follow the different components,
so that the evolutionary behavior for fixed radial bins may return unstable results, unless
the size and shape of each different component is properly identified.

Finally, the most discrepant result with respect to our findings is \citet{maughan2008}, 
where $\gamma=1.63\pm 0.35$ is obtained based on a sample of 111 
clusters\footnote{The sample of \citet{maughan2008} consists of 115 clusters, 
in which 111 clusters have measurements of metallicity within the extraction 
radius [0.15--1]$r_{500}$}, suggesting a strong evolution at $>4\sigma$ c.l. 
However, we notice that a 
number of high-redshift clusters in the sample of \citet{maughan2008} have only upper limits
for the measured abundance, often with small uncertainties. 
This is probably due to the low S/N of the data. 
For example, \citet{maughan2008} report an abundance $Z=0.00_{-0.00}^{+0.21}\, Z_\odot$ \citep[adopting the solar abundance of][]{asplund2009} for  
CLJ1415.1+3612 at redshift 1. However, recent studies using deeper observations have 
presented much higher abundance of this cluster. For example, \citet{santos2012} present
a result of $Z=0.88\pm 0.11$ within $\sim 0.35 r_{500}$ \citep[see also][]{degrandi2014}, 
and in this work we consistently measure $Z=0.76\pm 0.12$ within 0.4$r_{500}$. 
If we remove this single data point, the best-fit slope decreases immediately from 
$1.63\pm 0.35$ to $0.76\pm0.35$. Therefore, we suppose that the evolutionary signal detected 
by \citet{maughan2008} is likely to be caused, at least partially, 
by several low S/N clusters at high redshifts which return unreliable low values of metallicity.

Moreover, we also explore the 
impact of the use of emission-weighted abundance on the evolutionary signal found 
in \citet{maughan2008}. Among the 115 clusters in the sample of \citet{maughan2008}, 
48 are also in our sample. We therefore make a comparison between the gas mass-weighted abundance 
measured in this work, and the emission-weighted value provided in \citet{maughan2008}, 
as shown in Figure \ref{m08}. As expected, no significance evolution is found in the gas
mass-weighted abundance, with $\gamma=0.25\pm0.28$, in perfect agreement with the full 
sample. On the other hand, for the emission-weighted abundance  
we obtain a best-fit $\gamma$ of $0.95\pm0.39$. This result indicates that the use 
of emission-weighted abundance also contributes to the claim of evolution previously reported. 
In conclusion, past claims of evolution (or no evolution) are most likely due to the combined 
effect of the use of different sample selections (mostly X-ray versus SZ selected samples 
with different cool-core fractions) and emission-weighted abundance values, with the latter 
amplifying the effect of sample selection, especially in small samples. 
The use of larger samples, with a mixed selection function, and of gas mass-weighted measurements, 
shows a small amount of evolution within $r_{500}$, if any, which is, according to our
results, limited to the iron peak.

\subsection{Projection effects}

\begin{figure}
\begin{center}
\includegraphics[width=0.49\textwidth, trim=20 125 30 145, clip]{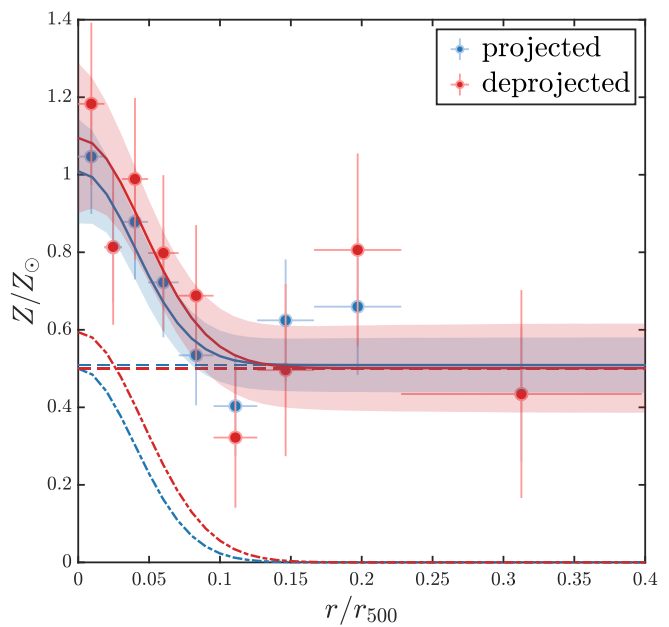}
\caption{The projected and deprojected iron abundance profiles of Abell 383: an example to show the impact of projection effect. The iron peak and iron plateau components in the best-fit model are plotted with dash-dotted lines and dashed lines, respectively.
 }
\label{a383}
\end{center}
\end{figure}

In this section we will estimate the impact of projection effect on the iron abundance profiles, 
and consequently the final results. We repeat our analysis in a fraction (about 10\%) of 
our sample including clusters with the steepest abundance gradient, to maximize the 
effects of projection.  We find that, as expected, the plateau is almost
unaffected by deprojection.  Instead, the deprojected peak values are found, in a few 
cases, 10--20\% higher than the projected ones.  In particular, the normalization of the 
deprojected abundance in the iron peak is about 10--20\% larger than the projected values, 
and a larger effect, up to a maximum of $\sim$40\%, is found for the deprojected iron peak 
mass. This is because a small change in the extension of the peak component is amplified 
by the volume effect. 

We show the case of Abell 383, that 
hosts a strong cool core, a steep iron peak and an intermediate S/N, therefore
it suffers a strongest projection effects among the clusters in our sample. 
The projected and deprojected 
abundance profiles of Abell 383 and the best-fit models are shown in Figure \ref{a383}. 
We find that the deprojection of the profile only induces a negligible change on the iron 
plateau. The main reason for this is that the plateau is dominated by the outer regions, 
where the projection effects are weak if not absent due to the flat temperature
profile. Therefore, since the mass of the iron plateau contributes $>95\%$ of 
the total iron mass budget, the impact of the projection effect on the total iron mass and 
the gas mass-weighted iron abundance within $r_{500}$, is also negligible. 

Instead, the iron peak is somewhat amplified by deprojection. In the case of 
Abell 383, the differences in the total iron mass, the iron plateau mass, and the gas
mass-weighted abundance within $r_{500}$ between projected and deprojected 
values are all lower than 3\%. On the other hand, the normalization of the iron peak measured 
from the deprojected profile is $0.59\pm 0.21$, $\sim$18\% higher than that measured from the 
projected profile $0.50\pm 0.14$. The deprojected iron peak mass within $r_{500}$ is $\sim$43\% 
higher than the projected value, since the slightly larger extension of the peak are 
amplified by the volume weighting. This increase, however, is smaller than the 1$\sigma$ 
statistical errors, which, averaged over all the clusters in our sample, are equal to 48\% and 
67\% for  the normalization of the iron peak and the iron peak mass within $r_{500}$, respectively. 
Since Abell 383 hosts one of the strongest iron peaks in our sample, these results can be 
conservatively taken as an upper limit of the magnitude of projection effect. 

To summarize, we acknowledge the fact that the strongest iron peaks in our analysis may be biased
towards lower values by $\sim few\times 10$\%, mostly because of the uncertainty in the extension of 
the iron peak, which is a key quantity in determining $M_{\rm peak}$.  We are also aware that 
any unnoticed irregular morphology in the central region may alter the deprojection results, 
which are based on a perfect spherical asymmetry, and, even in the absence of such features, 
it is prone to amplification of noise. In other words, by applying deprojection, we risk to 
introduce large random uncertainties possibly larger than the bias we aim to correct.
Clearly, accurate and stable quantification of projection effect is possible only by taking 
into account many aspects, including the 3D morphology of the cluster emission, and the 3D 
distributions of temperature and iron abundance, which is far beyond the goal of this paper, 
as well as the current state of the art. Therefore, we decide to ignore the deprojection correction
in this work, and to present results based on the projected values of iron abundance.

\subsection{The impact of $n_{\rm H}$ on the measurement of iron abundance}

\begin{figure}
\begin{center}
\includegraphics[width=0.49\textwidth, trim=15 120 30 145, clip]{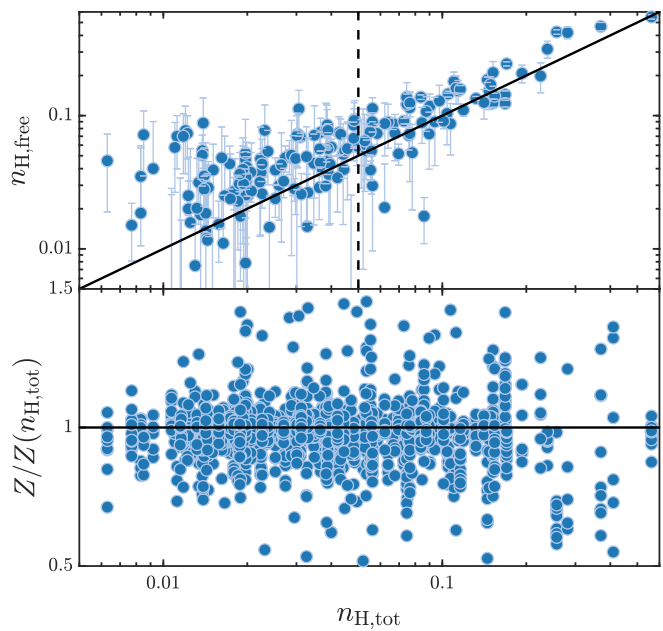}
\caption{{\sl Upper panel:} Comparison of the measured $n_{\rm H,free}$  and $n_{\rm H,tot}$. 
The $n_{\rm H}$ values are in units of $10^{22}~{\rm cm}^{-2}$. 
The vertical dashed line denotes 0.05$\times 10^{22}~{\rm cm}^{-2}$.  
The solid line corresponds to  $n_{\rm H,free}=
n_{\rm H,tot}$. {\sl Lower panel:} the ratio of all the best-fit abundance values 
obtained by adopting our strategy in $n_{\rm H}$ (free to vary in a limited interval around $n_{\rm H,free}$) and by fixing $n_{\rm H}$ to $n_{\rm H,tot}$, as a function of $n_{\rm H,tot}$. 
}
\label{nh}
\end{center}
\end{figure}

The HI column density which quantifies the absorption of X-rays is an important factor 
that affects the fitting of X-ray spectra, in particular the thermal continuum, 
and, therefore, it also affects the measurement of ICM metallicity. 
Usually there are three strategies to set the value of $n_{\rm H}$: i) $n_{\rm H,LAB}$, the 
measurement from the Leiden/Argentine/Bonn (LAB) survey \citep{2005Kalberla}, which only 
takes into account the neutral hydrogen; ii) $n_{\rm H,tot}$ from \citet{2013Willingale}, 
which also calculates the contribution of molecular and ionized hydrogen; iii) $n_{\rm H,free}$, 
which is obtained directly by fitting the X-ray spectrum with $n_{\rm H}$ set as a free parameter. 
The impact of these different values of $n_{\rm H}$ on the measurement of metal abundance 
has been discussed in detail in \citet{lovisari2019}. In general, $n_{\rm H,tot}$ provides a 
better fit to the spectrum than $n_{\rm H,LAB}$. However, \citet{lovisari2019} also find in 
a few cases that using $n_{\rm H,tot}$ may not be accurate in several cases.  
A safer strategy is obtained by setting $n_{\rm H}$ free, and putting some constraint 
on the range of possible 
values, to avoid strong degeneracies with other parameters. This allows us to identify 
the preferred $n_{\rm H}$ value without being too far from the reference value $n_{\rm H,tot}$. 
However, a caveat here is that the 
time-dependent contamination corrections of ACIS at low energies (below 2 keV) 
introduces some uncertainties in the {\sl Chandra} calibration, and this may also bias 
the measurement of $n_{\rm H}$. Therefore, we stress that leaving $n_{\rm H}$ free and 
setting a constraint according to $n_{\rm H,tot}$, despite being an already very 
conservative strategy, does not always guarantee to return the correct
$n_{\rm H}$ value. In this framework, we adopt a two-step strategy: 
we fit the spectrum of the global emission by setting $n_{\rm H}$ free to vary below a 
very loose upper limit at 10$\times n_{\rm H,tot}$, and obtain the $n_{\rm H,free}$; 
then for the spatially-resolved analysis, we adopt the measured $n_{\rm H,free}$, and 
allow it to vary within its statistical confidence interval, or within $\pm$50\% when its 
uncertainty is too small. With this strategy, we take into account not only the possible 
discrepancy between $n_{\rm H,tot}$ and $n_{\rm H,free}$, but also the fluctuation of 
$n_{\rm H}$ within the field of view of the cluster. We show in the upper panel of Figure \ref{nh} the comparison 
between $n_{\rm H,tot}$ and $n_{\rm H,free}$. Similarly to the result of 
\citet{lovisari2019}, we find a general agreement between $n_{\rm H,tot}$ and 
$n_{\rm H,free}$ above 0.05$\times 10^{22} {\rm cm}^{-2}$, 
while the discrepancy is relatively large for low column densities, for which the best fit values
are systematically larger than $n_{\rm H,tot}$, despite still consistent within $1\sigma$. 
In the lower panel of Figure \ref{nh}, we plot the ratio of all the abundance measured by 
adopting our strategy in $n_{\rm H}$ (free to vary in a limited interval around $n_{\rm H,free}$)
and by fixing $n_{\rm H}$ to $n_{\rm H,tot}$, as a function of $n_{\rm H,tot}$. We find only 
a slight bias of few percent, $Z/Z(n_{\rm H,tot}) \sim 0.98$, with no dependence on $n_{\rm H,tot}$.  
Despite it is hard to decide whether $n_{\rm H,free}$ or $n_{\rm H,tot}$ is better to describe the 
absorption effects of the HI Galactic column density, we conclude that any effect related to 
$n_{\rm H}$ is under control and it does not bias our results.  

\subsection{Physical interpretation of the results and future perspectives}

Our results are in general agreement with the well established picture that the bulk of iron 
in the ICM is produced at early epochs, approximately at $z>2$, well before cluster virialization
\citep[see][]{2020willis}. On the other hand, the signal of evolution of the iron peak, despite 
statistically weak, shows the complex effects of more recent processes, occurring after cluster
virialization (roughly $z<1$), including the star formation and 
supernova explosion in member galaxies, the galaxy-scale dynamical activities which eject metals 
from the galaxies to the ICM, and the ICM motions induced by AGN feedback activities that 
continuously transport the metals from the cluster center to the outer regions. In particular, 
in \citet{liu2018} we have shown the spatial broadening of the iron peak in a sample of 41 
most relaxed {\sl Chandra} clusters. However, in order to measure the size of the iron peak 
accurately, the sample of \citet{liu2018} was selected with a strict requirement on cluster 
relaxation, and we were not able to obtain any significant constraint on the evolution of 
the iron peak in mass. In this work, using a $>4$ times larger sample, we  
find weak evidence for the evolution of the normalized mass content iron peak. 
We stress, however, that both the samples 
of \citet{liu2018} and this work, suffer by a not-well-defined selection in the total cluster mass. 
A coherent evolutionary picture of iron in the ICM will be reached only with a 
complete and large sample of galaxy clusters.

At present, the best sample is provided by the combination of the {\sl Chandra} and 
XMM-{\sl Newton} archives. Depending on the mass selection one could adopt, the total number 
of clusters may reach $\sim 1000$.  Clearly, a proper spatially resolved analysis is feasible only
for about 1/3 of the sample, while most of the clusters could be characterized by
the X-ray morphology, a single temperature, and a single, average abundance.  
The abundance measurement would be emission-weighted, unless some priors on 
the abundance profile are assumed, to be combined with the observed surface brightness profile. 
In practice, the application of what we have learned with this sample to the largest sample 
that can be assembled today, would provide us with the most comprehensive study on ICM chemical 
evolution.  Clearly, the ultimate test on the robustness of this approach would be available 
only when a mission like {\sl Athena} \citep[e.g.,][]{2019Barret}, {\sl Lynx} \citep{2019Vikhlinin}
or AXIS \citep{2019Mushotzky}
will provide the access to a large number of well-characterized, spatially resolved cluster in a 
wide range of halo mass and redshift. In particular, 
we stress the importance of keeping in line with a {\sl Chandra}-like resolution, which is the 
only mean to sample the X-ray emission at scales below 10 kpc at any redshift, a mandatory requirement 
to properly investigate the core properties of medium and high-$z$ clusters. 

The present-day perspective allows a significant improvement only for relatively nearby clusters.  
XMM-{\sl Newton} and, more slowly, {\sl Chandra}, can provide a steady growth in the number
of spatially resolved clusters through pointed observations. At the same time, the survey 
mission eROSITA \citep[see][]{2012Merloni} will dramatically increase the number of known 
clusters.  However, its moderate resolution, but mostly the limited energy range of its 
spectral response, implies that abundance profiles can be attempted only for massive, medium 
and high-$z$ clusters, for which the hydrogen and helium-like
iron line complex enters in the observed energy range.  This however, occurs for clusters 
that would require a better angular resolution for the approach outlined here, so that the 
amount of eROSITA data that can be used here is necessarily limited.  Finally, in the next future, 
XRISM will be able to add a substantial piece of information by observing the outer regions of 
local clusters with poor angular resolution (with a PSF of $\sim 1.5$ arcmin)
but 10 times better spectral resolution thanks to the soft X-ray calorimeter {\tt Resolve}.  
Despite XRISM will be most efficient in tracing the global (i.e., not spatially resolved)
evolution of various metal abundance ratios (e.g., Si/Fe) over cosmic time, the outer regions
of at least a few bright, nearby clusters, constitute a potentially interesting target for
XRISM.

In summary, the only perspective before the advent of {\sl Athena}, or {\sl Lynx}, 
is to invest on the characterization 
of local and moderate redshift cluster to improve the educated guess applied to the high-redshift 
cluster sample available to date. This step would constitute, in our view, the 
state-of-the-art picture that could be ever achieved before the year 2030.

\section{Conclusions}

\begin{figure}
\begin{center}
\includegraphics[width=0.49\textwidth, trim=10 80 10 100, clip]{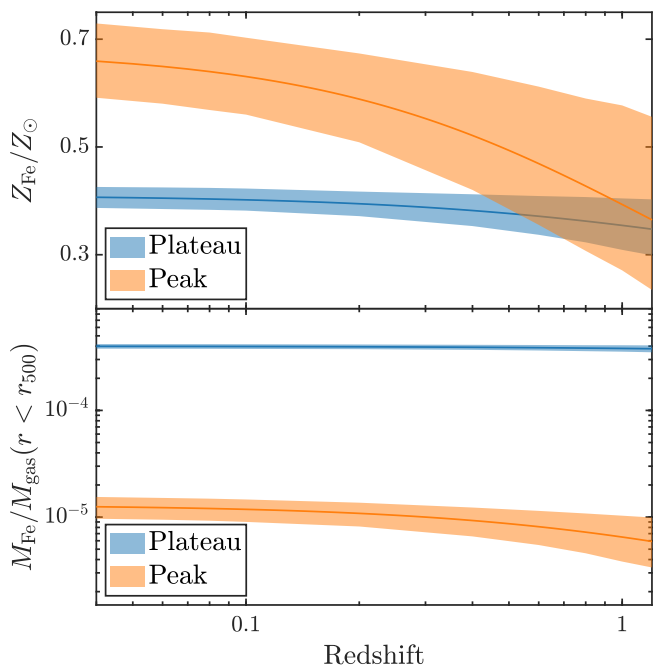}
\caption{{\sl Upper panel}: the dependence of $Z_{\rm Fe}^{\rm peak}$ and $Z_{\rm Fe}^{\rm plateau}$ 
on redshift 
as described by the best-fit power-law $X=n\cdot(1+z)^{-\gamma}$, with its $1\sigma$ uncertainty. 
{\sl Lower panel}: the dependence of $M_{\rm Fe}^{\rm peak}/M_{\rm gas}$ and $M_{\rm Fe}^{\rm plateau}/M_{\rm gas}$ 
on redshift.  These two plots represent a synthetic description of the cosmic evolution of the
iron abundance obtained in this work.  The curves and shaded areas have been already 
shown in the right panels of Figure \ref{Z1} and the upper panels of Figure \ref{ratio}.  }
\label{conclusion_fig}
\end{center}
\end{figure}

We measure the amount of mass in iron in a sample of galaxy clusters observed with {\sl Chandra}. 
We select 186 morphologically regular clusters in the redshift range [0.04, 1.07], 
from deep and medium-deep {\sl Chandra} archival observations.  Most of the clusters 
in the sample are found at $z<0.6$, so that any evolutionary behavior would reflect
this range of redshift, corresponding to about 5 Gyr.  The mass range of $M_{500}$  is 
[1--16]$\times 10^{14}M_{\odot}$, with the large majority of the clusters spanning the
[3--10]$\times 10^{14}M_{\odot}$ interval.  For each cluster we compute the azimuthally-averaged 
iron abundance and gas density profiles.  We fit the iron abundance profile with a two-component
model, namely a peak in the center, and an approximately constant plateau across the entire 
cluster.  In a few cases, we need to model a central drop in the iron abundance, as we 
already explored in a limited sample of nearby clusters \citep{2019Liu}.
This approach is physically motivated by a picture in which the central peak is 
associated with relatively recent ($z<1$) star formation in the BCG, occurring after the
virialization of the cluster.  Moreover, the almost constant plateau extending to large radii 
is possibly associated with uniform, early enrichment before cluster virialization ($z>2)$.
With this approach we are able to derive the total iron mass (and therefore the gas mass-weighted 
average iron abundance of the ICM) in each component separately out to a typical extraction radius 
$\sim r_{500}$. Therefore, we can investigate the chemical 
evolution of the ICM across cosmic epochs separately in the
central regions and at large radii.  Our conclusions are summarized as follows:

\begin{itemize}

    \item We find that at least two components (a central peak and a constant plateau) are statistically preferred to model the iron 
    distribution within 0.4$r_{500}$ in at least 39 clusters (more than 1/5 of the sample). 
    Most of the remaining clusters are also well described with a single component ($\beta$ model). 
    
    \item By fitting the distribution of the global, average gas mass-weighted iron abundance 
    within $r_{500}$ with a power-law in the form 
    $Z_{\rm mw}=Z_{\rm mw,0}\cdot (1+z)^{-\gamma}$, we 
    obtain $Z_{\rm mw,0}=0.38\pm 0.03\, Z_\odot$, and $\gamma=0.28\pm 0.31$, consistent with no 
    significant evolution of $Z_{\rm mw}$ across our sample. 

    \item The iron mass included in the central peak component is typically a fraction 
    of $\sim 1$\% with respect to the total iron mass included within $r_{500}$. The large 
    majority of iron in the ICM is therefore in the iron plateau. 

    \item We find an approximately constant distribution of the normalization of the iron plateau, 
    centered around $\langle Z_{\rm plateau}\rangle =0.38 \, Z_\odot$ with a total scatter of 
    about $0.14\, Z_\odot$, implying an intrinsic (physical) scatter of  $0.09\, Z_\odot$   
    \citep[adopting the solar abundance table of][]{asplund2009}.  This supports 
    a pristine, approximately uniform enrichment of the diffuse baryons before the cluster 
    virialization.
    
    \item On the other hand, the normalization of the iron peak component shows a larger spread, and a marginal decrease ($<2\sigma$ c.l.) with redshift, in line with the fact that the 
    peak is produced after the virialization of the halo and depends on the formation of a 
    cool core and the strength of the feedback processes, which leave their imprint in a 
    larger variance. We find $\langle Z_{\rm peak}\rangle =0.52 \, Z_\odot$ 
    with a total scatter of $0.49\, Z_\odot$, implying an intrinsic (physical) scatter of 
    $0.26\, Z_\odot$. 

    \item We also quantify the evolution of the two components using the ratio of the iron mass in each component to the 
    total ICM mass within $r_{500}$. We find that $M_{\rm Fe}^{\rm plateau}/M_{\rm gas}\, (r<r_{500})$ scales as $(1+z)^{-0.07\pm 0.07} $, while 
    $M_{\rm Fe}^{\rm peak}/M_{\rm gas}\, (r<r_{500})$ scales as $(1+z)^{-1.00\pm 0.61} $.  Therefore, while the plateau mass does not evolve with redshift, the peak mass is consistent with an evolution of a factor of 2 from $z=1$ to local, despite with a significance lower than $2 \sigma$. The redshift dependence of the two components are summarized
    in the upper and lower panels of Figure \ref{conclusion_fig}, where we show the best-fit 
    relations and their $1 \sigma$ uncertainties for the quantities 
    $Z_{\rm Fe}^{\rm peak}$-$Z_{\rm Fe}^{\rm plateau}$ and $M_{\rm Fe}^{\rm peak}$-$M_{\rm Fe}^{\rm plateau}$ as a function 
    of redshift. 
        
    \item We find that the average emission-weighted abundance within $0.4r_{500}$ is higher 
    than the average gas mass-weighted abundance within the same radius by $\sim$22\% in cool-core 
    clusters, and by $\sim 4$\% in non-cool-core clusters. Quantifying this
    well-known effect is not only a mere exercise, but is also helpful to estimate 
    the impact of the iron peak on the measurement of the global ICM abundance, 
    particularly for cool-core clusters, where the emission-weighted abundance is
    significantly amplified by the presence of the iron peak. 
    
    \item We are able to explain the previous claims of evolution in the average iron abundance 
    in the ICM between $z\sim 1.3$ and $z\sim 0$ as the combined effect of sample selection 
    and the use of emission-weighted abundance, possibly amplified by some evolution in the
    cool-core fraction with redshift across the sample.
  
\end{itemize}

\begin{table}
\caption{The best-fit parameters describing the evolution of the quantities investigated in 
this paper, assuming the power-law behavior $X=n\cdot(1+z)^{-\gamma}$. }
\begin{center}
\begin{tabular}[width=0.5\textwidth]{lcc}
\hline
$X$ & $n$ & $\gamma$ \\
\hline
$Z_{\rm mw} (r<r_{500})$ & $(0.38\pm0.03)\, Z_{\odot} $ & $0.28\pm0.31 $ \\
$Z_{\rm plateau}$ & $(0.41\pm0.02)\, Z_{\odot} $ & $0.21\pm0.18 $ \\
$Z_{\rm peak}$ & $(0.68\pm0.07)\, Z_{\odot} $ & $0.79\pm0.53 $ \\
$M_{\rm Fe}^{\rm plateau}/M_{\rm gas}\, (r<r_{500})$ & $(4.0\pm0.2)\times10^{-4} $ & $0.07\pm0.07 $ \\
$M_{\rm Fe}^{\rm peak}/M_{\rm gas}\, (r<r_{500})$ &$(1.3\pm0.3)\times10^{-5} $ & $1.00\pm0.61 $ \\
\hline
\end{tabular}
\end{center}
\label{table_summary}
\end{table}

\noindent
In Table \ref{table_summary} we summarize the best-fit parameters describing the evolution 
of the quantities investigated in this paper, assuming the power-law behavior
$X=n\cdot(1+z)^{-\gamma}$, which are also shown in Figure \ref{conclusion_fig}. Our results confirm 
the early-enrichment scenario suggested by recent works, with the majority of iron mass in the 
ICM of massive galaxy  clusters produced at epochs earlier than $z\sim 1$. Significant evolution 
is limited to the central peak component that, despite contributing a minor fraction of the total 
iron mass, shows a $\sim 2\sigma$ significant decrease with redshift and a large intrinsic scatter. 
The overall picture of the iron distribution and its evolution as we obtained in this work, will 
be important to extend our analysis to the total {\sl Chandra} 
archive including lower S/N data, therefore reaching a larger mass and redshift range.  
Another important extension of this work will be provided by 
the XMM-{\sl Newton} data, already available for a fraction of our sample. This addition will
be important to have a better handle on the iron plateau at large radii.  Next future
facilities, like XRISM, may provide further relevant information on the outer regions, 
despite the limited effective area and the small field of view ($3\times 3$ arcmin$^2$),
but only with a significant investment of observing time on a few, selected nearby targets.
Only future ($\sim$ 2030) X-ray facilities with an X-ray bolometer on board, like {\sl Athena}
with X-IFU, or, with a far better angular resolution, {\sl Lynx},
will provide high-quality observations for a large number of high redshift clusters, 
and will promisingly provide a coherent picture of the chemical evolution of the ICM.

\begin{acknowledgements}
We thank the anonymous referee for
a constructive report that helped in improving the paper. 
A.L., P.T. and S.E. acknowledge financial contribution from the agreement ASI-INAF n.2017-14-H.0.
PT acknowledges financial contribution from the Istituto Nazionale di Astrofisica (INAF) PRIN-SKA
2017 program 1.05.01.88.04 (ESKAPE). S.E. acknowledges financial contribution from the 
contracts ASI 2015-046-R.0, and from INAF ``Call per interventi aggiuntivi a sostegno della 
ricerca di main stream di INAF''.
\end{acknowledgements}

\bibliography{ironmass}

\begin{thebibliography}{93}
\expandafter\ifx\csname natexlab\endcsname\relax\def\natexlab#1{#1}\fi

\bibitem[{{Allen} {et~al.}(2011){Allen}, {Evrard}, \& {Mantz}}]{allen2011}
{Allen}, S.~W., {Evrard}, A.~E., \& {Mantz}, A.~B. 2011, ARA\&A, 49, 409

\bibitem[{{Anders} \& {Grevesse}(1989)}]{1989Anders}
{Anders}, E. \& {Grevesse}, N. 1989, \gca, 53, 197

\bibitem[{{Anderson} {et~al.}(2009){Anderson}, {Bregman}, {Butler}, \&
  {Mullis}}]{anderson2009}
{Anderson}, M.~E., {Bregman}, J.~N., {Butler}, S.~C., \& {Mullis}, C.~R. 2009,
  ApJ, 698, 317

\bibitem[{{Andrade-Santos} {et~al.}(2017){Andrade-Santos}, {Jones}, {Forman},
  {Lovisari}, {Vikhlinin}, {van Weeren}, {Murray}, {Arnaud}, {Pratt},
  {D{\'e}mocl{\`e}s}, {Kraft}, {Mazzotta}, {B{\"o}hringer}, {Chon},
  {Giacintucci}, {Clarke}, {Borgani}, {David}, {Douspis}, {Pointecouteau},
  {Dahle}, {Brown}, {Aghanim}, \& {Rasia}}]{2017Andrade}
{Andrade-Santos}, F., {Jones}, C., {Forman}, W.~R., {et~al.} 2017, \apj, 843,
  76

\bibitem[{{Arnaud}(1996)}]{1996Arnaud}
{Arnaud}, K.~A. 1996, in Astronomical Society of the Pacific Conference Series,
  Vol. 101, Astronomical Data Analysis Software and Systems V, ed. G.~H.
  {Jacoby} \& J.~{Barnes}, 17

\bibitem[{{Asplund} {et~al.}(2009){Asplund}, {Grevesse}, {Sauval}, \&
  {Scott}}]{asplund2009}
{Asplund}, M., {Grevesse}, N., {Sauval}, A.~J., \& {Scott}, P. 2009, ARA\&A,
  47, 481

\bibitem[{{Baldi} {et~al.}(2007){Baldi}, {Ettori}, {Mazzotta}, {Tozzi}, \&
  {Borgani}}]{baldi2007}
{Baldi}, A., {Ettori}, S., {Mazzotta}, P., {Tozzi}, P., \& {Borgani}, S. 2007,
  ApJ, 666, 835

\bibitem[{{Baldi} {et~al.}(2012){Baldi}, {Ettori}, {Molendi}, {Balestra},
  {Gastaldello}, \& {Tozzi}}]{2012Baldi}
{Baldi}, A., {Ettori}, S., {Molendi}, S., {et~al.} 2012, \aap, 537, A142

\bibitem[{{Balestra} {et~al.}(2007){Balestra}, {Tozzi}, {Ettori}, {Rosati},
  {Borgani}, {Mainieri}, {Norman}, \& {Viola}}]{2007Balestra}
{Balestra}, I., {Tozzi}, P., {Ettori}, S., {et~al.} 2007, A\&A, 462, 429

\bibitem[{{Balucinska-Church} \& {McCammon}(1992)}]{phabs1992}
{Balucinska-Church}, M. \& {McCammon}, D. 1992, \apj, 400, 699

\bibitem[{{Barret} {et~al.}(2019){Barret}, {Decourchelle}, {Fabian},
  {Guainazzi}, {Nandra}, {Smith}, \& {den Herder}}]{2019Barret}
{Barret}, D., {Decourchelle}, A., {Fabian}, A., {et~al.} 2019, arXiv e-prints,
  arXiv:1912.04615

\bibitem[{{Biffi} {et~al.}(2018{\natexlab{a}}){Biffi}, {Mernier}, \&
  {Medvedev}}]{biffi2018}
{Biffi}, V., {Mernier}, F., \& {Medvedev}, P. 2018{\natexlab{a}}, \ssr, 214,
  123

\bibitem[{{Biffi} {et~al.}(2018{\natexlab{b}}){Biffi}, {Planelles}, {Borgani},
  {Rasia}, {Murante}, {Fabjan}, \& {Gaspari}}]{2018biffi}
{Biffi}, V., {Planelles}, S., {Borgani}, S., {et~al.} 2018{\natexlab{b}},
  \mnras, 476, 2689

\bibitem[{{B{\"o}hringer} {et~al.}(2004){B{\"o}hringer}, {Matsushita},
  {Churazov}, {Finoguenov}, \& {Ikebe}}]{2004Boringer}
{B{\"o}hringer}, H., {Matsushita}, K., {Churazov}, E., {Finoguenov}, A., \&
  {Ikebe}, Y. 2004, \aap, 416, L21

\bibitem[{{B{\"o}hringer} {et~al.}(2007){B{\"o}hringer}, {Schuecker}, {Pratt},
  {Arnaud}, {Ponman}, {Croston}, {Borgani}, {Bower}, {Briel}, {Collins},
  {Donahue}, {Forman}, {Finoguenov}, {Geller}, {Guzzo}, {Henry}, {Kneissl},
  {Mohr}, {Matsushita}, {Mullis}, {Ohashi}, {Pedersen}, {Pierini}, {Quintana},
  {Raychaudhury}, {Reiprich}, {Romer}, {Rosati}, {Sabirli}, {Temple}, {Viana},
  {Vikhlinin}, {Voit}, \& {Zhang}}]{bohringer2007}
{B{\"o}hringer}, H., {Schuecker}, P., {Pratt}, G.~W., {et~al.} 2007, \aap, 469,
  363

\bibitem[{{B{\"o}hringer} \& {Werner}(2010)}]{2010_Bohringer}
{B{\"o}hringer}, H. \& {Werner}, N. 2010, \aapr, 18, 127

\bibitem[{{Buote} \& {Tsai}(1995)}]{buote1995}
{Buote}, D.~A. \& {Tsai}, J.~C. 1995, \apj, 452, 522

\bibitem[{{Buote} \& {Tsai}(1996)}]{buote1996}
{Buote}, D.~A. \& {Tsai}, J.~C. 1996, \apj, 458, 27

\bibitem[{{Cash}(1979)}]{cash1979}
{Cash}, W. 1979, \apj, 228, 939

\bibitem[{{Cassano} {et~al.}(2010){Cassano}, {Ettori}, {Giacintucci},
  {Brunetti}, {Markevitch}, {Venturi}, \& {Gitti}}]{cassano2010}
{Cassano}, R., {Ettori}, S., {Giacintucci}, S., {et~al.} 2010, \apjl, 721, L82

\bibitem[{{De Grandi} {et~al.}(2004){De Grandi}, {Ettori}, {Longhetti}, \&
  {Molendi}}]{degrandi2004}
{De Grandi}, S., {Ettori}, S., {Longhetti}, M., \& {Molendi}, S. 2004, A\&A,
  419, 7

\bibitem[{{De Grandi} \& {Molendi}(2001)}]{degrandi2001}
{De Grandi}, S. \& {Molendi}, S. 2001, ApJ, 551, 153

\bibitem[{{De Grandi} \& {Molendi}(2009)}]{degrandi2009}
{De Grandi}, S. \& {Molendi}, S. 2009, A\&A, 508, 565

\bibitem[{{De Grandi} {et~al.}(2014){De Grandi}, {Santos}, {Nonino}, {Molendi},
  {Tozzi}, {Rossetti}, {Fritz}, \& {Rosati}}]{degrandi2014}
{De Grandi}, S., {Santos}, J.~S., {Nonino}, M., {et~al.} 2014, A\&A, 567, A102

\bibitem[{{de Plaa}(2013)}]{2013dePlaa}
{de Plaa}, J. 2013, Astronomische Nachrichten, 334, 416

\bibitem[{{Eckert} {et~al.}(2011){Eckert}, {Molendi}, \&
  {Paltani}}]{eckert2011}
{Eckert}, D., {Molendi}, S., \& {Paltani}, S. 2011, \aap, 526, A79

\bibitem[{{Eckert} {et~al.}(2015){Eckert}, {Roncarelli}, {Ettori}, {Molendi},
  {Vazza}, {Gastaldello}, \& {Rossetti}}]{2015Eckert}
{Eckert}, D., {Roncarelli}, M., {Ettori}, S., {et~al.} 2015, \mnras, 447, 2198

\bibitem[{{Ettori}(2000)}]{ettori2000}
{Ettori}, S. 2000, \mnras, 318, 1041

\bibitem[{{Ettori} {et~al.}(2015){Ettori}, {Baldi}, {Balestra}, {Gastaldello},
  {Molendi}, \& {Tozzi}}]{2015Ettori}
{Ettori}, S., {Baldi}, A., {Balestra}, I., {et~al.} 2015, A\&A, 578, A46

\bibitem[{{Ettori} {et~al.}(2013){Ettori}, {Donnarumma}, {Pointecouteau},
  {Reiprich}, {Giodini}, {Lovisari}, \& {Schmidt}}]{Ettori2013}
{Ettori}, S., {Donnarumma}, A., {Pointecouteau}, E., {et~al.} 2013, \ssr, 177,
  119

\bibitem[{{Foreman-Mackey} {et~al.}(2013){Foreman-Mackey}, {Hogg}, {Lang}, \&
  {Goodman}}]{Foreman2013}
{Foreman-Mackey}, D., {Hogg}, D.~W., {Lang}, D., \& {Goodman}, J. 2013, \pasp,
  125, 306

\bibitem[{{Guainazzi} \& {Tashiro}(2018)}]{2018Guainazzi}
{Guainazzi}, M. \& {Tashiro}, M.~S. 2018, arXiv e-prints, arXiv:1807.06903

\bibitem[{{Hudson} {et~al.}(2010){Hudson}, {Mittal}, {Reiprich}, {Nulsen},
  {Andernach}, \& {Sarazin}}]{hudson2010}
{Hudson}, D.~S., {Mittal}, R., {Reiprich}, T.~H., {et~al.} 2010, A\&A, 513, A37

\bibitem[{{Kaastra} {et~al.}(2004){Kaastra}, {Tamura}, {Peterson}, {Bleeker},
  {Ferrigno}, {Kahn}, {Paerels}, {Piffaretti}, {Branduardi-Raymont}, \&
  {B{\"o}hringer}}]{kaastra2004}
{Kaastra}, J.~S., {Tamura}, T., {Peterson}, J.~R., {et~al.} 2004, \aap, 413,
  415

\bibitem[{{Kalberla} {et~al.}(2005){Kalberla}, {Burton}, {Hartmann}, {Arnal},
  {Bajaja}, {Morras}, \& {P{\"o}ppel}}]{2005Kalberla}
{Kalberla}, P.~M.~W., {Burton}, W.~B., {Hartmann}, D., {et~al.} 2005, \aap,
  440, 775

\bibitem[{{Kirkpatrick} {et~al.}(2009){Kirkpatrick}, {Gitti}, {Cavagnolo},
  {McNamara}, {David}, {Nulsen}, \& {Wise}}]{kirkpatrick2009}
{Kirkpatrick}, C.~C., {Gitti}, M., {Cavagnolo}, K.~W., {et~al.} 2009, \apjl,
  707, L69

\bibitem[{{Kitayama} {et~al.}(2014){Kitayama}, {Bautz}, {Markevitch},
  {Matsushita}, {Allen}, {Kawaharada}, {McNamara}, {Ota}, {Akamatsu}, {de
  Plaa}, {Galeazzi}, {Madejski}, {Main}, {Miller}, {Nakazawa}, {Russell},
  {Sato}, {Sekiya}, {Simionescu}, {Tamura}, {Uchida}, {Ursino}, {Werner},
  {Zhuravleva}, \& {ZuHone}}]{2014Kitayama}
{Kitayama}, T., {Bautz}, M., {Markevitch}, M., {et~al.} 2014, arXiv e-prints,
  arXiv:1412.1176

\bibitem[{{Komatsu} {et~al.}(2011){Komatsu}, {Smith}, {Dunkley}, {Bennett},
  {Gold}, {Hinshaw}, {Jarosik}, {Larson}, {Nolta}, {Page}, {Spergel},
  {Halpern}, {Hill}, {Kogut}, {Limon}, {Meyer}, {Odegard}, {Tucker}, {Weiland},
  {Wollack}, \& {Wright}}]{2011Komatsu}
{Komatsu}, E., {Smith}, K.~M., {Dunkley}, J., {et~al.} 2011, \apjs, 192, 18

\bibitem[{{Lakhchaura} {et~al.}(2019){Lakhchaura}, {Mernier}, \&
  {Werner}}]{lakhchaura2019}
{Lakhchaura}, K., {Mernier}, F., \& {Werner}, N. 2019, \aap, 623, A17

\bibitem[{{Leccardi} \& {Molendi}(2008)}]{leccardi2008}
{Leccardi}, A. \& {Molendi}, S. 2008, A\&A, 487, 461

\bibitem[{{Lin} {et~al.}(2012){Lin}, {Stanford}, {Eisenhardt}, {Vikhlinin},
  {Maughan}, \& {Kravtsov}}]{2012Lin}
{Lin}, Y.-T., {Stanford}, S.~A., {Eisenhardt}, P. R.~M., {et~al.} 2012, \apjl,
  745, L3

\bibitem[{{Liu} {et~al.}(2018){Liu}, {Tozzi}, {Yu}, {De Grandi}, \&
  {Ettori}}]{liu2018}
{Liu}, A., {Tozzi}, P., {Yu}, H., {De Grandi}, S., \& {Ettori}, S. 2018,
  \mnras, 481, 361

\bibitem[{{Liu} {et~al.}(2015){Liu}, {Yu}, {Tozzi}, \& {Zhu}}]{2015Liu}
{Liu}, A., {Yu}, H., {Tozzi}, P., \& {Zhu}, Z.-H. 2015, \apj, 809, 27

\bibitem[{{Liu} {et~al.}(2019){Liu}, {Zhai}, \& {Tozzi}}]{2019Liu}
{Liu}, A., {Zhai}, M., \& {Tozzi}, P. 2019, \mnras, 485, 1651

\bibitem[{{Lovisari} {et~al.}(2017){Lovisari}, {Forman}, {Jones}, {Ettori},
  {Andrade-Santos}, {Arnaud}, {D{\'e}mocl{\`e}s}, {Pratt}, {Randall}, \&
  {Kraft}}]{lovisari2017}
{Lovisari}, L., {Forman}, W.~R., {Jones}, C., {et~al.} 2017, \apj, 846, 51

\bibitem[{{Lovisari} \& {Reiprich}(2019)}]{lovisari2019}
{Lovisari}, L. \& {Reiprich}, T.~H. 2019, \mnras, 483, 540

\bibitem[{{Mantz} {et~al.}(2018){Mantz}, {Abdulla}, {Allen}, {Carlstrom},
  {Logan}, {Marrone}, {Maughan}, {Willis}, {Pacaud}, \& {Pierre}}]{mantz2018}
{Mantz}, A.~B., {Abdulla}, Z., {Allen}, S.~W., {et~al.} 2018, \aap, 620, A2

\bibitem[{{Mantz} {et~al.}(2017){Mantz}, {Allen}, {Morris}, {Simionescu},
  {Urban}, {Werner}, \& {Zhuravleva}}]{2017Mantz}
{Mantz}, A.~B., {Allen}, S.~W., {Morris}, R.~G., {et~al.} 2017, \mnras, 472,
  2877

\bibitem[{{Maughan} {et~al.}(2008){Maughan}, {Jones}, {Forman}, \& {Van
  Speybroeck}}]{maughan2008}
{Maughan}, B.~J., {Jones}, C., {Forman}, W., \& {Van Speybroeck}, L. 2008,
  ApJS, 174, 117

\bibitem[{{Mazzotta} {et~al.}(2004){Mazzotta}, {Rasia}, {Moscardini}, \&
  {Tormen}}]{2004Mazzotta}
{Mazzotta}, P., {Rasia}, E., {Moscardini}, L., \& {Tormen}, G. 2004, \mnras,
  354, 10

\bibitem[{{McDonald} {et~al.}(2016){McDonald}, {Bulbul}, {de Haan}, {Miller},
  {Benson}, {Bleem}, {Brodwin}, {Carlstrom}, {Chiu}, {Forman},
  {Hlavacek-Larrondo}, {Garmire}, {Gupta}, {Mohr}, {Reichardt}, {Saro},
  {Stalder}, {Stark}, \& {Vieira}}]{mcdonald2016}
{McDonald}, M., {Bulbul}, E., {de Haan}, T., {et~al.} 2016, ApJ, 826, 124

\bibitem[{{Merloni}(2012)}]{2012Merloni}
{Merloni}, A. 2012, in Science from the Next Generation Imaging and
  Spectroscopic Surveys, 43

\bibitem[{{Mernier} {et~al.}(2018){Mernier}, {Biffi}, {Yamaguchi}, {Medvedev},
  {Simionescu}, {Ettori}, {Werner}, {Kaastra}, {de Plaa}, \&
  {Gu}}]{mernier2018}
{Mernier}, F., {Biffi}, V., {Yamaguchi}, H., {et~al.} 2018, \ssr, 214, 129

\bibitem[{{Mernier} {et~al.}(2017){Mernier}, {de Plaa}, {Kaastra}, {Zhang},
  {Akamatsu}, {Gu}, {Kosec}, {Mao}, {Pinto}, {Reiprich}, {Sanders},
  {Simionescu}, \& {Werner}}]{mernier2017}
{Mernier}, F., {de Plaa}, J., {Kaastra}, J.~S., {et~al.} 2017, A\&A, 603, A80

\bibitem[{{Mernier} {et~al.}(2019){Mernier}, {Werner}, {Bagchi}, {Simionescu},
  {B{\"o}hringer}, {Allen}, \& {Jacob}}]{2019Mernier}
{Mernier}, F., {Werner}, N., {Bagchi}, J., {et~al.} 2019, \mnras, 486, 5430

\bibitem[{{Molendi} {et~al.}(2016){Molendi}, {Eckert}, {De Grandi}, {Ettori},
  {Gastaldello}, {Ghizzardi}, {Pratt}, \& {Rossetti}}]{2016Molendi}
{Molendi}, S., {Eckert}, D., {De Grandi}, S., {et~al.} 2016, \aap, 586, A32

\bibitem[{{Mushotzky} {et~al.}(2019){Mushotzky}, {Aird}, {Barger},
  {Cappelluti}, {Chartas}, {Corrales}, {Eufrasio}, {Fabian}, {Falcone},
  {Gallo}, {Gilli}, {Grant}, {Hardcastle}, {Hodges-Kluck}, {Kara}, {Koss},
  {Li}, {Lisse}, {Loewenstein}, {Markevitch}, {Meyer}, {Miller}, {Mulchaey},
  {Petre}, {Ptak}, {Reynolds}, {Russell}, {Safi-Harb}, {Smith}, {Snios},
  {Tombesi}, {Valencic}, {Walker}, {Williams}, {Winter}, {Yamaguchi}, {Zhang},
  {Arenberg}, {Brand t}, {Burrows}, {Georganopoulos}, {Miller}, {Norman}, \&
  {Rosati}}]{2019Mushotzky}
{Mushotzky}, R., {Aird}, J., {Barger}, A.~J., {et~al.} 2019, in \baas, Vol.~51,
  107

\bibitem[{{Mushotzky} {et~al.}(1996){Mushotzky}, {Loewenstein}, {Arnaud},
  {Tamura}, {Fukazawa}, {Matsushita}, {Kikuchi}, \&
  {Hatsukade}}]{1996Mushotzky}
{Mushotzky}, R., {Loewenstein}, M., {Arnaud}, K.~A., {et~al.} 1996, \apj, 466,
  686

\bibitem[{{Mushotzky} \& {Loewenstein}(1997)}]{mushotzky1997}
{Mushotzky}, R.~F. \& {Loewenstein}, M. 1997, ApJL, 481, L63

\bibitem[{{O'Hara} {et~al.}(2006){O'Hara}, {Mohr}, {Bialek}, \&
  {Evrard}}]{ohara2006}
{O'Hara}, T.~B., {Mohr}, J.~J., {Bialek}, J.~J., \& {Evrard}, A.~E. 2006, \apj,
  639, 64

\bibitem[{{Panagoulia} {et~al.}(2015){Panagoulia}, {Sanders}, \&
  {Fabian}}]{panagoulia2015}
{Panagoulia}, E.~K., {Sanders}, J.~S., \& {Fabian}, A.~C. 2015, \mnras, 447,
  417

\bibitem[{{Piffaretti} {et~al.}(2011){Piffaretti}, {Arnaud}, {Pratt},
  {Pointecouteau}, \& {Melin}}]{piffaretti2011}
{Piffaretti}, R., {Arnaud}, M., {Pratt}, G.~W., {Pointecouteau}, E., \&
  {Melin}, J.-B. 2011, \aap, 534, A109

\bibitem[{{Pratt} {et~al.}(2009){Pratt}, {Croston}, {Arnaud}, \&
  {B{\"o}hringer}}]{2009Pratt}
{Pratt}, G.~W., {Croston}, J.~H., {Arnaud}, M., \& {B{\"o}hringer}, H. 2009,
  \aap, 498, 361

\bibitem[{{Presotto} {et~al.}(2014){Presotto}, {Girardi}, {Nonino}, {Mercurio},
  {Grillo}, {Rosati}, {Biviano}, {Annunziatella}, {Balestra}, {Cui},
  {Sartoris}, {Lemze}, {Ascaso}, {Moustakas}, {Ford}, {Fritz}, {Czoske},
  {Ettori}, {Kuchner}, {Lombardi}, {Maier}, {Medezinski}, {Molino},
  {Scodeggio}, {Strazzullo}, {Tozzi}, {Ziegler}, {Bartelmann}, {Benitez},
  {Bradley}, {Brescia}, {Broadhurst}, {Coe}, {Donahue}, {Gobat}, {Graves},
  {Kelson}, {Koekemoer}, {Melchior}, {Meneghetti}, {Merten}, {Moustakas},
  {Munari}, {Postman}, {Reg{\H{o}}s}, {Seitz}, {Umetsu}, {Zheng}, \&
  {Zitrin}}]{2014Presotto}
{Presotto}, V., {Girardi}, M., {Nonino}, M., {et~al.} 2014, \aap, 565, A126

\bibitem[{{Rosati} {et~al.}(2009){Rosati}, {Tozzi}, {Gobat}, {Santos},
  {Nonino}, {Demarco}, {Lidman}, {Mullis}, {Strazzullo}, {B{\"o}hringer},
  {Fassbender}, {Dawson}, {Tanaka}, {Jee}, {Ford}, {Lamer}, \&
  {Schwope}}]{rosati2009}
{Rosati}, P., {Tozzi}, P., {Gobat}, R., {et~al.} 2009, A\&A, 508, 583

\bibitem[{{Rossetti} {et~al.}(2017){Rossetti}, {Gastaldello}, {Eckert}, {Della
  Torre}, {Pantiri}, {Cazzoletti}, \& {Molendi}}]{rossetti2017}
{Rossetti}, M., {Gastaldello}, F., {Eckert}, D., {et~al.} 2017, MNRAS, 468,
  1917

\bibitem[{{Russell} {et~al.}(2008){Russell}, {Sanders}, \&
  {Fabian}}]{2008Russell}
{Russell}, H.~R., {Sanders}, J.~S., \& {Fabian}, A.~C. 2008, \mnras, 390, 1207

\bibitem[{{Sanders} \& {Fabian}(2007)}]{sanders2007}
{Sanders}, J.~S. \& {Fabian}, A.~C. 2007, \mnras, 381, 1381

\bibitem[{{Sanderson} {et~al.}(2009){Sanderson}, {O'Sullivan}, \&
  {Ponman}}]{Sanderson2009}
{Sanderson}, A. J.~R., {O'Sullivan}, E., \& {Ponman}, T.~J. 2009, \mnras, 395,
  764

\bibitem[{{Santos} {et~al.}(2008){Santos}, {Rosati}, {Tozzi}, {B{\"o}hringer},
  {Ettori}, \& {Bignamini}}]{santos2008}
{Santos}, J.~S., {Rosati}, P., {Tozzi}, P., {et~al.} 2008, A\&A, 483, 35

\bibitem[{{Santos} {et~al.}(2010){Santos}, {Tozzi}, {Rosati}, \&
  {B{\"o}hringer}}]{santos2010}
{Santos}, J.~S., {Tozzi}, P., {Rosati}, P., \& {B{\"o}hringer}, H. 2010, A\&A,
  521, A64

\bibitem[{{Santos} {et~al.}(2012){Santos}, {Tozzi}, {Rosati}, {Nonino}, \&
  {Giovannini}}]{santos2012}
{Santos}, J.~S., {Tozzi}, P., {Rosati}, P., {Nonino}, M., \& {Giovannini}, G.
  2012, A\&A, 539, A105

\bibitem[{{Schellenberger} {et~al.}(2015){Schellenberger}, {Reiprich},
  {Lovisari}, {Nevalainen}, \& {David}}]{Schellenberger2015}
{Schellenberger}, G., {Reiprich}, T.~H., {Lovisari}, L., {Nevalainen}, J., \&
  {David}, L. 2015, \aap, 575, A30

\bibitem[{{Serlemitsos} {et~al.}(1977){Serlemitsos}, {Smith}, {Boldt}, {Holt},
  \& {Swank}}]{1977Serlemitsos}
{Serlemitsos}, P.~J., {Smith}, B.~W., {Boldt}, E.~A., {Holt}, S.~S., \&
  {Swank}, J.~H. 1977, \apjl, 211, L63

\bibitem[{{Simionescu} {et~al.}(2009){Simionescu}, {Werner}, {B{\"o}hringer},
  {Kaastra}, {Finoguenov}, {Br{\"u}ggen}, \& {Nulsen}}]{simionescu2009}
{Simionescu}, A., {Werner}, N., {B{\"o}hringer}, H., {et~al.} 2009, \aap, 493,
  409

\bibitem[{{Simionescu} {et~al.}(2017){Simionescu}, {Werner}, {Mantz}, {Allen},
  \& {Urban}}]{Simionescu2017}
{Simionescu}, A., {Werner}, N., {Mantz}, A., {Allen}, S.~W., \& {Urban}, O.
  2017, \mnras, 469, 1476

\bibitem[{{Simionescu} {et~al.}(2013){Simionescu}, {Werner}, {Urban}, {Allen},
  {Fabian}, {Mantz}, {Matsushita}, {Nulsen}, {Sanders}, {Sasaki}, {Sato},
  {Takei}, \& {Walker}}]{2013Simionescu}
{Simionescu}, A., {Werner}, N., {Urban}, O., {et~al.} 2013, \apj, 775, 4

\bibitem[{{Smith} {et~al.}(2001){Smith}, {Brickhouse}, {Liedahl}, \&
  {Raymond}}]{smith2001}
{Smith}, R.~K., {Brickhouse}, N.~S., {Liedahl}, D.~A., \& {Raymond}, J.~C.
  2001, \apjl, 556, L91

\bibitem[{{Sun} {et~al.}(2009){Sun}, {Voit}, {Donahue}, {Jones}, {Forman}, \&
  {Vikhlinin}}]{sun2009}
{Sun}, M., {Voit}, G.~M., {Donahue}, M., {et~al.} 2009, \apj, 693, 1142

\bibitem[{{Tamura} {et~al.}(2009){Tamura}, {Maeda}, {Mitsuda}, {Fabian},
  {Sanders}, {Furuzawa}, {Hughes}, {Iizuka}, {Matsushita}, \&
  {Tamagawa}}]{tamura2009}
{Tamura}, T., {Maeda}, Y., {Mitsuda}, K., {et~al.} 2009, ApJL, 705, L62

\bibitem[{{Th{\"o}lken} {et~al.}(2016){Th{\"o}lken}, {Lovisari}, {Reiprich}, \&
  {Hasenbusch}}]{2016tholken}
{Th{\"o}lken}, S., {Lovisari}, L., {Reiprich}, T.~H., \& {Hasenbusch}, J. 2016,
  \aap, 592, A37

\bibitem[{{Tozzi} {et~al.}(2003){Tozzi}, {Rosati}, {Ettori}, {Borgani},
  {Mainieri}, \& {Norman}}]{tozzi2003}
{Tozzi}, P., {Rosati}, P., {Ettori}, S., {et~al.} 2003, ApJ, 593, 705

\bibitem[{{Tozzi} {et~al.}(2015){Tozzi}, {Santos}, {Jee}, {Fassbender},
  {Rosati}, {Nastasi}, {Forman}, {Sartoris}, {Borgani}, {Boehringer},
  {Altieri}, {Pratt}, {Nonino}, \& {Jones}}]{2015Tozzi}
{Tozzi}, P., {Santos}, J.~S., {Jee}, M.~J., {et~al.} 2015, \apj, 799, 93

\bibitem[{{Tozzi} {et~al.}(2013){Tozzi}, {Santos}, {Nonino}, {Rosati},
  {Borgani}, {Sartoris}, {Altieri}, \& {Sanchez-Portal}}]{tozzi2013}
{Tozzi}, P., {Santos}, J.~S., {Nonino}, M., {et~al.} 2013, A\&A, 551, A45

\bibitem[{{Urban} {et~al.}(2017){Urban}, {Werner}, {Allen}, {Simionescu}, \&
  {Mantz}}]{urban2017}
{Urban}, O., {Werner}, N., {Allen}, S.~W., {Simionescu}, A., \& {Mantz}, A.
  2017, MNRAS, 470, 4583

\bibitem[{{Urdampilleta} {et~al.}(2019){Urdampilleta}, {Mernier}, {Kaastra},
  {Simionescu}, {de Plaa}, {Kara}, \& {Ercan}}]{2019Urdampilleta}
{Urdampilleta}, I., {Mernier}, F., {Kaastra}, J.~S., {et~al.} 2019, \aap, 629,
  A31

\bibitem[{{Vikhlinin}(2019)}]{2019Vikhlinin}
{Vikhlinin}, A. 2019, in \baas, Vol.~51, 30

\bibitem[{{Vikhlinin} {et~al.}(2009){Vikhlinin}, {Burenin}, {Ebeling},
  {Forman}, {Hornstrup}, {Jones}, {Kravtsov}, {Murray}, {Nagai}, {Quintana}, \&
  {Voevodkin}}]{2009Vikhlinin}
{Vikhlinin}, A., {Burenin}, R.~A., {Ebeling}, H., {et~al.} 2009, \apj, 692,
  1033

\bibitem[{{Vikhlinin} {et~al.}(2006){Vikhlinin}, {Kravtsov}, {Forman}, {Jones},
  {Markevitch}, {Murray}, \& {Van Speybroeck}}]{vikhlinin2006a}
{Vikhlinin}, A., {Kravtsov}, A., {Forman}, W., {et~al.} 2006, \apj, 640, 691

\bibitem[{{Vikhlinin} {et~al.}(2001){Vikhlinin}, {Markevitch}, {Forman}, \&
  {Jones}}]{2001Vikhlinin}
{Vikhlinin}, A., {Markevitch}, M., {Forman}, W., \& {Jones}, C. 2001, \apjl,
  555, L87

\bibitem[{{Werner} {et~al.}(2013){Werner}, {Urban}, {Simionescu}, \&
  {Allen}}]{2013Werner}
{Werner}, N., {Urban}, O., {Simionescu}, A., \& {Allen}, S.~W. 2013, \nat, 502,
  656

\bibitem[{{Willingale} {et~al.}(2013){Willingale}, {Starling}, {Beardmore},
  {Tanvir}, \& {O'Brien}}]{2013Willingale}
{Willingale}, R., {Starling}, R.~L.~C., {Beardmore}, A.~P., {Tanvir}, N.~R., \&
  {O'Brien}, P.~T. 2013, \mnras, 431, 394

\bibitem[{{Willis} {et~al.}(2020){Willis}, {Canning}, {Noordeh}, {Allen},
  {King}, {Mantz}, {Morris}, {Stanford}, \& {Brammer}}]{2020willis}
{Willis}, J.~P., {Canning}, R.~E.~A., {Noordeh}, E.~S., {et~al.} 2020, \nat,
  577, 39

\end{thebibliography}

\appendix
\section{Properties of the sample.}

In Table \ref{complete_sample} we list the results of our spectral analysis for the global quantities 
of the entire cluster sample. The 186 clusters (including the 16 clusters 
discarded from the final analysis because of the irregular abundance profile) 
are listed in alphabetical order according to the target name. The positions correspond to the 
X-ray centroid identified as described in Section \ref{reduction}. We report the X-ray redshift, which 
is the value used in our analysis, and the core-excised temperature, which is the emission-weighted
value obtained fitting the projected emission in the range (0.1--0.4) $r_{500}$ with a single
{\tt apec} model. $M_{500}$ is the total halo mass obtained from equations (2) and (3), while the ICM 
mass within $r_{500}$ is measured directly from the deprojected ICM density profile integrated over
the spherical volume within $r_{500}$. Finally, we list the three abundance measurements used in this
work: the emission-weighted iron abundance within 0.4$r_{500}$, and the gas mass-weighted iron abundance 
within 0.4$r_{500}$ and $r_{500}$.

\newpage

\onecolumn
\footnotesize
\begin{landscape}
\begin{longtable}{lccccccccc}
\caption{\label{complete_sample}The global properties we measured for the 186 clusters. Column 1: cluster name. Column 2--3: center of the cluster emission measured in Section \ref{reduction}. Column 4: X-ray redshift of the cluster. Column 5: temperature of the cluster measured within (0.1--0.4) $r_{500}$. Column 6: $M_{500}$ in units of $10^{14}M_{\odot}$. Column 7: gas mass within $r_{500}$ in units of $10^{14}M_{\odot}$. Column 8: Emission-weighted iron abundance within 0.4$r_{500}$. Column 9--10: Gas mass-weighted iron abundance within 0.4$r_{500}$ and $r_{500}$.}\\
\hline\hline
Name	&	RA	&	Dec	&	$z_{\rm X}$	&			$kT$	&			$M_{500}$ 	&			$M_{\rm gas,500}$ 	&			$Z_{\rm ew}$ 	&			$Z_{\rm mw}$ 	&			$Z_{\rm mw}$ 	 \\		
	&	[deg]	&	[deg]	&		&			[keV]\,  (0.1--0.4)$r_{500}$	&			$[10^{14}M_{\odot}]$ 	&			$[10^{14}M_{\odot}]$ 	&			$[Z_{\odot}]\, (r<0.4r_{500}) $ 	&			$[Z_{\odot}]\, (r<0.4r_{500}) $ 	&			$[Z_{\odot}]\, (r<r_{500}) $ 	 \\			
\hline
\endfirsthead
\caption{continued.}\\
\hline\hline
Name	&	RA	&	Dec	&	$z_{\rm X}$	&			$kT$	&			$M_{500}$ 	&			$M_{\rm gas,500}$ 	&			$Z_{\rm ew}$ 	&			$Z_{\rm mw}$ 	&			$Z_{\rm mw}$ 	 \\		
	&	[deg]	&	[deg]	&		&			[keV]\,  (0.1--0.4)$r_{500}$	&			$[10^{14}M_{\odot}]$ 	&			$[10^{14}M_{\odot}]$ 	&			$[Z_{\odot}]\, (r<0.4r_{500}) $ 	&			$[Z_{\odot}]\, (r<0.4r_{500}) $ 	&			$[Z_{\odot}]\, (r<r_{500}) $ 	 \\		
\hline
\endhead
\hline
\endfoot
3C186	&	  116.0728	&	   37.8882	& $	1.082	\pm	0.020	 $ & $	7.32	\pm	0.62	 $ & $	3.99	\pm	0.53	 $ & $	0.46	\pm	0.07	 $ & $	0.46	\pm	0.09	 $ & $	0.27	\pm	0.16	 $ & $	0.24	\pm	0.17	$ \\
4C+37.11	&	   61.4550	&	   38.0589	& $	0.058	\pm	0.001	 $ & $	4.62	\pm	0.09	 $ & $	3.51	\pm	0.10	 $ & $	0.75	\pm	0.26	 $ & $	0.54	\pm	0.02	 $ & $	0.37	\pm	0.10	 $ & $	0.30	\pm	0.18	$ \\
Abell0021	&	    5.1329	&	   28.6630	& $	0.079	\pm	0.006	 $ & $	6.77	\pm	0.42	 $ & $	6.35	\pm	0.62	 $ & $	0.54	\pm	0.25	 $ & $	0.71	\pm	0.14	 $ & $	0.45	\pm	0.20	 $ & $	0.45	\pm	0.38	$ \\
Abell0085	&	   10.4603	&	   -9.3033	& $	0.059	\pm	0.001	 $ & $	6.28	\pm	0.06	 $ & $	5.70	\pm	0.09	 $ & $	1.08	\pm	0.25	 $ & $	0.51	\pm	0.01	 $ & $	0.39	\pm	0.07	 $ & $	0.34	\pm	0.15	$ \\
Abell0119	&	   14.0596	&	   -1.2562	& $	0.050	\pm	0.002	 $ & $	6.15	\pm	0.13	 $ & $	5.53	\pm	0.18	 $ & $	1.05	\pm	0.40	 $ & $	0.32	\pm	0.03	 $ & $	0.60	\pm	0.32	 $ & $	0.60	\pm	0.40	$ \\
Abell0209	&	   22.9711	&	  -13.6110	& $	0.216	\pm	0.010	 $ & $	8.01	\pm	0.53	 $ & $	7.71	\pm	0.81	 $ & $	1.28	\pm	0.23	 $ & $	0.28	\pm	0.08	 $ & $	0.30	\pm	0.11	 $ & $	0.30	\pm	0.14	$ \\
Abell0267	&	   28.1764	&	    1.0125	& $	0.230	\pm	0.010	 $ & $	8.07	\pm	0.60	 $ & $	7.74	\pm	0.91	 $ & $	0.84	\pm	0.20	 $ & $	0.52	\pm	0.12	 $ & $	0.58	\pm	0.13	 $ & $	0.58	\pm	0.25	$ \\
Abell0383	&	   42.0142	&	   -3.5293	& $	0.190	\pm	0.003	 $ & $	5.11	\pm	0.14	 $ & $	3.84	\pm	0.17	 $ & $	0.35	\pm	0.09	 $ & $	0.57	\pm	0.04	 $ & $	0.53	\pm	0.10	 $ & $	0.52	\pm	0.24	$ \\
Abell0399	&	   44.4572	&	   13.0478	& $	0.076	\pm	0.003	 $ & $	7.32	\pm	0.19	 $ & $	7.20	\pm	0.30	 $ & $	0.85	\pm	0.21	 $ & $	0.32	\pm	0.04	 $ & $	0.49	\pm	0.13	 $ & $	0.49	\pm	0.22	$ \\
Abell0401	&	   44.7380	&	   13.5827	& $	0.086	\pm	0.005	 $ & $	8.02	\pm	0.22	 $ & $	8.28	\pm	0.36	 $ & $	1.62	\pm	0.71	 $ & $	0.65	\pm	0.06	 $ & $	0.36	\pm	0.13	 $ & $	0.33	\pm	0.29	$ \\
Abell0478	&	   63.3537	&	   10.4650	& $	0.088	\pm	0.002	 $ & $	7.36	\pm	0.24	 $ & $	7.21	\pm	0.37	 $ & $	1.06	\pm	0.37	 $ & $	0.56	\pm	0.04	 $ & $	0.42	\pm	0.09	 $ & $	0.41	\pm	0.26	$ \\
Abell0586	&	  113.0840	&	   31.6325	& $	0.181	\pm	0.002	 $ & $	7.30	\pm	0.17	 $ & $	6.79	\pm	0.25	 $ & $	0.87	\pm	0.14	 $ & $	0.46	\pm	0.03	 $ & $	0.45	\pm	0.05	 $ & $	0.45	\pm	0.13	$ \\
Abell0611	&	  120.2371	&	   36.0560	& $	0.285	\pm	0.004	 $ & $	8.20	\pm	0.43	 $ & $	7.70	\pm	0.64	 $ & $	0.77	\pm	0.17	 $ & $	0.47	\pm	0.08	 $ & $	0.37	\pm	0.09	 $ & $	0.36	\pm	0.16	$ \\
Abell0644	&	  124.3564	&	   -7.5082	& $	0.077	\pm	0.002	 $ & $	6.91	\pm	0.16	 $ & $	6.58	\pm	0.23	 $ & $	1.30	\pm	0.50	 $ & $	0.45	\pm	0.03	 $ & $	0.36	\pm	0.09	 $ & $	0.21	\pm	0.18	$ \\
Abell0697	&	  130.7395	&	   36.3662	& $	0.271	\pm	0.008	 $ & $	10.65	\pm	0.74	 $ & $	11.72	\pm	1.29	 $ & $	1.61	\pm	0.47	 $ & $	0.60	\pm	0.12	 $ & $	0.64	\pm	0.18	 $ & $	0.62	\pm	0.35	$ \\
Abell0744	&	  136.8359	&	   16.6519	& $	0.074	\pm	0.006	 $ & $	2.37	\pm	0.14	 $ & $	1.21	\pm	0.11	 $ & $	0.09	\pm	0.05	 $ & $	0.31	\pm	0.06	 $ & $	0.35	\pm	0.14	 $ & $	0.17	\pm	0.25	$ \\
Abell0750	&	  137.3031	&	   10.9747	& $	0.181	\pm	0.005	 $ & $	5.87	\pm	0.31	 $ & $	4.81	\pm	0.40	 $ & $	0.35	\pm	0.14	 $ & $	0.40	\pm	0.07	 $ & $	0.44	\pm	0.11	 $ & $	0.41	\pm	0.31	$ \\
Abell0773	&	  139.4695	&	   51.7273	& $	0.207	\pm	0.004	 $ & $	7.69	\pm	0.36	 $ & $	7.26	\pm	0.53	 $ & $	0.84	\pm	0.28	 $ & $	0.47	\pm	0.07	 $ & $	0.47	\pm	0.10	 $ & $	0.47	\pm	0.29	$ \\
Abell0795	&	  141.0239	&	   14.1737	& $	0.140	\pm	0.003	 $ & $	5.23	\pm	0.20	 $ & $	4.09	\pm	0.24	 $ & $	0.71	\pm	0.12	 $ & $	0.40	\pm	0.06	 $ & $	0.35	\pm	0.09	 $ & $	0.34	\pm	0.12	$ \\
Abell0907	&	  149.5915	&	  -11.0638	& $	0.163	\pm	0.002	 $ & $	5.95	\pm	0.12	 $ & $	4.96	\pm	0.16	 $ & $	0.65	\pm	0.13	 $ & $	0.57	\pm	0.03	 $ & $	0.48	\pm	0.07	 $ & $	0.46	\pm	0.17	$ \\
Abell0963	&	  154.2651	&	   39.0476	& $	0.203	\pm	0.002	 $ & $	8.42	\pm	0.29	 $ & $	8.40	\pm	0.45	 $ & $	1.12	\pm	0.20	 $ & $	0.66	\pm	0.07	 $ & $	0.42	\pm	0.11	 $ & $	0.40	\pm	0.16	$ \\
Abell1033	&	  157.9392	&	   35.0377	& $	0.115	\pm	0.004	 $ & $	6.62	\pm	0.21	 $ & $	6.02	\pm	0.29	 $ & $	0.54	\pm	0.18	 $ & $	0.45	\pm	0.05	 $ & $	0.41	\pm	0.10	 $ & $	0.41	\pm	0.24	$ \\
Abell1068	&	  160.1859	&	   39.9531	& $	0.139	\pm	0.002	 $ & $	4.82	\pm	0.14	 $ & $	3.60	\pm	0.16	 $ & $	0.41	\pm	0.20	 $ & $	0.54	\pm	0.04	 $ & $	0.32	\pm	0.10	 $ & $	0.28	\pm	0.25	$ \\
Abell1132	&	  164.6091	&	   56.7950	& $	0.140	\pm	0.006	 $ & $	9.54	\pm	0.58	 $ & $	10.59	\pm	1.02	 $ & $	1.10	\pm	0.42	 $ & $	0.44	\pm	0.08	 $ & $	0.41	\pm	0.12	 $ & $	0.41	\pm	0.28	$ \\
Abell1204	&	  168.3354	&	   17.5945	& $	0.174	\pm	0.003	 $ & $	3.92	\pm	0.14	 $ & $	2.55	\pm	0.14	 $ & $	0.34	\pm	0.06	 $ & $	0.47	\pm	0.05	 $ & $	0.31	\pm	0.10	 $ & $	0.29	\pm	0.14	$ \\
Abell1246	&	  170.9906	&	   21.4810	& $	0.188	\pm	0.012	 $ & $	8.48	\pm	0.79	 $ & $	8.56	\pm	1.26	 $ & $	0.99	\pm	0.23	 $ & $	0.42	\pm	0.12	 $ & $	0.21	\pm	0.12	 $ & $	0.21	\pm	0.23	$ \\
Abell1302	&	  173.3196	&	   66.3786	& $	0.122	\pm	0.005	 $ & $	5.25	\pm	0.34	 $ & $	4.16	\pm	0.43	 $ & $	0.43	\pm	0.12	 $ & $	0.57	\pm	0.11	 $ & $	0.44	\pm	0.14	 $ & $	0.43	\pm	0.13	$ \\
Abell1413	&	  178.8247	&	   23.4050	& $	0.145	\pm	0.002	 $ & $	7.91	\pm	0.16	 $ & $	7.85	\pm	0.25	 $ & $	1.37	\pm	0.23	 $ & $	0.44	\pm	0.03	 $ & $	0.40	\pm	0.06	 $ & $	0.39	\pm	0.13	$ \\
Abell1423	&	  179.3217	&	   33.6112	& $	0.219	\pm	0.006	 $ & $	6.69	\pm	0.36	 $ & $	5.80	\pm	0.49	 $ & $	0.80	\pm	0.12	 $ & $	0.50	\pm	0.08	 $ & $	0.35	\pm	0.11	 $ & $	0.32	\pm	0.94	$ \\
Abell1576	&	  189.2387	&	   63.1895	& $	0.296	\pm	0.010	 $ & $	7.98	\pm	0.55	 $ & $	7.32	\pm	0.79	 $ & $	1.03	\pm	0.22	 $ & $	0.36	\pm	0.08	 $ & $	0.16	\pm	0.09	 $ & $	0.04	\pm	0.10	$ \\
Abell1650	&	  194.6728	&	   -1.7623	& $	0.081	\pm	0.001	 $ & $	5.94	\pm	0.05	 $ & $	5.17	\pm	0.08	 $ & $	0.85	\pm	0.13	 $ & $	0.48	\pm	0.01	 $ & $	0.42	\pm	0.05	 $ & $	0.36	\pm	0.28	$ \\
Abell1651	&	  194.8427	&	   -4.1966	& $	0.090	\pm	0.002	 $ & $	7.17	\pm	0.35	 $ & $	6.91	\pm	0.54	 $ & $	0.78	\pm	0.26	 $ & $	0.61	\pm	0.08	 $ & $	0.47	\pm	0.12	 $ & $	0.47	\pm	0.13	$ \\
Abell1664	&	  195.9270	&	  -24.2455	& $	0.126	\pm	0.001	 $ & $	4.18	\pm	0.04	 $ & $	2.89	\pm	0.05	 $ & $	0.41	\pm	0.10	 $ & $	0.46	\pm	0.01	 $ & $	0.30	\pm	0.05	 $ & $	0.27	\pm	0.14	$ \\
Abell1682	&	  196.7088	&	   46.5579	& $	0.193	\pm	0.006	 $ & $	10.41	\pm	0.89	 $ & $	11.81	\pm	1.60	 $ & $	1.33	\pm	0.44	 $ & $	1.04	\pm	0.20	 $ & $	0.56	\pm	0.22	 $ & $	0.53	\pm	0.20	$ \\
Abell1689	&	  197.8731	&	   -1.3416	& $	0.186	\pm	0.002	 $ & $	10.39	\pm	0.15	 $ & $	11.82	\pm	0.27	 $ & $	1.64	\pm	0.24	 $ & $	0.45	\pm	0.02	 $ & $	0.48	\pm	0.05	 $ & $	0.48	\pm	0.15	$ \\
Abell1703	&	  198.7780	&	   51.8240	& $	0.271	\pm	0.008	 $ & $	8.38	\pm	0.39	 $ & $	8.02	\pm	0.59	 $ & $	1.16	\pm	0.23	 $ & $	0.35	\pm	0.05	 $ & $	0.35	\pm	0.08	 $ & $	0.35	\pm	0.42	$ \\
Abell1763	&	  203.8230	&	   41.0011	& $	0.221	\pm	0.008	 $ & $	7.64	\pm	0.53	 $ & $	7.13	\pm	0.78	 $ & $	1.13	\pm	0.25	 $ & $	0.44	\pm	0.09	 $ & $	0.44	\pm	0.13	 $ & $	0.44	\pm	0.29	$ \\
Abell1795	&	  207.2192	&	   26.5913	& $	0.066	\pm	0.002	 $ & $	5.94	\pm	0.12	 $ & $	5.20	\pm	0.17	 $ & $	0.93	\pm	0.23	 $ & $	0.46	\pm	0.03	 $ & $	0.33	\pm	0.08	 $ & $	0.29	\pm	0.21	$ \\
Abell1800	&	  207.3650	&	   28.1060	& $	0.070	\pm	0.005	 $ & $	4.63	\pm	0.30	 $ & $	3.50	\pm	0.36	 $ & $	0.33	\pm	0.15	 $ & $	0.50	\pm	0.10	 $ & $	0.52	\pm	0.17	 $ & $	0.52	\pm	0.32	$ \\
Abell1835	&	  210.2583	&	    2.8783	& $	0.250	\pm	0.002	 $ & $	9.98	\pm	0.34	 $ & $	10.70	\pm	0.58	 $ & $	1.63	\pm	0.35	 $ & $	0.52	\pm	0.04	 $ & $	0.46	\pm	0.07	 $ & $	0.46	\pm	0.15	$ \\
Abell1918	&	  216.3421	&	   63.1830	& $	0.155	\pm	0.005	 $ & $	5.73	\pm	0.42	 $ & $	4.70	\pm	0.54	 $ & $	0.41	\pm	0.15	 $ & $	0.55	\pm	0.09	 $ & $	0.46	\pm	0.14	 $ & $	0.43	\pm	0.31	$ \\
Abell1978	&	  222.7750	&	   14.6110	& $	0.149	\pm	0.006	 $ & $	5.44	\pm	0.43	 $ & $	4.34	\pm	0.54	 $ & $	0.43	\pm	0.11	 $ & $	0.48	\pm	0.10	 $ & $	0.42	\pm	0.13	 $ & $	0.42	\pm	0.21	$ \\
Abell2009	&	  225.0817	&	   21.3695	& $	0.158	\pm	0.004	 $ & $	6.78	\pm	0.27	 $ & $	6.11	\pm	0.38	 $ & $	0.71	\pm	0.26	 $ & $	0.57	\pm	0.07	 $ & $	0.48	\pm	0.11	 $ & $	0.48	\pm	0.18	$ \\
Abell2029	&	  227.7333	&	    5.7445	& $	0.079	\pm	0.002	 $ & $	8.32	\pm	0.21	 $ & $	8.81	\pm	0.36	 $ & $	1.35	\pm	0.35	 $ & $	0.65	\pm	0.05	 $ & $	0.25	\pm	0.10	 $ & $	0.14	\pm	0.16	$ \\
Abell2050	&	  229.0679	&	    0.0890	& $	0.137	\pm	0.009	 $ & $	5.74	\pm	0.52	 $ & $	4.75	\pm	0.68	 $ & $	0.57	\pm	0.11	 $ & $	0.31	\pm	0.10	 $ & $	0.26	\pm	0.12	 $ & $	0.26	\pm	0.12	$ \\
Abell2104	&	  235.0333	&	   -3.3049	& $	0.160	\pm	0.002	 $ & $	10.00	\pm	0.35	 $ & $	11.28	\pm	0.63	 $ & $	1.09	\pm	0.71	 $ & $	0.59	\pm	0.07	 $ & $	0.42	\pm	0.10	 $ & $	0.42	\pm	0.28	$ \\
Abell2107	&	  234.9100	&	   21.7890	& $	0.041	\pm	0.002	 $ & $	4.61	\pm	0.11	 $ & $	3.53	\pm	0.14	 $ & $	0.47	\pm	0.15	 $ & $	0.50	\pm	0.04	 $ & $	0.53	\pm	0.16	 $ & $	0.53	\pm	0.24	$ \\
Abell2111	&	  234.9242	&	   34.4167	& $	0.225	\pm	0.012	 $ & $	8.80	\pm	0.90	 $ & $	8.90	\pm	1.44	 $ & $	0.83	\pm	0.23	 $ & $	0.37	\pm	0.12	 $ & $	0.32	\pm	0.15	 $ & $	0.32	\pm	0.20	$ \\
Abell2204	&	  248.1951	&	    5.5757	& $	0.150	\pm	0.001	 $ & $	9.89	\pm	0.18	 $ & $	11.15	\pm	0.32	 $ & $	1.52	\pm	0.33	 $ & $	0.61	\pm	0.02	 $ & $	0.47	\pm	0.06	 $ & $	0.47	\pm	0.27	$ \\
Abell2218	&	  248.9625	&	   66.2105	& $	0.186	\pm	0.007	 $ & $	8.03	\pm	0.33	 $ & $	7.87	\pm	0.52	 $ & $	0.90	\pm	0.27	 $ & $	0.24	\pm	0.05	 $ & $	0.28	\pm	0.08	 $ & $	0.28	\pm	0.30	$ \\
Abell2219	&	  250.0827	&	   46.7109	& $	0.226	\pm	0.002	 $ & $	12.47	\pm	0.26	 $ & $	15.43	\pm	0.51	 $ & $	2.14	\pm	0.35	 $ & $	0.46	\pm	0.03	 $ & $	0.39	\pm	0.06	 $ & $	0.38	\pm	0.34	$ \\
Abell2244	&	  255.6773	&	   34.0609	& $	0.098	\pm	0.001	 $ & $	6.07	\pm	0.09	 $ & $	5.29	\pm	0.12	 $ & $	0.56	\pm	0.24	 $ & $	0.47	\pm	0.02	 $ & $	0.40	\pm	0.09	 $ & $	0.38	\pm	0.32	$ \\
Abell2255	&	  258.2055	&	   64.0654	& $	0.080	\pm	0.004	 $ & $	6.47	\pm	0.19	 $ & $	5.92	\pm	0.27	 $ & $	0.99	\pm	0.28	 $ & $	0.35	\pm	0.05	 $ & $	0.57	\pm	0.21	 $ & $	0.57	\pm	0.20	$ \\
Abell2259	&	  260.0345	&	   27.6698	& $	0.158	\pm	0.008	 $ & $	5.68	\pm	0.47	 $ & $	4.62	\pm	0.60	 $ & $	0.60	\pm	0.30	 $ & $	0.36	\pm	0.10	 $ & $	0.26	\pm	0.10	 $ & $	0.26	\pm	0.15	$ \\
Abell2261	&	  260.6136	&	   32.1331	& $	0.219	\pm	0.005	 $ & $	8.42	\pm	0.38	 $ & $	8.32	\pm	0.60	 $ & $	1.20	\pm	0.23	 $ & $	0.60	\pm	0.10	 $ & $	0.52	\pm	0.11	 $ & $	0.52	\pm	0.30	$ \\
Abell2294	&	  261.0594	&	   85.8868	& $	0.166	\pm	0.006	 $ & $	8.42	\pm	0.56	 $ & $	8.57	\pm	0.90	 $ & $	1.17	\pm	0.33	 $ & $	0.60	\pm	0.11	 $ & $	0.46	\pm	0.17	 $ & $	0.44	\pm	0.19	$ \\
Abell2409	&	  330.2200	&	   20.9695	& $	0.155	\pm	0.005	 $ & $	5.93	\pm	0.36	 $ & $	4.95	\pm	0.47	 $ & $	0.77	\pm	0.23	 $ & $	0.55	\pm	0.09	 $ & $	0.48	\pm	0.13	 $ & $	0.48	\pm	0.74	$ \\
Abell2415	&	  331.4109	&	   -5.5922	& $	0.057	\pm	0.005	 $ & $	2.87	\pm	0.16	 $ & $	1.65	\pm	0.15	 $ & $	0.23	\pm	0.07	 $ & $	0.48	\pm	0.07	 $ & $	0.54	\pm	0.20	 $ & $	0.54	\pm	6.98	$ \\
Abell2420	&	  332.5791	&	  -12.1732	& $	0.079	\pm	0.005	 $ & $	6.42	\pm	0.33	 $ & $	5.85	\pm	0.48	 $ & $	1.04	\pm	0.29	 $ & $	0.51	\pm	0.10	 $ & $	0.67	\pm	0.19	 $ & $	0.67	\pm	0.72	$ \\
Abell2426	&	  333.6400	&	  -10.3691	& $	0.100	\pm	0.007	 $ & $	5.17	\pm	0.26	 $ & $	4.11	\pm	0.33	 $ & $	0.53	\pm	0.19	 $ & $	0.39	\pm	0.08	 $ & $	0.51	\pm	0.14	 $ & $	0.51	\pm	0.27	$ \\
Abell2533	&	  346.8087	&	  -15.2242	& $	0.115	\pm	0.002	 $ & $	4.12	\pm	0.14	 $ & $	2.85	\pm	0.15	 $ & $	0.31	\pm	0.11	 $ & $	0.77	\pm	0.07	 $ & $	0.31	\pm	0.14	 $ & $	0.19	\pm	0.14	$ \\
Abell2537	&	  347.0922	&	   -2.1910	& $	0.298	\pm	0.005	 $ & $	8.37	\pm	0.37	 $ & $	7.88	\pm	0.55	 $ & $	0.85	\pm	0.14	 $ & $	0.51	\pm	0.07	 $ & $	0.36	\pm	0.10	 $ & $	0.35	\pm	0.34	$ \\
Abell2552	&	  347.8884	&	    3.6351	& $	0.303	\pm	0.010	 $ & $	8.99	\pm	0.74	 $ & $	8.80	\pm	1.15	 $ & $	1.14	\pm	0.46	 $ & $	0.49	\pm	0.09	 $ & $	0.39	\pm	0.13	 $ & $	0.39	\pm	0.83	$ \\
Abell2556	&	  348.2558	&	  -21.6346	& $	0.089	\pm	0.002	 $ & $	4.08	\pm	0.12	 $ & $	2.84	\pm	0.13	 $ & $	0.37	\pm	0.10	 $ & $	0.61	\pm	0.05	 $ & $	0.58	\pm	0.15	 $ & $	0.58	\pm	0.09	$ \\
Abell2566	&	  349.0213	&	  -20.4639	& $	0.085	\pm	0.002	 $ & $	2.90	\pm	0.10	 $ & $	1.66	\pm	0.09	 $ & $	0.21	\pm	0.08	 $ & $	0.67	\pm	0.07	 $ & $	0.65	\pm	0.13	 $ & $	0.65	\pm	0.29	$ \\
Abell2631	&	  354.4064	&	    0.2680	& $	0.275	\pm	0.013	 $ & $	9.20	\pm	0.90	 $ & $	9.28	\pm	1.44	 $ & $	1.37	\pm	0.36	 $ & $	0.42	\pm	0.11	 $ & $	0.30	\pm	0.15	 $ & $	0.27	\pm	0.28	$ \\
Abell2665	&	  357.7110	&	    6.1485	& $	0.053	\pm	0.004	 $ & $	4.14	\pm	0.20	 $ & $	2.96	\pm	0.23	 $ & $	0.16	\pm	0.13	 $ & $	0.54	\pm	0.10	 $ & $	0.52	\pm	0.22	 $ & $	0.51	\pm	0.25	$ \\
Abell2667	&	  357.9142	&	  -26.0842	& $	0.236	\pm	0.003	 $ & $	6.90	\pm	0.43	 $ & $	6.02	\pm	0.59	 $ & $	0.91	\pm	0.23	 $ & $	0.50	\pm	0.08	 $ & $	0.15	\pm	0.11	 $ & $	0.05	\pm	0.21	$ \\
Abell2717	&	    0.8029	&	  -35.9356	& $	0.047	\pm	0.002	 $ & $	2.20	\pm	0.09	 $ & $	1.09	\pm	0.07	 $ & $	0.11	\pm	0.06	 $ & $	0.47	\pm	0.06	 $ & $	0.83	\pm	0.30	 $ & $	0.81	\pm	0.25	$ \\
Abell2734	&	    2.8404	&	  -28.8548	& $	0.051	\pm	0.014	 $ & $	5.27	\pm	0.18	 $ & $	4.34	\pm	0.24	 $ & $	0.53	\pm	0.23	 $ & $	0.37	\pm	0.06	 $ & $	0.35	\pm	0.14	 $ & $	0.35	\pm	0.18	$ \\
Abell3112	&	   49.4899	&	  -44.2384	& $	0.075	\pm	0.001	 $ & $	5.24	\pm	0.08	 $ & $	4.26	\pm	0.11	 $ & $	0.62	\pm	0.16	 $ & $	0.60	\pm	0.03	 $ & $	0.38	\pm	0.08	 $ & $	0.29	\pm	0.19	$ \\
Abell3158	&	   55.7225	&	  -53.6296	& $	0.061	\pm	0.001	 $ & $	5.44	\pm	0.09	 $ & $	4.53	\pm	0.11	 $ & $	0.77	\pm	0.25	 $ & $	0.51	\pm	0.03	 $ & $	0.60	\pm	0.12	 $ & $	0.60	\pm	0.57	$ \\
Abell3391	&	   96.5964	&	  -53.6962	& $	0.055	\pm	0.004	 $ & $	6.63	\pm	0.25	 $ & $	6.23	\pm	0.37	 $ & $	0.48	\pm	0.46	 $ & $	0.37	\pm	0.07	 $ & $	0.54	\pm	0.47	 $ & $	0.48	\pm	0.27	$ \\
Abell3444	&	  155.9592	&	  -27.2563	& $	0.260	\pm	0.002	 $ & $	7.19	\pm	0.29	 $ & $	6.34	\pm	0.41	 $ & $	1.25	\pm	0.15	 $ & $	0.43	\pm	0.04	 $ & $	0.36	\pm	0.06	 $ & $	0.35	\pm	0.23	$ \\
Abell3532	&	  194.3404	&	  -30.3696	& $	0.051	\pm	0.009	 $ & $	5.10	\pm	0.26	 $ & $	4.12	\pm	0.33	 $ & $	1.04	\pm	0.25	 $ & $	0.41	\pm	0.09	 $ & $	0.63	\pm	0.23	 $ & $	0.63	\pm	0.32	$ \\
Abell3562	&	  203.3985	&	  -31.6721	& $	0.049	\pm	0.003	 $ & $	4.80	\pm	0.14	 $ & $	3.75	\pm	0.17	 $ & $	0.65	\pm	0.18	 $ & $	0.40	\pm	0.05	 $ & $	0.51	\pm	0.15	 $ & $	0.51	\pm	0.25	$ \\
Abell3695	&	  308.7049	&	  -35.8230	& $	0.078	\pm	0.012	 $ & $	6.46	\pm	0.44	 $ & $	5.91	\pm	0.63	 $ & $	0.60	\pm	0.25	 $ & $	0.24	\pm	0.11	 $ & $	0.27	\pm	0.18	 $ & $	0.27	\pm	0.20	$ \\
Abell3827	&	  330.4726	&	  -59.9461	& $	0.101	\pm	0.001	 $ & $	7.67	\pm	0.16	 $ & $	7.65	\pm	0.26	 $ & $	1.84	\pm	0.61	 $ & $	0.43	\pm	0.04	 $ & $	0.36	\pm	0.10	 $ & $	0.36	\pm	0.17	$ \\
Abell3866	&	  335.1400	&	  -35.1650	& $	0.160	\pm	0.005	 $ & $	4.37	\pm	0.29	 $ & $	3.05	\pm	0.32	 $ & $	0.35	\pm	0.10	 $ & $	0.50	\pm	0.08	 $ & $	0.39	\pm	0.11	 $ & $	0.38	\pm	0.17	$ \\
Abell3921	&	  342.4893	&	  -64.4294	& $	0.096	\pm	0.002	 $ & $	6.34	\pm	0.22	 $ & $	5.69	\pm	0.31	 $ & $	0.77	\pm	0.27	 $ & $	0.43	\pm	0.05	 $ & $	0.40	\pm	0.11	 $ & $	0.40	\pm	0.19	$ \\
CIZAJ0107.7+5408	&	   16.9138	&	   54.1375	& $	0.119	\pm	0.005	 $ & $	9.44	\pm	0.59	 $ & $	10.53	\pm	1.04	 $ & $	1.38	\pm	0.55	 $ & $	0.38	\pm	0.07	 $ & $	0.22	\pm	0.10	 $ & $	0.21	\pm	0.17	$ \\
CLJ1415+3612	&	  213.7963	&	   36.2010	& $	1.038	\pm	0.011	 $ & $	6.30	\pm	0.49	 $ & $	3.23	\pm	0.40	 $ & $	0.41	\pm	0.07	 $ & $	0.76	\pm	0.12	 $ & $	0.55	\pm	0.14	 $ & $	0.53	\pm	0.19	$ \\
G000.44-41.83	&	  316.0750	&	  -41.3300	& $	0.151	\pm	0.007	 $ & $	6.06	\pm	0.51	 $ & $	5.14	\pm	0.68	 $ & $	0.63	\pm	0.28	 $ & $	0.63	\pm	0.13	 $ & $	0.70	\pm	0.23	 $ & $	0.70	\pm	0.17	$ \\
G002.74-56.18	&	  334.6721	&	  -38.9047	& $	0.141	\pm	0.008	 $ & $	6.06	\pm	0.38	 $ & $	5.17	\pm	0.52	 $ & $	0.74	\pm	0.23	 $ & $	0.41	\pm	0.09	 $ & $	0.48	\pm	0.13	 $ & $	0.48	\pm	0.19	$ \\
G003.90-59.41	&	  338.6120	&	  -37.7400	& $	0.149	\pm	0.007	 $ & $	9.77	\pm	0.55	 $ & $	10.94	\pm	0.98	 $ & $	1.32	\pm	0.41	 $ & $	0.41	\pm	0.08	 $ & $	0.39	\pm	0.12	 $ & $	0.38	\pm	0.46	$ \\
G008.44-56.35	&	  334.4421	&	  -35.7228	& $	0.148	\pm	0.008	 $ & $	6.04	\pm	0.51	 $ & $	5.12	\pm	0.69	 $ & $	0.57	\pm	0.23	 $ & $	0.43	\pm	0.11	 $ & $	0.41	\pm	0.18	 $ & $	0.40	\pm	0.29	$ \\
G049.33+44.38	&	  245.1258	&	   29.8897	& $	0.104	\pm	0.008	 $ & $	6.29	\pm	0.48	 $ & $	5.59	\pm	0.68	 $ & $	0.53	\pm	0.30	 $ & $	0.41	\pm	0.11	 $ & $	0.24	\pm	0.14	 $ & $	0.22	\pm	0.19	$ \\
G086.45+15.29	&	  294.5821	&	   54.1573	& $	0.273	\pm	0.008	 $ & $	7.71	\pm	0.60	 $ & $	7.03	\pm	0.86	 $ & $	1.20	\pm	0.25	 $ & $	0.51	\pm	0.09	 $ & $	0.48	\pm	0.12	 $ & $	0.47	\pm	0.18	$ \\
G114.33+64.87	&	  198.7775	&	   51.8242	& $	0.272	\pm	0.006	 $ & $	8.36	\pm	0.39	 $ & $	7.99	\pm	0.59	 $ & $	1.25	\pm	0.30	 $ & $	0.35	\pm	0.05	 $ & $	0.38	\pm	0.09	 $ & $	0.38	\pm	0.14	$ \\
G115.71+17.52	&	  336.6142	&	   78.3200	& $	0.360	\pm	0.006	 $ & $	7.11	\pm	0.56	 $ & $	5.88	\pm	0.73	 $ & $	0.95	\pm	0.14	 $ & $	0.49	\pm	0.07	 $ & $	0.50	\pm	0.12	 $ & $	0.49	\pm	0.16	$ \\
G139.59+24.18	&	   95.4502	&	   74.7013	& $	0.274	\pm	0.005	 $ & $	8.50	\pm	0.52	 $ & $	8.20	\pm	0.80	 $ & $	1.04	\pm	0.19	 $ & $	0.45	\pm	0.07	 $ & $	0.34	\pm	0.10	 $ & $	0.26	\pm	0.18	$ \\
G163.72+53.53	&	  155.6138	&	   50.1045	& $	0.166	\pm	0.008	 $ & $	6.58	\pm	0.49	 $ & $	5.81	\pm	0.68	 $ & $	0.73	\pm	0.20	 $ & $	0.35	\pm	0.10	 $ & $	0.27	\pm	0.12	 $ & $	0.26	\pm	0.25	$ \\
G165.08+54.11	&	  155.9338	&	   49.1381	& $	0.145	\pm	0.010	 $ & $	6.39	\pm	0.44	 $ & $	5.61	\pm	0.61	 $ & $	0.77	\pm	0.18	 $ & $	0.40	\pm	0.10	 $ & $	0.30	\pm	0.14	 $ & $	0.26	\pm	0.15	$ \\
G167.65+17.64	&	   99.5200	&	   47.7917	& $	0.188	\pm	0.008	 $ & $	7.70	\pm	0.72	 $ & $	7.35	\pm	1.08	 $ & $	1.18	\pm	0.28	 $ & $	0.42	\pm	0.11	 $ & $	0.30	\pm	0.14	 $ & $	0.19	\pm	0.41	$ \\
G171.94-40.65	&	   48.2420	&	    8.3708	& $	0.293	\pm	0.016	 $ & $	12.65	\pm	1.25	 $ & $	15.19	\pm	2.38	 $ & $	2.09	\pm	0.47	 $ & $	0.43	\pm	0.10	 $ & $	0.36	\pm	0.13	 $ & $	0.36	\pm	0.13	$ \\
G172.88+65.32	&	  167.9046	&	   40.8339	& $	0.070	\pm	0.006	 $ & $	4.32	\pm	0.31	 $ & $	3.14	\pm	0.35	 $ & $	0.13	\pm	0.08	 $ & $	0.31	\pm	0.10	 $ & $	0.39	\pm	0.18	 $ & $	0.39	\pm	0.31	$ \\
G226.17-21.91	&	   88.2113	&	  -21.0660	& $	0.105	\pm	0.008	 $ & $	5.81	\pm	0.42	 $ & $	4.93	\pm	0.56	 $ & $	0.83	\pm	0.29	 $ & $	0.33	\pm	0.09	 $ & $	0.46	\pm	0.13	 $ & $	0.46	\pm	0.30	$ \\
G229.21-17.24	&	   94.1020	&	  -21.9430	& $	0.157	\pm	0.014	 $ & $	8.32	\pm	0.74	 $ & $	8.45	\pm	1.18	 $ & $	1.08	\pm	0.31	 $ & $	0.29	\pm	0.12	 $ & $	0.29	\pm	0.14	 $ & $	0.29	\pm	0.36	$ \\
G241.74-30.88	&	   83.2475	&	  -37.0277	& $	0.260	\pm	0.011	 $ & $	9.48	\pm	0.86	 $ & $	9.81	\pm	1.41	 $ & $	1.29	\pm	0.34	 $ & $	0.26	\pm	0.09	 $ & $	0.28	\pm	0.12	 $ & $	0.27	\pm	0.21	$ \\
G241.77-24.00	&	   91.4663	&	  -35.3073	& $	0.142	\pm	0.004	 $ & $	5.22	\pm	0.29	 $ & $	4.08	\pm	0.35	 $ & $	0.73	\pm	0.13	 $ & $	0.49	\pm	0.08	 $ & $	0.36	\pm	0.11	 $ & $	0.27	\pm	0.20	$ \\
G244.69+32.49	&	  146.3592	&	   -8.6683	& $	0.163	\pm	0.010	 $ & $	5.40	\pm	0.36	 $ & $	4.26	\pm	0.45	 $ & $	0.65	\pm	0.17	 $ & $	0.31	\pm	0.09	 $ & $	0.37	\pm	0.12	 $ & $	0.29	\pm	0.21	$ \\
G250.90-36.25	&	   77.5542	&	  -45.3247	& $	0.202	\pm	0.009	 $ & $	6.22	\pm	0.49	 $ & $	5.21	\pm	0.65	 $ & $	0.84	\pm	0.17	 $ & $	0.33	\pm	0.10	 $ & $	0.36	\pm	0.14	 $ & $	0.36	\pm	0.24	$ \\
G253.47-33.72	&	   81.4540	&	  -47.2500	& $	0.188	\pm	0.011	 $ & $	6.31	\pm	0.47	 $ & $	5.37	\pm	0.64	 $ & $	0.72	\pm	0.19	 $ & $	0.37	\pm	0.11	 $ & $	0.44	\pm	0.15	 $ & $	0.44	\pm	0.17	$ \\
G263.66-22.53	&	  101.3713	&	  -54.2291	& $	0.153	\pm	0.009	 $ & $	8.32	\pm	0.61	 $ & $	8.47	\pm	0.98	 $ & $	1.34	\pm	0.41	 $ & $	0.30	\pm	0.09	 $ & $	0.19	\pm	0.13	 $ & $	0.17	\pm	0.42	$ \\
G264.41+19.48	&	  150.0087	&	  -30.2655	& $	0.193	\pm	0.009	 $ & $	7.51	\pm	0.68	 $ & $	7.05	\pm	1.01	 $ & $	0.62	\pm	0.19	 $ & $	0.52	\pm	0.12	 $ & $	0.70	\pm	0.22	 $ & $	0.70	\pm	0.22	$ \\
G266.56-27.31	&	   93.9667	&	  -57.7810	& $	0.958	\pm	0.020	 $ & $	11.63	\pm	0.66	 $ & $	8.91	\pm	0.80	 $ & $	1.58	\pm	0.15	 $ & $	0.40	\pm	0.07	 $ & $	0.27	\pm	0.11	 $ & $	0.25	\pm	0.22	$ \\
G269.31-49.87	&	   52.1579	&	  -55.7104	& $	0.076	\pm	0.006	 $ & $	5.12	\pm	0.27	 $ & $	4.09	\pm	0.34	 $ & $	0.45	\pm	0.24	 $ & $	0.43	\pm	0.10	 $ & $	0.38	\pm	0.16	 $ & $	0.30	\pm	0.18	$ \\
G275.21+43.92	&	  172.5875	&	  -14.6028	& $	0.100	\pm	0.005	 $ & $	6.66	\pm	0.31	 $ & $	6.12	\pm	0.46	 $ & $	1.07	\pm	0.33	 $ & $	0.46	\pm	0.08	 $ & $	0.53	\pm	0.16	 $ & $	0.53	\pm	0.16	$ \\
G280.19+47.81	&	  177.4400	&	  -12.3140	& $	0.150	\pm	0.007	 $ & $	7.28	\pm	0.78	 $ & $	6.86	\pm	1.16	 $ & $	0.80	\pm	0.32	 $ & $	0.42	\pm	0.13	 $ & $	0.47	\pm	0.20	 $ & $	0.47	\pm	0.15	$ \\
G284.99-23.70	&	  110.8230	&	  -73.4550	& $	0.386	\pm	0.011	 $ & $	9.25	\pm	1.03	 $ & $	8.77	\pm	1.54	 $ & $	1.68	\pm	0.37	 $ & $	0.51	\pm	0.11	 $ & $	0.33	\pm	0.16	 $ & $	0.32	\pm	0.13	$ \\
G294.66-37.02	&	   45.9712	&	  -77.8707	& $	0.284	\pm	0.011	 $ & $	10.08	\pm	0.88	 $ & $	10.67	\pm	1.48	 $ & $	1.48	\pm	0.43	 $ & $	0.35	\pm	0.09	 $ & $	0.33	\pm	0.12	 $ & $	0.33	\pm	1.09	$ \\
G295.33+23.33	&	  183.8700	&	  -39.0297	& $	0.119	\pm	0.008	 $ & $	5.95	\pm	0.45	 $ & $	5.07	\pm	0.60	 $ & $	0.85	\pm	0.25	 $ & $	0.28	\pm	0.08	 $ & $	0.33	\pm	0.16	 $ & $	0.33	\pm	0.15	$ \\
G313.87-17.10	&	  240.4588	&	  -75.7494	& $	0.160	\pm	0.007	 $ & $	9.83	\pm	0.64	 $ & $	10.98	\pm	1.13	 $ & $	1.46	\pm	0.36	 $ & $	0.45	\pm	0.08	 $ & $	0.51	\pm	0.12	 $ & $	0.51	\pm	0.24	$ \\
G325.70+17.31	&	  221.9054	&	  -40.3306	& $	0.302	\pm	0.014	 $ & $	10.24	\pm	1.10	 $ & $	10.82	\pm	1.84	 $ & $	1.28	\pm	0.31	 $ & $	0.35	\pm	0.11	 $ & $	0.36	\pm	0.11	 $ & $	0.36	\pm	0.19	$ \\
G332.88-19.28	&	  273.3396	&	  -61.4633	& $	0.144	\pm	0.007	 $ & $	9.23	\pm	0.93	 $ & $	10.02	\pm	1.59	 $ & $	0.93	\pm	0.43	 $ & $	0.60	\pm	0.14	 $ & $	0.49	\pm	0.18	 $ & $	0.49	\pm	0.13	$ \\
Hercules	&	  252.7838	&	    4.9925	& $	0.155	\pm	0.002	 $ & $	5.72	\pm	0.14	 $ & $	4.68	\pm	0.18	 $ & $	0.32	\pm	0.10	 $ & $	0.63	\pm	0.04	 $ & $	0.39	\pm	0.07	 $ & $	0.38	\pm	0.18	$ \\
Hydra	&	  139.5245	&	  -12.0949	& $	0.061	\pm	0.002	 $ & $	3.91	\pm	0.06	 $ & $	2.70	\pm	0.06	 $ & $	0.40	\pm	0.09	 $ & $	0.64	\pm	0.03	 $ & $	0.54	\pm	0.10	 $ & $	0.52	\pm	0.32	$ \\
IRAS09104+4109	&	  138.4397	&	   40.9415	& $	0.441	\pm	0.004	 $ & $	8.60	\pm	0.48	 $ & $	7.57	\pm	0.67	 $ & $	1.64	\pm	0.33	 $ & $	0.62	\pm	0.06	 $ & $	0.47	\pm	0.13	 $ & $	0.47	\pm	0.19	$ \\
MACSJ0011.7-1523	&	    2.9285	&	  -15.3890	& $	0.376	\pm	0.007	 $ & $	7.36	\pm	0.46	 $ & $	6.16	\pm	0.61	 $ & $	0.82	\pm	0.19	 $ & $	0.37	\pm	0.06	 $ & $	0.34	\pm	0.08	 $ & $	0.34	\pm	0.13	$ \\
MACSJ0035.4-2015	&	    8.8604	&	  -20.2632	& $	0.361	\pm	0.017	 $ & $	7.84	\pm	0.58	 $ & $	6.86	\pm	0.80	 $ & $	1.12	\pm	0.22	 $ & $	0.39	\pm	0.09	 $ & $	0.35	\pm	0.10	 $ & $	0.35	\pm	0.16	$ \\
MACSJ0159.8-0849	&	   29.9554	&	   -8.8333	& $	0.408	\pm	0.007	 $ & $	10.39	\pm	0.53	 $ & $	10.41	\pm	0.84	 $ & $	1.34	\pm	0.28	 $ & $	0.50	\pm	0.08	 $ & $	0.26	\pm	0.10	 $ & $	0.23	\pm	0.07	$ \\
MACSJ0242.5-2132	&	   40.6495	&	  -21.5407	& $	0.314	\pm	0.007	 $ & $	5.96	\pm	0.53	 $ & $	4.57	\pm	0.64	 $ & $	0.66	\pm	0.17	 $ & $	0.53	\pm	0.09	 $ & $	0.16	\pm	0.17	 $ & $	0.06	\pm	0.18	$ \\
MACSJ0257.1-2325	&	   44.2873	&	  -23.4348	& $	0.514	\pm	0.061	 $ & $	10.70	\pm	1.03	 $ & $	10.22	\pm	1.56	 $ & $	1.71	\pm	0.43	 $ & $	0.13	\pm	0.11	 $ & $	0.17	\pm	0.14	 $ & $	0.15	\pm	0.14	$ \\
MACSJ0257.6-2209	&	   44.4223	&	  -22.1549	& $	0.350	\pm	0.013	 $ & $	8.62	\pm	0.92	 $ & $	8.01	\pm	1.35	 $ & $	0.79	\pm	0.22	 $ & $	0.49	\pm	0.13	 $ & $	0.40	\pm	0.16	 $ & $	0.39	\pm	0.15	$ \\
MACSJ0308.9+2645	&	   47.2329	&	   26.7611	& $	0.330	\pm	0.013	 $ & $	10.11	\pm	0.82	 $ & $	10.44	\pm	1.34	 $ & $	1.43	\pm	0.28	 $ & $	0.39	\pm	0.10	 $ & $	0.35	\pm	0.14	 $ & $	0.34	\pm	0.18	$ \\
MACSJ0329.6-0211	&	   52.4234	&	   -2.1965	& $	0.457	\pm	0.005	 $ & $	8.05	\pm	0.45	 $ & $	6.75	\pm	0.60	 $ & $	1.26	\pm	0.17	 $ & $	0.59	\pm	0.06	 $ & $	0.46	\pm	0.10	 $ & $	0.45	\pm	0.30	$ \\
MACSJ0429.6-0253	&	   67.4000	&	   -2.8853	& $	0.400	\pm	0.007	 $ & $	7.14	\pm	0.73	 $ & $	5.78	\pm	0.94	 $ & $	0.80	\pm	0.12	 $ & $	0.59	\pm	0.10	 $ & $	0.47	\pm	0.14	 $ & $	0.46	\pm	0.10	$ \\
MACSJ0520.7-1328	&	   80.1750	&	  -13.4799	& $	0.342	\pm	0.010	 $ & $	8.67	\pm	1.05	 $ & $	8.13	\pm	1.56	 $ & $	0.92	\pm	0.35	 $ & $	0.64	\pm	0.13	 $ & $	0.40	\pm	0.19	 $ & $	0.39	\pm	0.17	$ \\
MACSJ0647.7+7015	&	  101.9603	&	   70.2483	& $	0.572	\pm	0.046	 $ & $	9.53	\pm	1.01	 $ & $	8.22	\pm	1.38	 $ & $	1.25	\pm	0.21	 $ & $	0.25	\pm	0.11	 $ & $	0.36	\pm	0.16	 $ & $	0.35	\pm	0.14	$ \\
MACSJ0744.8+3927	&	  116.2201	&	   39.4576	& $	0.693	\pm	0.012	 $ & $	8.50	\pm	0.51	 $ & $	6.38	\pm	0.61	 $ & $	1.10	\pm	0.13	 $ & $	0.42	\pm	0.08	 $ & $	0.22	\pm	0.10	 $ & $	0.19	\pm	0.15	$ \\
MACSJ0947.2+7623	&	  146.8029	&	   76.3874	& $	0.355	\pm	0.002	 $ & $	7.68	\pm	0.29	 $ & $	6.66	\pm	0.39	 $ & $	0.84	\pm	0.15	 $ & $	0.51	\pm	0.04	 $ & $	0.40	\pm	0.08	 $ & $	0.39	\pm	0.11	$ \\
MACSJ1115.8+0129	&	  168.9661	&	    1.4990	& $	0.360	\pm	0.006	 $ & $	8.15	\pm	0.37	 $ & $	7.30	\pm	0.53	 $ & $	1.07	\pm	0.19	 $ & $	0.45	\pm	0.05	 $ & $	0.41	\pm	0.10	 $ & $	0.39	\pm	0.27	$ \\
MACSJ1149.5+2223	&	  177.3970	&	   22.4027	& $	0.528	\pm	0.007	 $ & $	9.85	\pm	0.36	 $ & $	8.90	\pm	0.51	 $ & $	1.59	\pm	0.15	 $ & $	0.33	\pm	0.04	 $ & $	0.30	\pm	0.06	 $ & $	0.29	\pm	0.18	$ \\
MACSJ1206.2-0847	&	  181.5511	&	   -8.8006	& $	0.468	\pm	0.016	 $ & $	11.29	\pm	1.30	 $ & $	11.45	\pm	2.09	 $ & $	2.18	\pm	0.44	 $ & $	0.44	\pm	0.11	 $ & $	0.39	\pm	0.12	 $ & $	0.39	\pm	0.21	$ \\
MACSJ1311.0-0310	&	  197.7565	&	   -3.1771	& $	0.491	\pm	0.006	 $ & $	5.95	\pm	0.30	 $ & $	4.10	\pm	0.33	 $ & $	0.50	\pm	0.08	 $ & $	0.48	\pm	0.06	 $ & $	0.45	\pm	0.08	 $ & $	0.44	\pm	0.22	$ \\
MACSJ1423.8+2404	&	  215.9496	&	   24.0784	& $	0.545	\pm	0.003	 $ & $	7.90	\pm	0.33	 $ & $	6.22	\pm	0.41	 $ & $	0.66	\pm	0.11	 $ & $	0.56	\pm	0.04	 $ & $	0.39	\pm	0.09	 $ & $	0.31	\pm	0.29	$ \\
MACSJ1427.2+4407	&	  216.8174	&	   44.1251	& $	0.477	\pm	0.008	 $ & $	8.57	\pm	0.70	 $ & $	7.36	\pm	0.95	 $ & $	0.89	\pm	0.15	 $ & $	0.55	\pm	0.08	 $ & $	0.46	\pm	0.13	 $ & $	0.45	\pm	0.08	$ \\
MACSJ1427.6-2521	&	  216.9143	&	  -25.3508	& $	0.313	\pm	0.007	 $ & $	5.84	\pm	0.50	 $ & $	4.44	\pm	0.60	 $ & $	0.36	\pm	0.12	 $ & $	0.48	\pm	0.09	 $ & $	0.51	\pm	0.12	 $ & $	0.51	\pm	0.41	$ \\
MACSJ1532.8+3021	&	  233.2244	&	   30.3498	& $	0.360	\pm	0.001	 $ & $	6.40	\pm	0.13	 $ & $	4.97	\pm	0.17	 $ & $	0.91	\pm	0.12	 $ & $	0.49	\pm	0.03	 $ & $	0.40	\pm	0.05	 $ & $	0.39	\pm	0.13	$ \\
MACSJ1621.3+3810	&	  245.3536	&	   38.1690	& $	0.475	\pm	0.007	 $ & $	9.96	\pm	0.58	 $ & $	9.35	\pm	0.86	 $ & $	0.78	\pm	0.16	 $ & $	0.52	\pm	0.06	 $ & $	0.36	\pm	0.09	 $ & $	0.30	\pm	0.12	$ \\
MACSJ1720.2+3536	&	  260.0700	&	   35.6071	& $	0.387	\pm	0.005	 $ & $	7.15	\pm	0.44	 $ & $	5.83	\pm	0.57	 $ & $	0.77	\pm	0.12	 $ & $	0.50	\pm	0.06	 $ & $	0.40	\pm	0.11	 $ & $	0.39	\pm	0.10	$ \\
MACSJ1931.8-2634	&	  292.9569	&	  -26.5760	& $	0.351	\pm	0.002	 $ & $	7.44	\pm	0.22	 $ & $	6.35	\pm	0.30	 $ & $	0.99	\pm	0.16	 $ & $	0.51	\pm	0.03	 $ & $	0.48	\pm	0.08	 $ & $	0.47	\pm	0.12	$ \\
MACSJ2046.0-3430	&	  311.5022	&	  -34.5049	& $	0.425	\pm	0.006	 $ & $	5.15	\pm	0.30	 $ & $	3.40	\pm	0.31	 $ & $	0.49	\pm	0.08	 $ & $	0.45	\pm	0.07	 $ & $	0.29	\pm	0.09	 $ & $	0.27	\pm	0.30	$ \\
MACSJ2129.4-0741	&	  322.3591	&	   -7.6908	& $	0.577	\pm	0.017	 $ & $	9.59	\pm	1.04	 $ & $	8.27	\pm	1.42	 $ & $	1.34	\pm	0.24	 $ & $	0.57	\pm	0.13	 $ & $	0.64	\pm	0.18	 $ & $	0.64	\pm	0.22	$ \\
MACSJ2135.2-0102	&	  323.7976	&	   -1.0479	& $	0.315	\pm	0.012	 $ & $	9.35	\pm	0.87	 $ & $	9.30	\pm	1.36	 $ & $	0.81	\pm	0.14	 $ & $	0.58	\pm	0.12	 $ & $	0.42	\pm	0.14	 $ & $	0.42	\pm	0.47	$ \\
MACSJ2211.7-0349	&	  332.9413	&	   -3.8301	& $	0.347	\pm	0.011	 $ & $	8.79	\pm	0.75	 $ & $	8.28	\pm	1.12	 $ & $	1.48	\pm	0.27	 $ & $	0.53	\pm	0.10	 $ & $	0.44	\pm	0.14	 $ & $	0.44	\pm	0.17	$ \\
MACSJ2214.9-1359	&	  333.7385	&	  -14.0030	& $	0.484	\pm	0.016	 $ & $	8.66	\pm	0.85	 $ & $	7.45	\pm	1.16	 $ & $	1.33	\pm	0.22	 $ & $	0.46	\pm	0.11	 $ & $	0.52	\pm	0.16	 $ & $	0.52	\pm	0.30	$ \\
MACSJ2229.7-2755	&	  337.4382	&	  -27.9264	& $	0.331	\pm	0.005	 $ & $	5.19	\pm	0.25	 $ & $	3.63	\pm	0.28	 $ & $	0.50	\pm	0.13	 $ & $	0.66	\pm	0.07	 $ & $	0.62	\pm	0.13	 $ & $	0.62	\pm	0.22	$ \\
MACSJ2243.3-0935	&	  340.8393	&	   -9.5958	& $	0.443	\pm	0.016	 $ & $	10.48	\pm	0.64	 $ & $	10.34	\pm	1.00	 $ & $	1.61	\pm	0.17	 $ & $	0.28	\pm	0.07	 $ & $	0.24	\pm	0.07	 $ & $	0.24	\pm	0.19	$ \\
MACSJ2245.0+2637	&	  341.2695	&	   26.6345	& $	0.297	\pm	0.007	 $ & $	6.20	\pm	0.52	 $ & $	4.92	\pm	0.65	 $ & $	0.51	\pm	0.20	 $ & $	0.73	\pm	0.12	 $ & $	0.58	\pm	0.17	 $ & $	0.57	\pm	0.32	$ \\
MS0015.9+1609	&	    4.6396	&	   16.4358	& $	0.558	\pm	0.008	 $ & $	8.90	\pm	0.64	 $ & $	7.45	\pm	0.85	 $ & $	1.48	\pm	0.18	 $ & $	0.47	\pm	0.08	 $ & $	0.39	\pm	0.11	 $ & $	0.28	\pm	0.19	$ \\
MS2137.3-2353	&	  325.0633	&	  -23.6612	& $	0.314	\pm	0.001	 $ & $	6.68	\pm	0.14	 $ & $	5.48	\pm	0.19	 $ & $	0.69	\pm	0.11	 $ & $	0.54	\pm	0.03	 $ & $	0.44	\pm	0.06	 $ & $	0.44	\pm	0.21	$ \\
PKS0745-191	&	  116.8798	&	  -19.2946	& $	0.103	\pm	0.003	 $ & $	8.30	\pm	0.30	 $ & $	8.66	\pm	0.49	 $ & $	1.42	\pm	0.31	 $ & $	0.49	\pm	0.04	 $ & $	0.20	\pm	0.04	 $ & $	0.07	\pm	0.19	$ \\
RCS2327.4-0204	&	  351.8653	&	   -2.0772	& $	0.698	\pm	0.018	 $ & $	9.07	\pm	0.36	 $ & $	7.05	\pm	0.45	 $ & $	1.18	\pm	0.16	 $ & $	0.32	\pm	0.06	 $ & $	0.21	\pm	0.08	 $ & $	0.19	\pm	0.41	$ \\
RXJ0043.4-2037	&	   10.8523	&	  -20.6247	& $	0.294	\pm	0.008	 $ & $	8.51	\pm	0.55	 $ & $	8.11	\pm	0.82	 $ & $	1.16	\pm	0.39	 $ & $	0.49	\pm	0.09	 $ & $	0.47	\pm	0.15	 $ & $	0.47	\pm	0.26	$ \\
RXJ0118.1-2658	&	   19.5472	&	  -26.9662	& $	0.218	\pm	0.013	 $ & $	7.61	\pm	0.69	 $ & $	7.11	\pm	1.01	 $ & $	0.91	\pm	0.23	 $ & $	0.49	\pm	0.12	 $ & $	0.40	\pm	0.15	 $ & $	0.38	\pm	0.08	$ \\
RXJ0220.9-3829	&	   35.2357	&	  -38.4802	& $	0.229	\pm	0.006	 $ & $	5.13	\pm	0.35	 $ & $	3.79	\pm	0.41	 $ & $	0.36	\pm	0.14	 $ & $	0.81	\pm	0.13	 $ & $	0.63	\pm	0.21	 $ & $	0.62	\pm	0.30	$ \\
RXJ0232.2-4420	&	   38.0774	&	  -44.3467	& $	0.299	\pm	0.009	 $ & $	11.44	\pm	1.19	 $ & $	12.92	\pm	2.11	 $ & $	1.72	\pm	0.28	 $ & $	0.66	\pm	0.14	 $ & $	0.45	\pm	0.14	 $ & $	0.45	\pm	0.15	$ \\
RXJ0237.4-2630	&	   39.3651	&	  -26.5079	& $	0.224	\pm	0.005	 $ & $	6.51	\pm	0.41	 $ & $	5.52	\pm	0.55	 $ & $	0.69	\pm	0.17	 $ & $	0.73	\pm	0.11	 $ & $	0.68	\pm	0.14	 $ & $	0.68	\pm	0.16	$ \\
RXJ0307.0-2840	&	   46.7582	&	  -28.6657	& $	0.245	\pm	0.008	 $ & $	8.92	\pm	0.95	 $ & $	8.98	\pm	1.52	 $ & $	1.26	\pm	0.25	 $ & $	0.43	\pm	0.10	 $ & $	0.44	\pm	0.16	 $ & $	0.43	\pm	0.26	$ \\
RXJ0331.1-2100	&	   52.7747	&	  -21.0087	& $	0.193	\pm	0.004	 $ & $	5.49	\pm	0.35	 $ & $	4.30	\pm	0.43	 $ & $	0.56	\pm	0.11	 $ & $	0.67	\pm	0.08	 $ & $	0.46	\pm	0.14	 $ & $	0.40	\pm	0.20	$ \\
RXJ0336.3-4037	&	   54.0644	&	  -40.6291	& $	0.176	\pm	0.005	 $ & $	5.77	\pm	0.35	 $ & $	4.70	\pm	0.45	 $ & $	0.56	\pm	0.16	 $ & $	0.59	\pm	0.10	 $ & $	0.62	\pm	0.14	 $ & $	0.62	\pm	0.14	$ \\
RXJ0439.0+0520	&	   69.7592	&	    5.3455	& $	0.203	\pm	0.004	 $ & $	4.88	\pm	0.23	 $ & $	3.55	\pm	0.26	 $ & $	0.38	\pm	0.07	 $ & $	0.66	\pm	0.07	 $ & $	0.57	\pm	0.12	 $ & $	0.56	\pm	0.16	$ \\
RXJ0439.0+0715	&	   69.7529	&	    7.2684	& $	0.254	\pm	0.007	 $ & $	7.25	\pm	0.55	 $ & $	6.45	\pm	0.77	 $ & $	1.08	\pm	0.23	 $ & $	0.54	\pm	0.10	 $ & $	0.48	\pm	0.12	 $ & $	0.46	\pm	0.10	$ \\
RXJ0547.6-3152	&	   86.9058	&	  -31.8688	& $	0.166	\pm	0.012	 $ & $	7.31	\pm	0.48	 $ & $	6.86	\pm	0.71	 $ & $	0.80	\pm	0.24	 $ & $	0.35	\pm	0.08	 $ & $	0.34	\pm	0.11	 $ & $	0.34	\pm	0.31	$ \\
RXJ1144.0+0547	&	  176.0283	&	    5.7982	& $	0.095	\pm	0.007	 $ & $	4.53	\pm	0.34	 $ & $	3.34	\pm	0.39	 $ & $	0.25	\pm	0.13	 $ & $	0.40	\pm	0.10	 $ & $	0.44	\pm	0.19	 $ & $	0.43	\pm	0.26	$ \\
RXJ1459.4-1811	&	  224.8706	&	  -18.1793	& $	0.233	\pm	0.002	 $ & $	6.37	\pm	0.22	 $ & $	5.31	\pm	0.29	 $ & $	0.95	\pm	0.22	 $ & $	0.65	\pm	0.05	 $ & $	0.62	\pm	0.14	 $ & $	0.57	\pm	0.13	$ \\
RXJ1504.1-0248	&	  226.0308	&	   -2.8041	& $	0.219	\pm	0.001	 $ & $	7.52	\pm	0.10	 $ & $	6.97	\pm	0.15	 $ & $	1.14	\pm	0.12	 $ & $	0.41	\pm	0.01	 $ & $	0.40	\pm	0.04	 $ & $	0.39	\pm	0.14	$ \\
RXJ1524.2-3154	&	  231.0534	&	  -31.9061	& $	0.102	\pm	0.001	 $ & $	4.55	\pm	0.13	 $ & $	3.36	\pm	0.15	 $ & $	0.49	\pm	0.13	 $ & $	0.56	\pm	0.03	 $ & $	0.14	\pm	0.07	 $ & $	0.04	\pm	0.18	$ \\
RXJ1558.3-1410	&	  239.5908	&	  -14.1666	& $	0.098	\pm	0.001	 $ & $	5.21	\pm	0.09	 $ & $	4.17	\pm	0.12	 $ & $	0.85	\pm	0.15	 $ & $	0.64	\pm	0.03	 $ & $	0.47	\pm	0.08	 $ & $	0.47	\pm	0.19	$ \\
RXJ1720.1+2638	&	  260.0414	&	   26.6257	& $	0.162	\pm	0.002	 $ & $	6.63	\pm	0.16	 $ & $	5.89	\pm	0.22	 $ & $	0.72	\pm	0.12	 $ & $	0.53	\pm	0.03	 $ & $	0.50	\pm	0.07	 $ & $	0.50	\pm	0.16	$ \\
RXJ1750.2+3505	&	  267.5705	&	   35.0829	& $	0.162	\pm	0.004	 $ & $	5.65	\pm	0.43	 $ & $	4.57	\pm	0.55	 $ & $	0.36	\pm	0.13	 $ & $	0.58	\pm	0.09	 $ & $	0.39	\pm	0.13	 $ & $	0.38	\pm	0.26	$ \\
RXJ2014.8-2430	&	  303.7156	&	  -24.5062	& $	0.153	\pm	0.002	 $ & $	6.74	\pm	0.22	 $ & $	6.07	\pm	0.31	 $ & $	0.88	\pm	0.20	 $ & $	0.53	\pm	0.04	 $ & $	0.38	\pm	0.12	 $ & $	0.37	\pm	0.20	$ \\
RXJ2129.6+0005	&	  322.4158	&	    0.0895	& $	0.246	\pm	0.005	 $ & $	7.68	\pm	0.36	 $ & $	7.10	\pm	0.52	 $ & $	1.19	\pm	0.17	 $ & $	0.46	\pm	0.05	 $ & $	0.45	\pm	0.09	 $ & $	0.45	\pm	0.14	$ \\
SPT-CLJ0000-5748	&	    0.2499	&	  -57.8095	& $	0.695	\pm	0.008	 $ & $	7.06	\pm	0.52	 $ & $	4.75	\pm	0.56	 $ & $	0.56	\pm	0.09	 $ & $	0.62	\pm	0.08	 $ & $	0.43	\pm	0.11	 $ & $	0.39	\pm	0.16	$ \\
SPT-CLJ2043-5035	&	  310.8234	&	  -50.5921	& $	0.724	\pm	0.008	 $ & $	5.75	\pm	0.33	 $ & $	3.37	\pm	0.30	 $ & $	0.62	\pm	0.08	 $ & $	0.48	\pm	0.06	 $ & $	0.42	\pm	0.09	 $ & $	0.42	\pm	0.12	$ \\
SPT-CLJ2331-5051	&	  352.9580	&	  -50.8640	& $	0.599	\pm	0.009	 $ & $	7.98	\pm	0.58	 $ & $	6.11	\pm	0.70	 $ & $	0.76	\pm	0.10	 $ & $	0.43	\pm	0.07	 $ & $	0.16	\pm	0.10	 $ & $	0.12	\pm	0.10	$ \\
SPT-CLJ2344-4242	&	  356.1834	&	  -42.7202	& $	0.602	\pm	0.002	 $ & $	11.52	\pm	0.26	 $ & $	10.89	\pm	0.39	 $ & $	1.71	\pm	0.24	 $ & $	0.56	\pm	0.02	 $ & $	0.54	\pm	0.05	 $ & $	0.54	\pm	0.14	$ \\
Triangulum	&	  249.5710	&	  -64.3579	& $	0.049	\pm	0.001	 $ & $	9.45	\pm	0.15	 $ & $	10.92	\pm	0.28	 $ & $	1.41	\pm	0.78	 $ & $	0.24	\pm	0.02	 $ & $	0.31	\pm	0.09	 $ & $	0.31	\pm	0.31	$ \\
Zwicky0808	&	   45.4091	&	    1.9205	& $	0.172	\pm	0.004	 $ & $	5.08	\pm	0.30	 $ & $	3.84	\pm	0.36	 $ & $	0.36	\pm	0.14	 $ & $	0.52	\pm	0.08	 $ & $	0.36	\pm	0.13	 $ & $	0.35	\pm	0.27	$ \\
Zwicky1358	&	  209.9605	&	   62.5179	& $	0.325	\pm	0.004	 $ & $	10.76	\pm	0.65	 $ & $	11.54	\pm	1.09	 $ & $	1.16	\pm	0.31	 $ & $	0.69	\pm	0.11	 $ & $	0.49	\pm	0.13	 $ & $	0.49	\pm	0.26	$ \\
Zwicky2089	&	  135.1537	&	   20.8943	& $	0.238	\pm	0.002	 $ & $	4.51	\pm	0.16	 $ & $	3.08	\pm	0.17	 $ & $	0.41	\pm	0.07	 $ & $	0.47	\pm	0.04	 $ & $	0.30	\pm	0.09	 $ & $	0.27	\pm	0.13	$ \\
Zwicky2701	&	  148.2050	&	   51.8848	& $	0.214	\pm	0.001	 $ & $	5.15	\pm	0.11	 $ & $	3.84	\pm	0.13	 $ & $	0.74	\pm	0.13	 $ & $	0.57	\pm	0.03	 $ & $	0.43	\pm	0.07	 $ & $	0.42	\pm	0.14	$ \\
Zwicky3146	&	  155.9151	&	    4.1865	& $	0.296	\pm	0.003	 $ & $	7.93	\pm	0.20	 $ & $	7.25	\pm	0.28	 $ & $	1.12	\pm	0.26	 $ & $	0.45	\pm	0.03	 $ & $	0.41	\pm	0.07	 $ & $	0.40	\pm	0.18	$ \\
Zwicky5029	&	  184.4280	&	    3.6610	& $	0.087	\pm	0.006	 $ & $	6.71	\pm	0.26	 $ & $	6.24	\pm	0.39	 $ & $	0.91	\pm	0.30	 $ & $	0.37	\pm	0.07	 $ & $	0.33	\pm	0.11	 $ & $	0.33	\pm	0.20	$ \\
Zwicky7160	&	  224.3128	&	   22.3429	& $	0.258	\pm	0.002	 $ & $	5.15	\pm	0.11	 $ & $	3.74	\pm	0.12	 $ & $	0.47	\pm	0.12	 $ & $	0.53	\pm	0.03	 $ & $	0.41	\pm	0.07	 $ & $	0.38	\pm	0.19	$ \\
\hline
\end{longtable}
\end{landscape}

\end{document}